\documentclass[longauth]{aa}
\usepackage[utf8]{inputenc}
\usepackage{txfonts}

\usepackage{graphicx}
\graphicspath{{Figures/}}
\usepackage{scalerel}
\usepackage[dvipsnames]{xcolor}
\usepackage{booktabs}
\usepackage{multirow}

\usepackage{siunitx}            % Units
\usepackage{natbib}
\usepackage[nolist]{acronym}    % Don't include acronym list nor hyperlinks

\usepackage[pdfencoding=auto,psdextra]{hyperref}
\hypersetup{
    colorlinks=true,
    linkcolor=blue,
    filecolor=magenta,
    urlcolor=blue,
    citecolor=blue
}
\makeatletter
\renewcommand*\aa@pageof{, page \thepage{} of \pageref*{LastPage}}
\makeatother

\usepackage[switch, modulo]{lineno} % needed by euclid.sty
\linenumbers

\usepackage[normalem]{ulem}     % Strike through (\sout)

\usepackage{euclid}             % Euclid commands

\usepackage{orcidlink}          % This has to come after hyperref
\newcommand{\orcid}[1]{\orcidlink{#1}}

\newcommand{\JMASS}{\ensuremath{J_{\sfont{2MASS}}}\xspace}
\newcommand{\XSIR}{\ensuremath{X_{\text{SIR}}}}
\newcommand{\YSIR}{\ensuremath{Y_{\text{SIR}}}}

\defcitealias{Q1-TP006}{SIR-Q1}%
\defcitealias{DR1-TP023}{SIR-DR1}%

\begin{document}

% Title for a Key Project paper
\title{Euclid data release 1 (DR1)}
\subtitle{SIR spectroscopic calibration}    

%% please do not edit the author list once you copy it from the Publication
%% Portal -- contact ECEB Bureau for changes
%\input{OLDFILES/authors_old}
%%%% Version Wednesday 11th of November 2026 03:19:49 PM UT
%%%% Assumes the new A&A style file from Oct 2025 or later
%%%% Please do not edit the author list -- contact ECEB Bureau for changes
%\newcommand{\orcid}[1]{} %% if already defined in aa.cls: comment, or use renewcommand			   
\author{Euclid Collaboration: M.~Fumana\orcid{0000-0001-6787-5950}\thanks{\email{marco.fumana@inaf.it}}\inst{\ref{aff1}}
\and Y.~Copin\orcid{0000-0002-5317-7518}\inst{\ref{aff2}}
\and P.~N.~Appleton\orcid{0000-0002-7607-8766}\inst{\ref{aff3}}
\and J.~Jacobson\inst{\ref{aff3}}
\and D.~C.~Masters\orcid{0000-0001-5382-6138}\inst{\ref{aff4}}
\and S.~Hemmati\orcid{0000-0003-2226-5395}\inst{\ref{aff3}}
\and D.~Tavagnacco\orcid{0000-0001-7475-9894}\inst{\ref{aff5}}
\and A.~Gargiulo\orcid{0000-0002-3351-1216}\inst{\ref{aff1}}
\and C.~Mancini\orcid{0000-0002-4297-0561}\inst{\ref{aff1}}
\and A.~Carle\inst{\ref{aff2}}
\and G.~Herzog\orcid{0000-0001-5909-4054}\inst{\ref{aff1}}
\and A.~Alavi\orcid{0000-0002-8630-6435}\inst{\ref{aff3}}
\and R.~Chary\orcid{0000-0001-7583-0621}\inst{\ref{aff4},\ref{aff6}}
\and S.~Conseil\orcid{0000-0002-3657-4191}\inst{\ref{aff7},\ref{aff2}}
\and I.~Das\orcid{0009-0007-7088-2044}\inst{\ref{aff3}}
\and A.~L.~Faisst\orcid{0000-0002-9382-9832}\inst{\ref{aff3}}
\and W.~Gillard\orcid{0000-0003-4744-9748}\inst{\ref{aff8}}
\and V.~Le~Brun\orcid{0000-0002-5027-1939}\inst{\ref{aff9}}
\and C.~Scarlata\orcid{0000-0002-9136-8876}\inst{\ref{aff10}}
\and N.~R.~Stickley\orcid{0000-0003-0987-5738}\inst{\ref{aff11}}
\and H.~I.~Teplitz\orcid{0000-0002-7064-5424}\inst{\ref{aff4}}
\and Q.~Xie\inst{\ref{aff3}}
\and M.~Scodeggio\inst{\ref{aff1}}
\and B.~Altieri\orcid{0000-0003-3936-0284}\inst{\ref{aff12}}
\and S.~Andreon\orcid{0000-0002-2041-8784}\inst{\ref{aff13}}
\and N.~Auricchio\orcid{0000-0003-4444-8651}\inst{\ref{aff14}}
\and H.~Aussel\orcid{0000-0002-1371-5705}\inst{\ref{aff15}}
\and C.~Baccigalupi\orcid{0000-0002-8211-1630}\inst{\ref{aff16},\ref{aff5},\ref{aff17},\ref{aff18}}
\and M.~Baldi\orcid{0000-0003-4145-1943}\inst{\ref{aff19},\ref{aff14},\ref{aff20}}
\and A.~Balestra\orcid{0000-0002-6967-261X}\inst{\ref{aff21}}
\and S.~Bardelli\orcid{0000-0002-8900-0298}\inst{\ref{aff14}}
\and P.~Battaglia\orcid{0000-0002-7337-5909}\inst{\ref{aff14}}
\and A.~N.~Belikov\inst{\ref{aff22},\ref{aff23}}
\and A.~Biviano\orcid{0000-0002-0857-0732}\inst{\ref{aff5},\ref{aff16}}
\and E.~Branchini\orcid{0000-0002-0808-6908}\inst{\ref{aff24},\ref{aff25},\ref{aff13}}
\and M.~Brescia\orcid{0000-0001-9506-5680}\inst{\ref{aff26},\ref{aff27}}
\and J.~Brinchmann\orcid{0000-0003-4359-8797}\inst{\ref{aff28},\ref{aff29},\ref{aff30}}
\and S.~Camera\orcid{0000-0003-3399-3574}\inst{\ref{aff31},\ref{aff32},\ref{aff33}}
\and G.~Ca\~nas-Herrera\orcid{0000-0003-2796-2149}\inst{\ref{aff34}}
\and V.~Capobianco\orcid{0000-0002-3309-7692}\inst{\ref{aff33}}
\and C.~Carbone\orcid{0000-0003-0125-3563}\inst{\ref{aff1}}
\and J.~Carretero\orcid{0000-0002-3130-0204}\inst{\ref{aff35},\ref{aff36}}
\and F.~J.~Castander\orcid{0000-0001-7316-4573}\inst{\ref{aff37},\ref{aff38}}
\and M.~Castellano\orcid{0000-0001-9875-8263}\inst{\ref{aff39}}
\and G.~Castignani\orcid{0000-0001-6831-0687}\inst{\ref{aff14}}
\and S.~Cavuoti\orcid{0000-0002-3787-4196}\inst{\ref{aff27},\ref{aff40}}
\and K.~C.~Chambers\orcid{0000-0001-6965-7789}\inst{\ref{aff41}}
\and A.~Cimatti\inst{\ref{aff42}}
\and C.~Colodro-Conde\inst{\ref{aff43}}
\and G.~Congedo\orcid{0000-0003-2508-0046}\inst{\ref{aff44}}
\and C.~J.~Conselice\orcid{0000-0003-1949-7638}\inst{\ref{aff45}}
\and L.~Conversi\orcid{0000-0002-6710-8476}\inst{\ref{aff46},\ref{aff12}}
\and A.~Costille\inst{\ref{aff9}}
\and F.~Courbin\orcid{0000-0003-0758-6510}\inst{\ref{aff47},\ref{aff48},\ref{aff49}}
\and H.~M.~Courtois\orcid{0000-0003-0509-1776}\inst{\ref{aff50}}
\and J.-G.~Cuby\orcid{0000-0002-8767-1442}\inst{\ref{aff51},\ref{aff9}}
\and A.~Da~Silva\orcid{0000-0002-6385-1609}\inst{\ref{aff52},\ref{aff53}}
\and R.~da~Silva\orcid{0000-0003-4788-677X}\inst{\ref{aff39},\ref{aff54}}
\and H.~Degaudenzi\orcid{0000-0002-5887-6799}\inst{\ref{aff55}}
\and S.~de~la~Torre\inst{\ref{aff9}}
\and G.~De~Lucia\orcid{0000-0002-6220-9104}\inst{\ref{aff5}}
\and A.~M.~Di~Giorgio\orcid{0000-0002-4767-2360}\inst{\ref{aff56}}
\and H.~Dole\orcid{0000-0002-9767-3839}\inst{\ref{aff57}}
\and F.~Dubath\orcid{0000-0002-6533-2810}\inst{\ref{aff55}}
\and X.~Dupac\inst{\ref{aff12}}
\and S.~Dusini\orcid{0000-0002-1128-0664}\inst{\ref{aff58}}
\and S.~Escoffier\orcid{0000-0002-2847-7498}\inst{\ref{aff8}}
\and M.~Farina\orcid{0000-0002-3089-7846}\inst{\ref{aff56}}
\and R.~Farinelli\inst{\ref{aff14}}
\and S.~Ferriol\inst{\ref{aff2}}
\and S.~Fotopoulou\orcid{0000-0002-9686-254X}\inst{\ref{aff59}}
\and N.~Fourmanoit\orcid{0009-0005-6816-6925}\inst{\ref{aff8}}
\and M.~Frailis\orcid{0000-0002-7400-2135}\inst{\ref{aff5}}
\and E.~Franceschi\orcid{0000-0002-0585-6591}\inst{\ref{aff14}}
\and P.~Franzetti\inst{\ref{aff1}}
\and L.~Gabarra\orcid{0000-0002-8486-8856}\inst{\ref{aff60}}
\and S.~Galeotta\orcid{0000-0002-3748-5115}\inst{\ref{aff5}}
\and K.~George\orcid{0000-0002-1734-8455}\inst{\ref{aff61}}
\and B.~Gillis\orcid{0000-0002-4478-1270}\inst{\ref{aff44}}
\and C.~Giocoli\orcid{0000-0002-9590-7961}\inst{\ref{aff14},\ref{aff20}}
\and J.~Gracia-Carpio\orcid{0000-0003-4689-3134}\inst{\ref{aff62}}
\and B.~R.~Granett\orcid{0000-0003-2694-9284}\inst{\ref{aff13}}
\and A.~Grazian\orcid{0000-0002-5688-0663}\inst{\ref{aff21}}
\and F.~Grupp\inst{\ref{aff62},\ref{aff63}}
\and L.~Guzzo\orcid{0000-0001-8264-5192}\inst{\ref{aff64},\ref{aff13},\ref{aff65}}
\and W.~G.~Hartley\orcid{0000-0002-4563-5101}\inst{\ref{aff55}}
\and S.~V.~H.~Haugan\orcid{0000-0001-9648-7260}\inst{\ref{aff66}}
\and G.~Helou\orcid{0000-0003-3367-3415}\inst{\ref{aff11},\ref{aff3}}
\and H.~Hoekstra\orcid{0000-0002-0641-3231}\inst{\ref{aff34}}
\and W.~Holmes\orcid{0009-0007-8554-4646}\inst{\ref{aff67}}
\and F.~Hormuth\inst{\ref{aff68}}
\and A.~Hornstrup\orcid{0000-0002-3363-0936}\inst{\ref{aff69},\ref{aff70}}
\and M.~Huertas-Company\orcid{0000-0002-1416-8483}\inst{\ref{aff43},\ref{aff71},\ref{aff72}}
\and K.~Jahnke\orcid{0000-0003-3804-2137}\inst{\ref{aff73}}
\and M.~Jhabvala\inst{\ref{aff74}}
\and S.~Kermiche\orcid{0000-0002-0302-5735}\inst{\ref{aff8}}
\and A.~Kiessling\orcid{0000-0002-2590-1273}\inst{\ref{aff67}}
\and R.~Kohley\inst{\ref{aff12}}
\and B.~Kubik\orcid{0009-0006-5823-4880}\inst{\ref{aff2}}
\and M.~K\"ummel\orcid{0000-0003-2791-2117}\inst{\ref{aff63}}
\and M.~Kunz\orcid{0000-0002-3052-7394}\inst{\ref{aff75}}
\and H.~Kurki-Suonio\orcid{0000-0002-4618-3063}\inst{\ref{aff76},\ref{aff77}}
\and A.~M.~C.~Le~Brun\orcid{0000-0002-0936-4594}\inst{\ref{aff78}}
\and S.~Ligori\orcid{0000-0003-4172-4606}\inst{\ref{aff33}}
\and P.~B.~Lilje\orcid{0000-0003-4324-7794}\inst{\ref{aff66}}
\and V.~Lindholm\orcid{0000-0003-2317-5471}\inst{\ref{aff76},\ref{aff77}}
\and I.~Lloro\orcid{0000-0001-5966-1434}\inst{\ref{aff79}}
\and M.~Magliocchetti\orcid{0000-0001-9158-4838}\inst{\ref{aff56}}
\and G.~Mainetti\orcid{0000-0003-2384-2377}\inst{\ref{aff80}}
\and O.~Mansutti\orcid{0000-0001-5758-4658}\inst{\ref{aff5}}
\and S.~Marcin\inst{\ref{aff81}}
\and O.~Marggraf\orcid{0000-0001-7242-3852}\inst{\ref{aff82}}
\and K.~Markovic\orcid{0000-0001-6764-073X}\inst{\ref{aff67}}
\and M.~Martinelli\orcid{0000-0002-6943-7732}\inst{\ref{aff39},\ref{aff83}}
\and N.~Martinet\orcid{0000-0003-2786-7790}\inst{\ref{aff9}}
\and F.~Marulli\orcid{0000-0002-8850-0303}\inst{\ref{aff84},\ref{aff14},\ref{aff20}}
\and R.~J.~Massey\orcid{0000-0002-6085-3780}\inst{\ref{aff85}}
\and N.~Mauri\orcid{0000-0001-8196-1548}\inst{\ref{aff42},\ref{aff20}}
\and E.~Medinaceli\orcid{0000-0002-4040-7783}\inst{\ref{aff14}}
\and S.~Mei\orcid{0000-0002-2849-559X}\inst{\ref{aff86},\ref{aff87}}
\and M.~Meneghetti\orcid{0000-0003-1225-7084}\inst{\ref{aff14},\ref{aff20}}
\and E.~Merlin\orcid{0000-0001-6870-8900}\inst{\ref{aff21}}
\and G.~Meylan\orcid{0000-0001-6503-0209}\inst{\ref{aff88}}
\and P.~Monaco\orcid{0000-0003-2083-7564}\inst{\ref{aff89},\ref{aff5},\ref{aff17},\ref{aff16}}
\and A.~Mora\orcid{0000-0002-1922-8529}\inst{\ref{aff90}}
\and M.~Moresco\orcid{0000-0002-7616-7136}\inst{\ref{aff84},\ref{aff14}}
\and C.~Moretti\orcid{0000-0003-3314-8936}\inst{\ref{aff5},\ref{aff16},\ref{aff17}}
\and L.~Moscardini\orcid{0000-0002-3473-6716}\inst{\ref{aff84},\ref{aff14},\ref{aff20}}
\and C.~Neissner\orcid{0000-0001-8524-4968}\inst{\ref{aff91},\ref{aff36}}
\and R.~C.~Nichol\orcid{0000-0003-0939-6518}\inst{\ref{aff92}}
\and S.-M.~Niemi\orcid{0009-0005-0247-0086}\inst{\ref{aff93}}
\and C.~Padilla\orcid{0000-0001-7951-0166}\inst{\ref{aff91}}
\and S.~Paltani\orcid{0000-0002-8108-9179}\inst{\ref{aff55}}
\and F.~Pasian\orcid{0000-0002-4869-3227}\inst{\ref{aff5}}
\and W.~J.~Percival\orcid{0000-0002-0644-5727}\inst{\ref{aff94},\ref{aff95},\ref{aff96}}
\and V.~Pettorino\orcid{0000-0002-4203-9320}\inst{\ref{aff93}}
\and A.~Pezzotta\orcid{0000-0003-0726-2268}\inst{\ref{aff13}}
\and S.~Pires\orcid{0000-0002-0249-2104}\inst{\ref{aff15}}
\and G.~Polenta\orcid{0000-0003-4067-9196}\inst{\ref{aff54}}
\and M.~Poncet\inst{\ref{aff97}}
\and L.~A.~Popa\inst{\ref{aff98}}
\and L.~Pozzetti\orcid{0000-0001-7085-0412}\inst{\ref{aff14}}
\and F.~Raison\orcid{0000-0002-7819-6918}\inst{\ref{aff62}}
\and A.~Refregier\inst{\ref{aff99}}
\and A.~Renzi\orcid{0000-0001-9856-1970}\inst{\ref{aff100},\ref{aff58},\ref{aff14}}
\and J.~Rhodes\orcid{0000-0002-4485-8549}\inst{\ref{aff67}}
\and G.~Riccio\inst{\ref{aff27}}
\and I.~Risso\orcid{0000-0003-2525-7761}\inst{\ref{aff24},\ref{aff25},\ref{aff13}}
\and F.~Rizzo\orcid{0000-0002-9407-585X}\inst{\ref{aff5}}
\and E.~Romelli\orcid{0000-0003-3069-9222}\inst{\ref{aff5}}
\and M.~Roncarelli\orcid{0000-0001-9587-7822}\inst{\ref{aff14}}
\and C.~Rosset\orcid{0000-0003-0286-2192}\inst{\ref{aff86}}
\and B.~Rusholme\orcid{0000-0001-7648-4142}\inst{\ref{aff3}}
\and R.~Saglia\orcid{0000-0003-0378-7032}\inst{\ref{aff63},\ref{aff62}}
\and Z.~Sakr\orcid{0000-0002-4823-3757}\inst{\ref{aff101},\ref{aff102},\ref{aff103}}
\and A.~G.~S\'anchez\orcid{0000-0003-1198-831X}\inst{\ref{aff62}}
\and D.~Sapone\orcid{0000-0001-7089-4503}\inst{\ref{aff104}}
\and B.~Sartoris\orcid{0000-0003-1337-5269}\inst{\ref{aff105}}
\and M.~Sauvage\orcid{0000-0002-0809-2574}\inst{\ref{aff15}}
\and M.~Schirmer\orcid{0000-0003-2568-9994}\inst{\ref{aff73}}
\and P.~Schneider\orcid{0000-0001-8561-2679}\inst{\ref{aff82}}
\and T.~Schrabback\orcid{0000-0002-6987-7834}\inst{\ref{aff106}}
\and A.~Secroun\orcid{0000-0003-0505-3710}\inst{\ref{aff8}}
\and E.~Sefusatti\orcid{0000-0003-0473-1567}\inst{\ref{aff5},\ref{aff16},\ref{aff17}}
\and E.~Sihvola\orcid{0000-0003-1804-7715}\inst{\ref{aff107}}
\and P.~Simon\inst{\ref{aff82}}
\and C.~Sirignano\orcid{0000-0002-0995-7146}\inst{\ref{aff100},\ref{aff58}}
\and G.~Sirri\orcid{0000-0003-2626-2853}\inst{\ref{aff20}}
\and L.~Stanco\orcid{0000-0002-9706-5104}\inst{\ref{aff58}}
\and P.~Tallada-Cresp\'{i}\orcid{0000-0002-1336-8328}\inst{\ref{aff35},\ref{aff36}}
\and A.~N.~Taylor\inst{\ref{aff44}}
\and I.~Tereno\orcid{0000-0002-4537-6218}\inst{\ref{aff52},\ref{aff53}}
\and N.~Tessore\orcid{0000-0002-9696-7931}\inst{\ref{aff108}}
\and S.~Toft\orcid{0000-0003-3631-7176}\inst{\ref{aff109},\ref{aff110}}
\and R.~Toledo-Moreo\orcid{0000-0002-2997-4859}\inst{\ref{aff111},\ref{aff112}}
\and F.~Torradeflot\orcid{0000-0003-1160-1517}\inst{\ref{aff36},\ref{aff35}}
\and I.~Tutusaus\orcid{0000-0002-3199-0399}\inst{\ref{aff37},\ref{aff38},\ref{aff102}}
\and J.~Valiviita\orcid{0000-0001-6225-3693}\inst{\ref{aff76},\ref{aff77}}
\and T.~Vassallo\orcid{0000-0001-6512-6358}\inst{\ref{aff5},\ref{aff61}}
\and A.~Veropalumbo\orcid{0000-0003-2387-1194}\inst{\ref{aff13},\ref{aff25},\ref{aff24}}
\and D.~Vibert\orcid{0009-0008-0607-631X}\inst{\ref{aff9}}
\and Y.~Wang\orcid{0000-0002-4749-2984}\inst{\ref{aff3}}
\and J.~Weller\orcid{0000-0002-8282-2010}\inst{\ref{aff63},\ref{aff62}}
\and A.~Zacchei\orcid{0000-0003-0396-1192}\inst{\ref{aff5},\ref{aff16}}
\and G.~Zamorani\orcid{0000-0002-2318-301X}\inst{\ref{aff14}}
\and F.~M.~Zerbi\orcid{0000-0002-9996-973X}\inst{\ref{aff13}}
\and I.~A.~Zinchenko\orcid{0000-0002-2944-2449}\inst{\ref{aff113}}
\and E.~Zucca\orcid{0000-0002-5845-8132}\inst{\ref{aff14}}
\and V.~Allevato\orcid{0000-0001-7232-5152}\inst{\ref{aff27}}
\and I.~T.~Andika\orcid{0000-0001-6102-9526}\inst{\ref{aff63}}
\and M.~Ballardini\orcid{0000-0003-4481-3559}\inst{\ref{aff114},\ref{aff115},\ref{aff14}}
\and P.~Bergamini\orcid{0000-0003-1383-9414}\inst{\ref{aff14}}
\and S.~Borgani\orcid{0000-0001-6151-6439}\inst{\ref{aff89},\ref{aff16},\ref{aff5},\ref{aff17},\ref{aff116}}
\and E.~Bozzo\orcid{0000-0002-8201-1525}\inst{\ref{aff55}}
\and C.~Burigana\orcid{0000-0002-3005-5796}\inst{\ref{aff117},\ref{aff118}}
\and R.~Cabanac\orcid{0000-0001-6679-2600}\inst{\ref{aff102}}
\and M.~Calabrese\orcid{0000-0002-2637-2422}\inst{\ref{aff119},\ref{aff1}}
\and A.~Cappi\inst{\ref{aff105},\ref{aff14}}
\and F.~Caro\orcid{0009-0003-1053-0507}\inst{\ref{aff39}}
\and T.~Castro\orcid{0000-0002-6292-3228}\inst{\ref{aff120},\ref{aff5},\ref{aff16}}
\and G.~Daste\inst{\ref{aff9}}
\and D.~Di~Ferdinando\inst{\ref{aff20}}
\and J.~A.~Escartin~Vigo\inst{\ref{aff62}}
\and G.~Fabbian\orcid{0000-0002-3255-4695}\inst{\ref{aff57}}
\and J.~Garc\'ia-Bellido\orcid{0000-0002-9370-8360}\inst{\ref{aff101}}
\and T.~Gasparetto\orcid{0000-0002-7913-4866}\inst{\ref{aff39}}
\and A.~Gregorio\orcid{0000-0003-4028-8785}\inst{\ref{aff89},\ref{aff5},\ref{aff17}}
\and C.~M.~Gutierrez\orcid{0000-0001-7854-783X}\inst{\ref{aff43},\ref{aff121}}
\and A.~Hall\orcid{0000-0002-3139-8651}\inst{\ref{aff44}}
\and Y.~Kang\orcid{0009-0000-8588-7250}\inst{\ref{aff106}}
\and A.~Loureiro\orcid{0000-0002-4371-0876}\inst{\ref{aff122},\ref{aff123}}
\and J.~Macias-Perez\orcid{0000-0002-5385-2763}\inst{\ref{aff124}}
\and R.~Maoli\orcid{0000-0002-6065-3025}\inst{\ref{aff125},\ref{aff39}}
\and R.~B.~Metcalf\orcid{0000-0003-3167-2574}\inst{\ref{aff84},\ref{aff14}}
\and M.~P\"ontinen\orcid{0000-0001-5442-2530}\inst{\ref{aff76}}
\and E.~Sarpa\orcid{0000-0002-1256-655X}\inst{\ref{aff5}}
\and V.~Scottez\orcid{0009-0008-3864-940X}\inst{\ref{aff126},\ref{aff127}}
\and M.~Sereno\orcid{0000-0003-0302-0325}\inst{\ref{aff14},\ref{aff20}}
\and K.~Tanidis\orcid{0000-0001-9843-5130}\inst{\ref{aff128}}
\and M.~Tenti\orcid{0000-0002-4254-5901}\inst{\ref{aff20}}
\and M.~Tucci\inst{\ref{aff55}}
\and M.~Viel\orcid{0000-0002-2642-5707}\inst{\ref{aff16},\ref{aff5},\ref{aff18},\ref{aff17},\ref{aff116}}
\and M.~Wiesmann\orcid{0009-0000-8199-5860}\inst{\ref{aff66}}
\and J.~A.~Acevedo~Barroso\orcid{0000-0002-9654-1711}\inst{\ref{aff67}}
\and Y.~Akrami\orcid{0000-0002-2407-7956}\inst{\ref{aff101},\ref{aff129}}
\and G.~Angora\orcid{0000-0002-0316-6562}\inst{\ref{aff27},\ref{aff114}}
\and S.~Anselmi\orcid{0000-0002-3579-9583}\inst{\ref{aff58},\ref{aff100},\ref{aff130}}
\and M.~Archidiacono\orcid{0000-0003-4952-9012}\inst{\ref{aff64},\ref{aff65}}
\and F.~Atrio-Barandela\orcid{0000-0002-2130-2513}\inst{\ref{aff131}}
\and M.~Aubert\orcid{0009-0002-7667-8814}\inst{\ref{aff2}}
\and M.~Baes\orcid{0000-0002-3930-2757}\inst{\ref{aff132}}
\and L.~Bazzanini\orcid{0000-0003-0727-0137}\inst{\ref{aff114},\ref{aff14}}
\and J.~R.~Bermejo-Climent\inst{\ref{aff133},\ref{aff134}}
\and D.~Bertacca\orcid{0000-0002-2490-7139}\inst{\ref{aff100},\ref{aff21},\ref{aff58}}
\and M.~Bethermin\orcid{0000-0002-3915-2015}\inst{\ref{aff135}}
\and F.~Beutler\orcid{0000-0003-0467-5438}\inst{\ref{aff44}}
\and A.~Blanchard\orcid{0000-0001-8555-9003}\inst{\ref{aff102}}
\and F.~Bouch\`e\orcid{0000-0002-4663-1786}\inst{\ref{aff136},\ref{aff40}}
\and M.~L.~Brown\orcid{0000-0002-0370-8077}\inst{\ref{aff45}}
\and S.~Bruton\orcid{0000-0002-6503-5218}\inst{\ref{aff11}}
\and A.~Calabro\orcid{0000-0003-2536-1614}\inst{\ref{aff39}}
\and B.~Camacho~Quevedo\orcid{0000-0002-8789-4232}\inst{\ref{aff16},\ref{aff18},\ref{aff5}}
\and P.~Carrilho\orcid{0000-0003-1339-0194}\inst{\ref{aff137}}
\and C.~S.~Carvalho\inst{\ref{aff138}}
\and Y.~Charles\orcid{0009-0000-3636-834X}\inst{\ref{aff9}}
\and F.~Cogato\orcid{0000-0003-4632-6113}\inst{\ref{aff14}}
\and T.~E.~Collett\orcid{0000-0001-5564-3140}\inst{\ref{aff139}}
\and A.~R.~Cooray\orcid{0000-0002-3892-0190}\inst{\ref{aff140}}
\and P.~Corcho-Caballero\orcid{0000-0001-6327-7080}\inst{\ref{aff22}}
\and M.~Correnti\orcid{0000-0001-6464-3257}\inst{\ref{aff39},\ref{aff54}}
\and B.~Csizi\orcid{0000-0003-3227-6581}\inst{\ref{aff106}}
\and O.~Cucciati\orcid{0000-0002-9336-7551}\inst{\ref{aff14}}
\and S.~Davini\orcid{0000-0003-3269-1718}\inst{\ref{aff25}}
\and T.~de~Boer\orcid{0000-0001-5486-2747}\inst{\ref{aff41}}
\and F.~De~Paolis\orcid{0000-0001-6460-7563}\inst{\ref{aff141},\ref{aff142},\ref{aff143}}
\and G.~Desprez\orcid{0000-0001-8325-1742}\inst{\ref{aff22}}
\and A.~D\'iaz-S\'anchez\orcid{0000-0003-0748-4768}\inst{\ref{aff144}}
\and S.~Di~Domizio\orcid{0000-0003-2863-5895}\inst{\ref{aff24},\ref{aff25}}
\and J.~M.~Diego\orcid{0000-0001-9065-3926}\inst{\ref{aff145}}
\and F.~Dufresne\orcid{0009-0009-0580-7886}\inst{\ref{aff9}}
\and V.~Duret\orcid{0009-0009-0383-4960}\inst{\ref{aff8}}
\and A.~Enia\orcid{0000-0002-0200-2857}\inst{\ref{aff146},\ref{aff14}}
\and Y.~Fang\orcid{0000-0002-0334-6950}\inst{\ref{aff63}}
\and A.~Farina\orcid{0009-0000-3420-929X}\inst{\ref{aff13},\ref{aff25}}
\and A.~M.~N.~Ferguson\orcid{0000-0001-7934-1278}\inst{\ref{aff44}}
\and A.~Finoguenov\orcid{0000-0002-4606-5403}\inst{\ref{aff76}}
\and A.~Franco\orcid{0000-0002-4761-366X}\inst{\ref{aff141},\ref{aff142},\ref{aff143}}
\and A.~Fumagalli\orcid{0009-0004-0300-2535}\inst{\ref{aff5}}
\and Y.~Fu\orcid{0000-0002-0759-0504}\inst{\ref{aff34},\ref{aff22}}
\and K.~Ganga\orcid{0000-0001-8159-8208}\inst{\ref{aff86}}
\and E.~Gaztanaga\orcid{0000-0001-9632-0815}\inst{\ref{aff37},\ref{aff38},\ref{aff139}}
\and Z.~Ghaffari\orcid{0000-0002-6467-8078}\inst{\ref{aff5},\ref{aff16}}
\and F.~Giacomini\orcid{0000-0002-3129-2814}\inst{\ref{aff20}}
\and F.~Gianotti\orcid{0000-0003-4666-119X}\inst{\ref{aff14}}
\and G.~Gozaliasl\orcid{0000-0002-0236-919X}\inst{\ref{aff147},\ref{aff76}}
\and A.~Gruppuso\orcid{0000-0001-9272-5292}\inst{\ref{aff14},\ref{aff20}}
\and M.~Guidi\orcid{0000-0001-9408-1101}\inst{\ref{aff19},\ref{aff14}}
\and C.~Hern\'andez-Monteagudo\orcid{0000-0001-5471-9166}\inst{\ref{aff121},\ref{aff43}}
\and T.~Heuberger\orcid{0009-0009-3474-8467}\inst{\ref{aff81}}
\and H.~Hildebrandt\orcid{0000-0002-9814-3338}\inst{\ref{aff148}}
\and J.~Hjorth\orcid{0000-0002-4571-2306}\inst{\ref{aff149}}
\and S.~Hosseini\inst{\ref{aff102}}
\and M.~Jauzac\orcid{0000-0003-1974-8732}\inst{\ref{aff150},\ref{aff85},\ref{aff151},\ref{aff152},\ref{aff102}}
\and S.~Joudaki\orcid{0000-0001-8820-673X}\inst{\ref{aff35},\ref{aff139}}
\and J.~J.~E.~Kajava\orcid{0000-0002-3010-8333}\inst{\ref{aff153},\ref{aff154},\ref{aff155}}
\and V.~Kansal\orcid{0000-0002-4008-6078}\inst{\ref{aff156},\ref{aff157}}
\and D.~Karagiannis\orcid{0000-0002-4927-0816}\inst{\ref{aff114},\ref{aff158},\ref{aff159}}
\and J.~Kim\orcid{0000-0003-2776-2761}\inst{\ref{aff60}}
\and C.~C.~Kirkpatrick\inst{\ref{aff107}}
\and A.~Kov\'acs\orcid{0000-0002-5825-579X}\inst{\ref{aff134},\ref{aff133}}
\and I.~Kova{\v{c}}i{\'{c}}\orcid{0000-0001-6751-3263}\inst{\ref{aff132}}
\and K.~Koyama\orcid{0000-0001-6727-6915}\inst{\ref{aff139}}
\and S.~Kruk\orcid{0000-0001-8010-8879}\inst{\ref{aff12}}
\and M.~C.~Lam\orcid{0000-0002-9347-2298}\inst{\ref{aff44}}
\and M.~Lattanzi\orcid{0000-0003-1059-2532}\inst{\ref{aff115}}
\and S.~Lee\orcid{0000-0002-8289-740X}\inst{\ref{aff160},\ref{aff67}}
\and L.~Legrand\orcid{0000-0003-0610-5252}\inst{\ref{aff161},\ref{aff57}}
\and M.~Lembo\orcid{0000-0002-5271-5070}\inst{\ref{aff162}}
\and M.~D.~Lepinzan\orcid{0000-0003-1287-9801}\inst{\ref{aff5}}
\and J.~Lesgourgues\orcid{0000-0001-7627-353X}\inst{\ref{aff163}}
\and T.~I.~Liaudat\orcid{0000-0002-9104-314X}\inst{\ref{aff164}}
\and L.~Linke\orcid{0000-0002-2622-8113}\inst{\ref{aff106}}
\and S.~J.~Liu\orcid{0000-0001-7680-2139}\inst{\ref{aff56}}
\and G.~Maggio\orcid{0000-0003-4020-4836}\inst{\ref{aff5}}
\and F.~Mannucci\orcid{0000-0002-4803-2381}\inst{\ref{aff165}}
\and F.~R.~Marleau\orcid{0000-0002-1442-2947}\inst{\ref{aff106}}
\and C.~J.~A.~P.~Martins\orcid{0000-0002-4886-9261}\inst{\ref{aff166},\ref{aff29}}
\and M.~Miluzio\inst{\ref{aff12},\ref{aff167}}
\and S.~Nadathur\orcid{0000-0001-9070-3102}\inst{\ref{aff139}}
\and K.~Naidoo\orcid{0000-0002-9182-1802}\inst{\ref{aff139},\ref{aff73}}
\and P.~Natoli\orcid{0000-0003-0126-9100}\inst{\ref{aff114},\ref{aff115}}
\and A.~Navarro-Alsina\orcid{0000-0002-3173-2592}\inst{\ref{aff82}}
\and S.~Nesseris\orcid{0000-0002-0567-0324}\inst{\ref{aff101}}
\and L.~Nicastro\orcid{0000-0001-8534-6788}\inst{\ref{aff14}}
\and F.~Oppizzi\orcid{0000-0003-3904-8370}\inst{\ref{aff25},\ref{aff58},\ref{aff100}}
\and F.~Pace\orcid{0000-0001-8039-0480}\inst{\ref{aff31},\ref{aff32},\ref{aff33},\ref{aff53}}
\and E.~Palazzi\orcid{0000-0002-8691-7666}\inst{\ref{aff14}}
\and D.~Paoletti\orcid{0000-0003-4761-6147}\inst{\ref{aff14},\ref{aff118}}
\and F.~Passalacqua\orcid{0000-0002-8606-4093}\inst{\ref{aff58}}
\and K.~Paterson\orcid{0000-0001-8340-3486}\inst{\ref{aff73}}
\and L.~Patrizii\inst{\ref{aff20}}
\and C.~Pattison\orcid{0000-0003-3272-2617}\inst{\ref{aff139}}
\and A.~Pisani\orcid{0000-0002-6146-4437}\inst{\ref{aff8}}
\and L.~Pizzuti\orcid{0000-0001-5654-7580}\inst{\ref{aff168}}
\and D.~Potter\orcid{0000-0002-0757-5195}\inst{\ref{aff169}}
\and M.~Poulain\orcid{0000-0002-7664-4510}\inst{\ref{aff170}}
\and G.~W.~Pratt\inst{\ref{aff15}}
\and S.~Quai\orcid{0000-0002-0449-8163}\inst{\ref{aff84},\ref{aff14}}
\and M.~Radovich\orcid{0000-0002-3585-866X}\inst{\ref{aff21}}
\and F.~Ricci\orcid{0000-0001-5742-5980}\inst{\ref{aff171},\ref{aff39}}
\and G.~Rodighiero\orcid{0000-0002-9415-2296}\inst{\ref{aff100},\ref{aff21}}
\and W.~Roster\orcid{0000-0002-9149-6528}\inst{\ref{aff62}}
\and M.~Sahl\'en\orcid{0000-0003-0973-4804}\inst{\ref{aff172}}
\and D.~B.~Sanders\orcid{0000-0002-1233-9998}\inst{\ref{aff41}}
\and A.~Schneider\orcid{0000-0001-7055-8104}\inst{\ref{aff169}}
\and M.~Schultheis\inst{\ref{aff105}}
\and D.~Sciotti\orcid{0009-0008-4519-2620}\inst{\ref{aff39},\ref{aff83}}
\and E.~Sellentin\orcid{0009-0002-2655-3458}\inst{\ref{aff173},\ref{aff34}}
\and S.~Serjeant\orcid{0000-0002-0517-7943}\inst{\ref{aff174}}
\and L.~C.~Smith\orcid{0000-0002-3259-2771}\inst{\ref{aff175}}
\and J.~G.~Sorce\orcid{0000-0002-2307-2432}\inst{\ref{aff176},\ref{aff57}}
\and I.~Szapudi\orcid{0000-0003-2274-0301}\inst{\ref{aff41}}
\and M.~Talia\orcid{0000-0003-4352-2063}\inst{\ref{aff84},\ref{aff14}}
\and G.~Testera\orcid{0000-0003-2970-766X}\inst{\ref{aff25}}
\and R.~Teyssier\orcid{0000-0001-7689-0933}\inst{\ref{aff177}}
\and S.~Tosi\orcid{0000-0002-7275-9193}\inst{\ref{aff24},\ref{aff13},\ref{aff25}}
\and A.~Troja\orcid{0000-0003-0239-4595}\inst{\ref{aff5}}
\and C.~Uhlemann\orcid{0000-0001-7831-1579}\inst{\ref{aff178},\ref{aff179}}
\and C.~Valieri\inst{\ref{aff20}}
\and A.~Venhola\orcid{0000-0001-6071-4564}\inst{\ref{aff170}}
\and D.~Vergani\orcid{0000-0003-0898-2216}\inst{\ref{aff14}}
\and E.~Vilenius\orcid{0000-0002-6184-7681}\inst{\ref{aff108}}
\and S.~Vinciguerra\orcid{0009-0005-4018-3184}\inst{\ref{aff9}}
\and M.~von~Wietersheim-Kramsta\orcid{0000-0003-4986-5091}\inst{\ref{aff85},\ref{aff150}}
\and N.~A.~Walton\orcid{0000-0003-3983-8778}\inst{\ref{aff175}}
\and A.~H.~Wright\orcid{0000-0001-7363-7932}\inst{\ref{aff148}}
\and L.~Zalesky\orcid{0000-0001-5680-2326}\inst{\ref{aff41}}
\and D.~Scott\orcid{0000-0002-6878-9840}\inst{\ref{aff180}}}
										   
%%%% please do not edit the affiliation list -- contact ECEB Bureau for changes
\institute{INAF-IASF Milano, Via Alfonso Corti 12, 20133 Milano, Italy\label{aff1}
\and
Universit\'e Lyon 1, CNRS/IN2P3, IP2I Lyon, Villeurbanne, France\label{aff2}
\and
Caltech/IPAC, 1200 E. California Blvd., Pasadena, CA 91125, USA\label{aff3}
\and
Infrared Processing and Analysis Center, California Institute of Technology, Pasadena, CA 91125, USA\label{aff4}
\and
INAF-Osservatorio Astronomico di Trieste, Via G. B. Tiepolo 11, 34143 Trieste, Italy\label{aff5}
\and
University of California, Los Angeles, CA 90095-1562, USA\label{aff6}
\and
Centre de Recherche Astrophysique de Lyon, UMR5574, CNRS, Universit\'e Claude Bernard Lyon 1, ENS de Lyon, 69230, Saint-Genis-Laval, France\label{aff7}
\and
Aix-Marseille Universit\'e, CNRS/IN2P3, CPPM, Marseille, France\label{aff8}
\and
Aix-Marseille Universit\'e, CNRS, CNES, LAM, Marseille, France\label{aff9}
\and
Minnesota Institute for Astrophysics, University of Minnesota, 116 Church St SE, Minneapolis, MN 55455, USA\label{aff10}
\and
California Institute of Technology, 1200 E California Blvd, Pasadena, CA 91125, USA\label{aff11}
\and
ESAC/ESA, Camino Bajo del Castillo, s/n., Urb. Villafranca del Castillo, 28692 Villanueva de la Ca\~nada, Madrid, Spain\label{aff12}
\and
INAF-Osservatorio Astronomico di Brera, Via Brera 28, 20122 Milano, Italy\label{aff13}
\and
INAF-Osservatorio di Astrofisica e Scienza dello Spazio di Bologna, Via Piero Gobetti 93/3, 40129 Bologna, Italy\label{aff14}
\and
Universit\'e Paris-Saclay, Universit\'e Paris Cit\'e, CEA, CNRS, AIM, 91191, Gif-sur-Yvette, France\label{aff15}
\and
IFPU, Institute for Fundamental Physics of the Universe, via Beirut 2, 34151 Trieste, Italy\label{aff16}
\and
INFN, Sezione di Trieste, Via Valerio 2, 34127 Trieste TS, Italy\label{aff17}
\and
SISSA, International School for Advanced Studies, Via Bonomea 265, 34136 Trieste TS, Italy\label{aff18}
\and
Dipartimento di Fisica e Astronomia, Universit\`a di Bologna, Via Gobetti 93/2, 40129 Bologna, Italy\label{aff19}
\and
INFN-Sezione di Bologna, Viale Berti Pichat 6/2, 40127 Bologna, Italy\label{aff20}
\and
INAF-Osservatorio Astronomico di Padova, Via dell'Osservatorio 5, 35122 Padova, Italy\label{aff21}
\and
Kapteyn Astronomical Institute, University of Groningen, PO Box 800, 9700 AV Groningen, The Netherlands\label{aff22}
\and
ATG Europe BV, Huygensstraat 34, 2201 DK Noordwijk, The Netherlands\label{aff23}
\and
Dipartimento di Fisica, Universit\`a di Genova, Via Dodecaneso 33, 16146, Genova, Italy\label{aff24}
\and
INFN-Sezione di Genova, Via Dodecaneso 33, 16146, Genova, Italy\label{aff25}
\and
Department of Physics "E. Pancini", University Federico II, Via Cinthia 6, 80126, Napoli, Italy\label{aff26}
\and
INAF-Osservatorio Astronomico di Capodimonte, Via Moiariello 16, 80131 Napoli, Italy\label{aff27}
\and
European Southern Observatory, Karl-Schwarzschild-Str.~2, 85748 Garching, Germany\label{aff28}
\and
Instituto de Astrof\'isica e Ci\^encias do Espa\c{c}o, Universidade do Porto, CAUP, Rua das Estrelas, PT4150-762 Porto, Portugal\label{aff29}
\and
Faculdade de Ci\^encias da Universidade do Porto, Rua do Campo de Alegre, 4150-007 Porto, Portugal\label{aff30}
\and
Dipartimento di Fisica, Universit\`a degli Studi di Torino, Via P. Giuria 1, 10125 Torino, Italy\label{aff31}
\and
INFN-Sezione di Torino, Via P. Giuria 1, 10125 Torino, Italy\label{aff32}
\and
INAF-Osservatorio Astrofisico di Torino, Via Osservatorio 20, 10025 Pino Torinese (TO), Italy\label{aff33}
\and
Leiden Observatory, Leiden University, Einsteinweg 55, 2333 CC Leiden, The Netherlands\label{aff34}
\and
Centro de Investigaciones Energ\'eticas, Medioambientales y Tecnol\'ogicas (CIEMAT), Avenida Complutense 40, 28040 Madrid, Spain\label{aff35}
\and
Port d'Informaci\'{o} Cient\'{i}fica, Campus UAB, C. Albareda s/n, 08193 Bellaterra (Barcelona), Spain\label{aff36}
\and
Institute of Space Sciences (ICE, CSIC), Campus UAB, Carrer de Can Magrans, s/n, 08193 Barcelona, Spain\label{aff37}
\and
Institut d'Estudis Espacials de Catalunya (IEEC),  Edifici RDIT, Campus UPC, 08860 Castelldefels, Barcelona, Spain\label{aff38}
\and
INAF-Osservatorio Astronomico di Roma, Via Frascati 33, 00078 Monteporzio Catone, Italy\label{aff39}
\and
INFN -- Sezione di Napoli, Via Cinthia 6, 80126, Napoli, Italy\label{aff40}
\and
Institute for Astronomy, University of Hawaii, 2680 Woodlawn Drive, Honolulu, HI 96822, USA\label{aff41}
\and
Dipartimento di Fisica e Astronomia "Augusto Righi" - Alma Mater Studiorum Universit\`a di Bologna, Viale Berti Pichat 6/2, 40127 Bologna, Italy\label{aff42}
\and
Instituto de Astrof\'{\i}sica de Canarias, E-38205 La Laguna, Tenerife, Spain\label{aff43}
\and
Institute for Astronomy, University of Edinburgh, Royal Observatory, Blackford Hill, Edinburgh EH9 3HJ, UK\label{aff44}
\and
Jodrell Bank Centre for Astrophysics, Department of Physics and Astronomy, University of Manchester, Oxford Road, Manchester M13 9PL, UK\label{aff45}
\and
European Space Agency/ESRIN, Largo Galileo Galilei 1, 00044 Frascati, Roma, Italy\label{aff46}
\and
Institut de Ci\`{e}ncies del Cosmos (ICCUB), Universitat de Barcelona (IEEC-UB), Mart\'{i} i Franqu\`{e}s 1, 08028 Barcelona, Spain\label{aff47}
\and
Instituci\'o Catalana de Recerca i Estudis Avan\c{c}ats (ICREA), Passeig de Llu\'{\i}s Companys 23, 08010 Barcelona, Spain\label{aff48}
\and
Institut de Ciencies de l'Espai (IEEC-CSIC), Campus UAB, Carrer de Can Magrans, s/n Cerdanyola del Vall\'es, 08193 Barcelona, Spain\label{aff49}
\and
Universit\'e Lyon 1, CNRS/IN2P3, IUF, IP2I Lyon, Villeurbanne, France \label{aff50}
\and
Canada-France-Hawaii Telescope, 65-1238 Mamalahoa Hwy, Kamuela, HI 96743, USA\label{aff51}
\and
Departamento de F\'isica, Faculdade de Ci\^encias, Universidade de Lisboa, Edif\'icio C8, Campo Grande, PT1749-016 Lisboa, Portugal\label{aff52}
\and
Instituto de Astrof\'isica e Ci\^encias do Espa\c{c}o, Faculdade de Ci\^encias, Universidade de Lisboa, Campo Grande, 1749-016 Lisboa, Portugal\label{aff53}
\and
Space Science Data Center, Italian Space Agency, via del Politecnico snc, 00133 Roma, Italy\label{aff54}
\and
Department of Astronomy, University of Geneva, ch. d'Ecogia 16, 1290 Versoix, Switzerland\label{aff55}
\and
INAF-Istituto di Astrofisica e Planetologia Spaziali, via del Fosso del Cavaliere, 100, 00100 Roma, Italy\label{aff56}
\and
Universit\'e Paris-Saclay, CNRS, Institut d'astrophysique spatiale, 91405, Orsay, France\label{aff57}
\and
INFN-Padova, Via Marzolo 8, 35131 Padova, Italy\label{aff58}
\and
School of Physics, HH Wills Physics Laboratory, University of Bristol, Tyndall Avenue, Bristol, BS8 1TL, UK\label{aff59}
\and
Department of Physics, University of Oxford, Keble Road, Oxford OX1 3RH, UK\label{aff60}
\and
University Observatory, LMU Faculty of Physics, Scheinerstr.~1, 81679 Munich, Germany\label{aff61}
\and
Max Planck Institute for Extraterrestrial Physics, Giessenbachstr. 1, 85748 Garching, Germany\label{aff62}
\and
Universit\"ats-Sternwarte M\"unchen, Fakult\"at f\"ur Physik, Ludwig-Maximilians-Universit\"at M\"unchen, Scheinerstr.~1, 81679 M\"unchen, Germany\label{aff63}
\and
Dipartimento di Fisica "Aldo Pontremoli", Universit\`a degli Studi di Milano, Via Celoria 16, 20133 Milano, Italy\label{aff64}
\and
INFN-Sezione di Milano, Via Celoria 16, 20133 Milano, Italy\label{aff65}
\and
Institute of Theoretical Astrophysics, University of Oslo, P.O. Box 1029 Blindern, 0315 Oslo, Norway\label{aff66}
\and
Jet Propulsion Laboratory, California Institute of Technology, 4800 Oak Grove Drive, Pasadena, CA, 91109, USA\label{aff67}
\and
Felix Hormuth Engineering, Goethestr. 17, 69181 Leimen, Germany\label{aff68}
\and
Technical University of Denmark, Elektrovej 327, 2800 Kgs. Lyngby, Denmark\label{aff69}
\and
Cosmic Dawn Center (DAWN), Denmark\label{aff70}
\and
Universit\'e PSL, Observatoire de Paris, Sorbonne Universit\'e, CNRS, LERMA, 75014, Paris, France\label{aff71}
\and
Universit\'e Paris-Cit\'e, 5 Rue Thomas Mann, 75013, Paris, France\label{aff72}
\and
Max-Planck-Institut f\"ur Astronomie, K\"onigstuhl 17, 69117 Heidelberg, Germany\label{aff73}
\and
NASA Goddard Space Flight Center, Greenbelt, MD 20771, USA\label{aff74}
\and
Universit\'e de Gen\`eve, D\'epartement de Physique Th\'eorique and Centre for Astroparticle Physics, 24 quai Ernest-Ansermet, CH-1211 Gen\`eve 4, Switzerland\label{aff75}
\and
Department of Physics, P.O. Box 64, University of Helsinki, 00014 Helsinki, Finland\label{aff76}
\and
Helsinki Institute of Physics, Gustaf H{\"a}llstr{\"o}min katu 2, University of Helsinki, 00014 Helsinki, Finland\label{aff77}
\and
Laboratoire d'etude de l'Univers et des phenomenes eXtremes, Observatoire de Paris, Universit\'e PSL, Sorbonne Universit\'e, CNRS, 92190 Meudon, France\label{aff78}
\and
SKAO, Jodrell Bank, Lower Withington, Macclesfield SK11 9FT, UK\label{aff79}
\and
Centre de Calcul de l'IN2P3/CNRS, 21 avenue Pierre de Coubertin 69627 Villeurbanne Cedex, France\label{aff80}
\and
University of Applied Sciences and Arts of Northwestern Switzerland, School of Computer Science, 5210 Windisch, Switzerland\label{aff81}
\and
Universit\"at Bonn, Argelander-Institut f\"ur Astronomie, Auf dem H\"ugel 71, 53121 Bonn, Germany\label{aff82}
\and
INFN-Sezione di Roma, Piazzale Aldo Moro, 2 - c/o Dipartimento di Fisica, Edificio G. Marconi, 00185 Roma, Italy\label{aff83}
\and
Dipartimento di Fisica e Astronomia "Augusto Righi" - Alma Mater Studiorum Universit\`a di Bologna, via Piero Gobetti 93/2, 40129 Bologna, Italy\label{aff84}
\and
Department of Physics, Institute for Computational Cosmology, Durham University, South Road, Durham, DH1 3LE, UK\label{aff85}
\and
Universit\'e Paris Cit\'e, CNRS, Astroparticule et Cosmologie, 75013 Paris, France\label{aff86}
\and
CNRS-UCB International Research Laboratory, Centre Pierre Bin\'etruy, IRL2007, CPB-IN2P3, Berkeley, USA\label{aff87}
\and
Institute of Physics, Laboratory of Astrophysics, Ecole Polytechnique F\'ed\'erale de Lausanne (EPFL), Observatoire de Sauverny, 1290 Versoix, Switzerland\label{aff88}
\and
Dipartimento di Fisica - Sezione di Astronomia, Universit\`a di Trieste, Via Tiepolo 11, 34131 Trieste, Italy\label{aff89}
\and
Telespazio UK S.L. for European Space Agency (ESA), Camino bajo del Castillo, s/n, Urbanizacion Villafranca del Castillo, Villanueva de la Ca\~nada, 28692 Madrid, Spain\label{aff90}
\and
Institut de F\'{i}sica d'Altes Energies (IFAE), The Barcelona Institute of Science and Technology, Campus UAB, 08193 Bellaterra (Barcelona), Spain\label{aff91}
\and
School of Mathematics and Physics, University of Surrey, Guildford, Surrey, GU2 7XH, UK\label{aff92}
\and
European Space Agency/ESTEC, Keplerlaan 1, 2201 AZ Noordwijk, The Netherlands\label{aff93}
\and
Waterloo Centre for Astrophysics, University of Waterloo, Waterloo, Ontario N2L 3G1, Canada\label{aff94}
\and
Department of Physics and Astronomy, University of Waterloo, Waterloo, Ontario N2L 3G1, Canada\label{aff95}
\and
Perimeter Institute for Theoretical Physics, Waterloo, Ontario N2L 2Y5, Canada\label{aff96}
\and
Centre National d'Etudes Spatiales -- Centre spatial de Toulouse, 18 avenue Edouard Belin, 31401 Toulouse Cedex 9, France\label{aff97}
\and
Institute of Space Science, Str. Atomistilor, nr. 409 M\u{a}gurele, Ilfov, 077125, Romania\label{aff98}
\and
Institute for Particle Physics and Astrophysics, Dept. of Physics, ETH Zurich, Wolfgang-Pauli-Strasse 27, 8093 Zurich, Switzerland\label{aff99}
\and
Dipartimento di Fisica e Astronomia "G. Galilei", Universit\`a di Padova, Via Marzolo 8, 35131 Padova, Italy\label{aff100}
\and
Instituto de F\'isica Te\'orica UAM-CSIC, Campus de Cantoblanco, 28049 Madrid, Spain\label{aff101}
\and
Institut de Recherche en Astrophysique et Plan\'etologie (IRAP), Universit\'e de Toulouse, CNRS, UPS, CNES, 14 Av. Edouard Belin, 31400 Toulouse, France\label{aff102}
\and
Universit\'e St Joseph; Faculty of Sciences, Beirut, Lebanon\label{aff103}
\and
Departamento de F\'isica, FCFM, Universidad de Chile, Blanco Encalada 2008, Santiago, Chile\label{aff104}
\and
Universit\'e C\^{o}te d'Azur, Observatoire de la C\^{o}te d'Azur, CNRS, Laboratoire Lagrange, Bd de l'Observatoire, CS 34229, 06304 Nice cedex 4, France\label{aff105}
\and
Universit\"at Innsbruck, Institut f\"ur Astro- und Teilchenphysik, Technikerstr. 25/8, 6020 Innsbruck, Austria\label{aff106}
\and
Department of Physics and Helsinki Institute of Physics, Gustaf H\"allstr\"omin katu 2, University of Helsinki, 00014 Helsinki, Finland\label{aff107}
\and
Mullard Space Science Laboratory, University College London, Holmbury St Mary, Dorking, Surrey RH5 6NT, UK\label{aff108}
\and
Cosmic Dawn Center (DAWN)\label{aff109}
\and
Niels Bohr Institute, University of Copenhagen, Jagtvej 128, 2200 Copenhagen, Denmark\label{aff110}
\and
Universidad Polit\'ecnica de Cartagena, Departamento de Electr\'onica y Tecnolog\'ia de Computadoras,  Plaza del Hospital 1, 30202 Cartagena, Spain\label{aff111}
\and
European University of Technology EUt+, European Union\label{aff112}
\and
Astronomisches Rechen-Institut, Zentrum f\"ur Astronomie der Universit\"at Heidelberg, M\"onchhofstr. 12-14, 69120 Heidelberg, Germany\label{aff113}
\and
Dipartimento di Fisica e Scienze della Terra, Universit\`a degli Studi di Ferrara, Via Giuseppe Saragat 1, 44122 Ferrara, Italy\label{aff114}
\and
Istituto Nazionale di Fisica Nucleare, Sezione di Ferrara, Via Giuseppe Saragat 1, 44122 Ferrara, Italy\label{aff115}
\and
ICSC - Centro Nazionale di Ricerca in High Performance Computing, Big Data e Quantum Computing, Via Magnanelli 2, Bologna, Italy\label{aff116}
\and
INAF, Istituto di Radioastronomia, Via Piero Gobetti 101, 40129 Bologna, Italy\label{aff117}
\and
INFN-Bologna, Via Irnerio 46, 40126 Bologna, Italy\label{aff118}
\and
Astronomical Observatory of the Autonomous Region of the Aosta Valley (OAVdA), Loc. Lignan 39, I-11020, Nus (Aosta Valley), Italy\label{aff119}
\and
Department of Mathematical Physics, Institute of Physics, University of S\~ao Paulo, R. do Mat\~ao 1371, 05508-090, S\~ao Paulo, SP, Brazil\label{aff120}
\and
Universidad de La Laguna, Dpto. Astrof\'\i sica, E-38206 La Laguna, Tenerife, Spain\label{aff121}
\and
Oskar Klein Centre for Cosmoparticle Physics, Department of Physics, Stockholm University, Stockholm, SE-106 91, Sweden\label{aff122}
\and
Astrophysics Group, Blackett Laboratory, Imperial College London, London SW7 2AZ, UK\label{aff123}
\and
Univ. Grenoble Alpes, CNRS, Grenoble INP, LPSC-IN2P3, 53, Avenue des Martyrs, 38000 Grenoble, France\label{aff124}
\and
Dipartimento di Fisica, Sapienza Universit\`a di Roma, Piazzale Aldo Moro 2, 00185 Roma, Italy\label{aff125}
\and
Institut d'Astrophysique de Paris, 98bis Boulevard Arago, 75014, Paris, France\label{aff126}
\and
ICL, Junia, Universit\'e Catholique de Lille, LITL, 59000 Lille, France\label{aff127}
\and
Center for Astrophysics and Cosmology, University of Nova Gorica, Nova Gorica, Slovenia\label{aff128}
\and
CERCA/ISO, Department of Physics, Case Western Reserve University, 10900 Euclid Avenue, Cleveland, OH 44106, USA\label{aff129}
\and
Laboratoire Univers et Th\'eorie, Observatoire de Paris, Universit\'e PSL, Universit\'e Paris Cit\'e, CNRS, 92190 Meudon, France\label{aff130}
\and
Departamento de F{\'\i}sica Fundamental. Universidad de Salamanca. Plaza de la Merced s/n. 37008 Salamanca, Spain\label{aff131}
\and
Universiteit Gent, Department of Physics and Astronomy, Proeftuinstraat 86 N3, 9000 Ghent, Belgium
\label{aff132}
\and
Konkoly Observatory, HUN-REN CSFK, MTA Centre of Excellence, Budapest, Konkoly Thege Mikl\'os {\'u}t 15-17. H-1121, Hungary\label{aff133}
\and
MTA-CSFK Lend\"ulet Large-Scale Structure Research Group, Konkoly-Thege Mikl\'os \'ut 15-17, H-1121 Budapest, Hungary\label{aff134}
\and
Universit\'e de Strasbourg, CNRS, Observatoire astronomique de Strasbourg, UMR 7550, 67000 Strasbourg, France\label{aff135}
\and
Scuola Superiore Meridionale, Via Mezzocannone 4, 80138, Napoli, Italy\label{aff136}
\and
Department of Physics, Astronomy and Mathematics, University of Hertfordshire, College Lane, Hatfield AL10 9AB, UK\label{aff137}
\and
Instituto de Astrof\'isica e Ci\^encias do Espa\c{c}o, Faculdade de Ci\^encias, Universidade de Lisboa, Tapada da Ajuda, 1349-018 Lisboa, Portugal\label{aff138}
\and
Institute of Cosmology and Gravitation, University of Portsmouth, Portsmouth PO1 3FX, UK\label{aff139}
\and
Department of Physics \& Astronomy, University of California Irvine, Irvine CA 92697, USA\label{aff140}
\and
Department of Mathematics and Physics E. De Giorgi, University of Salento, Via per Arnesano, CP-I93, 73100, Lecce, Italy\label{aff141}
\and
INFN, Sezione di Lecce, Via per Arnesano, CP-193, 73100, Lecce, Italy\label{aff142}
\and
INAF-Sezione di Lecce, c/o Dipartimento Matematica e Fisica, Via per Arnesano, 73100, Lecce, Italy\label{aff143}
\and
Departamento F\'isica Aplicada, Universidad Polit\'ecnica de Cartagena, Campus Muralla del Mar, 30202 Cartagena, Murcia, Spain\label{aff144}
\and
Instituto de F\'isica de Cantabria, Edificio Juan Jord\'a, Avenida de los Castros, 39005 Santander, Spain\label{aff145}
\and
CEA Saclay, DFR/IRFU, Service d'Astrophysique, Bat. 709, 91191 Gif-sur-Yvette, France\label{aff146}
\and
Department of Computer Science, Aalto University, PO Box 15400, Espoo, FI-00 076, Finland\label{aff147}
\and
Ruhr University Bochum, Faculty of Physics and Astronomy, Astronomical Institute (AIRUB), German Centre for Cosmological Lensing (GCCL), 44780 Bochum, Germany\label{aff148}
\and
DARK, Niels Bohr Institute, University of Copenhagen, Jagtvej 155, 2200 Copenhagen, Denmark\label{aff149}
\and
Department of Physics, Centre for Extragalactic Astronomy, Durham University, South Road, Durham, DH1 3LE, UK\label{aff150}
\and
Astrophysics Research Centre, University of KwaZulu-Natal, Westville Campus, Durban 4041, South Africa\label{aff151}
\and
School of Mathematics, Statistics \& Computer Science, University of KwaZulu-Natal, Westville Campus, Durban 4041, South Africa\label{aff152}
\and
Department of Physics and Astronomy, Vesilinnantie 5, University of Turku, 20014 Turku, Finland\label{aff153}
\and
Finnish Centre for Astronomy with ESO (FINCA), Quantum, Vesilinnantie 5, University of Turku, 20014 Turku, Finland\label{aff154}
\and
Serco for European Space Agency (ESA), Camino bajo del Castillo, s/n, Urbanizacion Villafranca del Castillo, Villanueva de la Ca\~nada, 28692 Madrid, Spain\label{aff155}
\and
ARC Centre of Excellence for Dark Matter Particle Physics, Melbourne, Australia\label{aff156}
\and
Centre for Astrophysics \& Supercomputing, Swinburne University of Technology,  Hawthorn, Victoria 3122, Australia\label{aff157}
\and
Department of Physics and Astronomy, University of the Western Cape, Bellville, Cape Town, 7535, South Africa\label{aff158}
\and
Van Swinderen Institute, University of Groningen, Nijenborgh 3, 9747 AG Groningen, The Netherlands\label{aff159}
\and
Ohio University, Physics \& Astronomy Department,1 Ohio University, Athens, OH 45701, USA\label{aff160}
\and
Brazilian Center for Research in Physics (CBPF), Dr. Xavier Sigaud st. 150, zip 22290-180, Rio de Janeiro, RJ, Brazil\label{aff161}
\and
Institut d'Astrophysique de Paris, UMR 7095, CNRS, and Sorbonne Universit\'e, 98 bis boulevard Arago, 75014 Paris, France\label{aff162}
\and
Institute for Theoretical Particle Physics and Cosmology (TTK), RWTH Aachen University, 52056 Aachen, Germany\label{aff163}
\and
IRFU, CEA, Universit\'e Paris-Saclay 91191 Gif-sur-Yvette Cedex, France\label{aff164}
\and
INAF-Osservatorio Astrofisico di Arcetri, Largo E. Fermi 5, 50125, Firenze, Italy\label{aff165}
\and
Centro de Astrof\'{\i}sica da Universidade do Porto, Rua das Estrelas, 4150-762 Porto, Portugal\label{aff166}
\and
HE Space for European Space Agency (ESA), Camino bajo del Castillo, s/n, Urbanizacion Villafranca del Castillo, Villanueva de la Ca\~nada, 28692 Madrid, Spain\label{aff167}
\and
Dipartimento di Fisica ``G. Occhialini", Universit\`a degli Studi di Milano Bicocca, Piazza della Scienza 3, 20126 Milano, Italy\label{aff168}
\and
Department of Astrophysics, University of Zurich, Winterthurerstrasse 190, 8057 Zurich, Switzerland\label{aff169}
\and
Space physics and astronomy research unit, University of Oulu, Pentti Kaiteran katu 1, FI-90014 Oulu, Finland\label{aff170}
\and
Department of Mathematics and Physics, Roma Tre University, Via della Vasca Navale 84, 00146 Rome, Italy\label{aff171}
\and
Theoretical astrophysics, Department of Physics and Astronomy, Uppsala University, Box 524, 751 20 Uppsala, Sweden\label{aff172}
\and
Mathematical Institute, University of Leiden, Einsteinweg 55, 2333 CA Leiden, The Netherlands\label{aff173}
\and
School of Physical Sciences, The Open University, Milton Keynes, MK7 6AA, UK\label{aff174}
\and
Institute of Astronomy, University of Cambridge, Madingley Road, Cambridge CB3 0HA, UK\label{aff175}
\and
Univ. Lille, CNRS, Centrale Lille, UMR 9189 CRIStAL, 59000 Lille, France\label{aff176}
\and
Department of Astrophysical Sciences, Peyton Hall, Princeton University, Princeton, NJ 08544, USA\label{aff177}
\and
Fakult\"at f\"ur Physik, Universit\"at Bielefeld, Postfach 100131, 33501 Bielefeld, Germany\label{aff178}
\and
School of Mathematics, Statistics and Physics, Newcastle University, Herschel Building, Newcastle-upon-Tyne, NE1 7RU, UK\label{aff179}
\and
Department of Physics and Astronomy, University of British Columbia, Vancouver, BC V6T 1Z1, Canada\label{aff180}}    

%
% Put your abstract here:
%
\abstract{ % Context
  The \Euclid space mission is dedicated to probing the nature of dark energy
  and dark matter, through a comprehensive mapping of the Universe's
  large-scale structure.  Central to these cosmological goals are the spectra
  produced by the \ac{NISP} and reduced by the scientific pipelines of the
    SIR (Spectral near InfraRed) processing function.
%}{ % Aims
  Due to the intrinsic complexity of slitless spectroscopic data, scientific
  observations cannot be self-calibrated by the science pipelines.
  Consequently, the development of dedicated SIR calibration pipelines,
  utilising specific pointing sequences and observation fields, is essential.
  The resulting calibration products are fundamental to ensuring that reduced
  spectra are characterised within requirements in both wavelength and flux.
%}{ % Methods
  This paper provides a detailed description of the SIR calibration pipelines
  responsible for generating the products utilised by the science pipelines for
  \ac{DR1}.  A significant challenge in these calibration tasks arises from the
  requirement to model the entire focal plane as a unified entity.  Such a
  global modelling approach is critical for maintaining astrometric and
  spectroscopic continuity across the entire field of view.
%}{ % Results
  \ac{DR1} provides more than 170~millions validated spectra in both red- and
  blue-grism modes.  The accuracy of the reduced spectra is strictly dependent
  on the performance of the calibration pipelines.
%}{ % Conclusions
  Comprehensive analyses of the calibration products and the resulting science
  spectra demonstrate that current outputs successfully meet the mission
  requirements.  Nevertheless, specific areas for optimisation have been
  identified; ongoing development is focused on addressing these critical
  points to further enhance data quality for future data releases.}

%
% Provide up to five key words from the list in
% https://www.aanda.org/for-authors/latex-issues/information-files#pop
%
\keywords{
  Cosmology: observations --
  Instrumentation: spectrographs --
  Techniques: imaging spectroscopy --
  Methods: data analysis
}
%
% Add short versions of title and author list for page headings
%
\titlerunning{The SIR calibration}
\authorrunning{Euclid Collaboration: M.\ Fumana et al.}

\maketitle
\nolinenumbers %Per upload in 
% -------------------------------------------------------------------
%
%   Start the main text of your paper here
%

\acresetall                     % Reset acronyms after abstract

\section{Introduction}
\label{sec:introduction}

One of the foremost scientific goals of the \Euclid space mission is to
constrain the expansion history of the Universe through precise measurements of
\ac{BAO} and \ac{RSD}.  Achieving these objectives requires a spectroscopic
data set that is simultaneously accurate, complete, and statistically
homogeneous over an exceptionally large volume.

This translates into three key requirements: (i) a wide-area coverage of
$\sim \SI{14 000}{deg^{2}}$ to minimise sample variance; (ii) high-precision
redshifts, with $\sigma (z)/(1+z)\leq 0.001$, necessitating a minimum resolving
power of $\mathcal{R} \geq 380$ for a \ang{;;0.5}-diameter reference source to
sufficiently resolve the H\,$\alpha$ line and preserve both the sharpness of
the \ac{BAO} feature and the small-scale anisotropies probed by redshift-space
distortions; and (iii) an accurate, spatially uniform flux calibration,
ensuring a position-independent H\,$\alpha$ line-flux limit of
$\sim \SI{2e-16}{erg.s^{-1}.cm^{-2}}$, which is essential for maintaining the
required completeness of the galaxy tracer population across the entire survey
footprint.

To achieve these survey-level requirements, \Euclid relies on the near-infrared
slitless spectroscopic channel of the \aclu{NISP}
\citep[\acs{NISP},][]{EuclidSkyNISP}, hereafter NISP-S.  This specific design
choice allows for the massive, simultaneous collection of spectra for all
sources within the field of view without the need for a physical slit, thereby
maximising the survey's efficiency and cosmological volume coverage.  A
comprehensive overview of the strategies and frameworks developed for the
calibration of these spectroscopic data can be found in \cite{DR1-TP026}.

However, slitless spectroscopy also introduces intrinsic complexities.  In the
absence of slits, spectral traces are affected by overlaps between neighbouring
spectra, self-contamination arising from the entanglement of spatial and
spectral information along the dispersion direction, and a strong dependence of
the effective resolving power on the intrinsic morphology of the source
\citep[hereafter \citetalias{Q1-TP006}]{Q1-TP006}.  The $4 \times 4$ detector
mosaic, with physical gaps of approximately \SI{6}{mm} and \SI{3}{mm} along and
across the dispersion direction respectively (see
e.g. Fig.~\ref{fig:ids_pn_cross}), further complicates the prediction of the
exact position and shape of each dispersed spectrum, particularly when part of
the trace falls in inter-detector regions.

Delivering the accuracy required for \ac{BAO} and \ac{RSD} therefore requires
spectra calibration and spectra reduction pipelines capable of modelling and
correcting all instrumental signatures in a globally consistent manner.  Such
pipelines must be able to reconstruct the geometric distortions, the dispersion
law, the \ac{PSF} over the entire \ac{FP}, and an accurate spectro-photometric
response, which is strictly dependent on the characterisation of the detectors
\citep{EU-Kubik}.  This effort builds upon the initial optical models and
results obtained during the ground-based calibration campaigns
\citep{EP-Gillard}, ensuring that the measured wavelengths, fluxes, and
uncertainties remain stable at the sub-per cent level across the whole
\Euclid~survey.

Within the \Euclid science ground segment \citep{DR1-TP033}, the SIR
science pipelines (\citetalias{Q1-TP006}, \citealt[hereafter
\citetalias{DR1-TP023}]{DR1-TP023}) are responsible for transforming the raw
NISP-S exposures from spacecraft into wavelength- and flux-calibrated spectra.
Their performance relies on a comprehensive set of calibration products that
describe the full spectroscopic behaviour of the NISP-S instrument.  This work
presents the SIR calibration pipelines developed to generate all the
spectroscopic calibration products required by the SIR science pipelines for
\citet{DR1cite}.  These include (i)~the geometric models (Sect.~\ref{sec:Geo})
most directly related to redshift accuracy requirements, and (ii)~the
spectro-photometric models (Sect.~\ref{sec:SpectroPhoto}), which ensure the
fulfilment of sample completeness requirements.

We present the calibration observing strategy, with a focus on the dedicated
calibration fields (Sect.~\ref{sec:NISPSpec}), the mathematical models and the
calibration algorithms (Sects.~\ref{sec:Geo} and \ref{sec:SpectroPhoto}), and
the validation procedures (Sect.~\ref{sec:DataQuality}) employed to derive
these products and assess their performance against \Euclid's mission-level
requirements within the \ac{DR1} framework \citep{DR1-TP026}.

The SIR science pipelines also incorporate a dedicated preprocessing stage
responsible for the removal of instrumental signatures from the raw data.
Except for the \ac{QE} correction (see Sect.~\ref{sec:DETScaling}), the
preprocessing algorithms and their associated calibrations have been developed
by the NIR processing function \citep{DR1-TP003} and are therefore excluded
from the scope of this paper.

\section{NISP spectroscopic observations}
\label{sec:NISPSpec}

NISP-S operates in the near-infrared regime, covering the
\SIrange{950}{2000}{nm} range using a \ac{GWA} hosting four grisms, three red
(RGS000/180/270\acused{RGS}) and one blue (BGS000\acused{BGS}).  Each grism
consists of a fused-silica grating-prism combination with a slightly curved
base designed to guarantee optimal focus across the entire~\ac{FP}.

The \aclu{EWS} \citep[\acs{EWS},][]{Scaramella-EP1, DR1-TP027} employs two out
of three\footnote{The third red grism (RGS270) is carried onboard, but cannot
  be used in the survey because of its non-conformity \citep{EP-Gillard}.}
\acp{RGS}, RGS000 and RGS180, which cover the same nominal spectral domain,
$\RGE = \text{\SIrange{1206}{1892}{nm}}$, with an effective dispersion of
\SI{1.372}{nm.px^{-1}} and a resolving power $\mathcal{R} > 480$ for a typical
\ang{;;0.5}-diameter source.  To support the spectral decontamination process,
the \ac{EWS} uses these two contra-dispersing \acp{RGS} as designed, as well as
two configurations where an additional \ang{\pm 4}~tilt is introduced in the
\ac{GWA} orientation to further diversify the dispersion angles.  Accordingly,
the \ac{EWS} acquires spectroscopic frames using four distinct red grism-tilt
configurations: RGS000\texttt{+}0, RGS180\texttt{+}4, RGS000\texttt{-}4, and
RGS180\texttt{+}0.  Hereafter, this specific sequence of four acquired frames
is referred to as a \ac{ROS}.

In addition to the \ac{EWS}, \Euclid carries out the \ac{EDS}, a distinct
observational programme targeting a limited number of specific fields with
significantly longer exposure times.  Unlike the \ac{EWS}, which exclusively
uses the \acp{RGS}, the \ac{EDS} incorporates observations from both the red
and blue grisms.  The \ac{BGS} covers the nominal pass-band,
$\BGE = \text{\SIrange{926}{1366}{nm}}$, with a dispersion of
\SI{1.239}{nm.px^{-1}} and a resolving power~$\mathcal{R} > 400$.

To process the data from both surveys, the SIR calibration pipelines
independently characterise five distinct and independent grism-tilt
configurations: the four \ac{RGS} configurations used in the \ac{EWS} and
\ac{EDS}, and the \ac{BGS} setup specific to the \ac{EDS}.

\subsection{NISP-S calibration observing strategy}
\label{sec:CalibObservingStrategy}

To ensure the high precision required by the NISP-S scientific objectives
\citep{EuclidSkyOverview}, a dedicated calibration campaign was designed,
targeting specific astronomical sources with well-known properties
\citep{DR1-TP026}.  Unlike the standard scientific observations characterised
by a single \ac{ROS}, the calibration strategy often involves multiple
\acp{ROS} or specialised dither sequences.  These are executed at different
positions within the same field to ensure an accurate and uniform coverage of
the \ac{FP}, allowing the same calibration source to be sampled across multiple
detector positions.  The geometric and spectro-photometric calibration
framework primarily relies on three dedicated fields obtained during the
\ac{PV} phase:
\begin{description}
\item[CALBLOCK-PV-001:] This self-calibration field is fundamental for
  monitoring the stability of the instrument's spectro-photometric response
  against space weathering or out-gassing.  The spectroscopic pipelines
  responsible for the optical, curvature, and \ac{PSF} models (see
  Sect.~\ref{sec:Geo}), as well as the relative flux model
  (Sect.~\ref{sec:RFX}), leverage the high density of bright, non-saturated
  point sources within this field.  These sources are utilised to map the
  spatial variations of the modelled quantities across the entire \ac{FP}.  The
  calibration pipelines rely on this field, which is acquired in four visits
  and each visit produces 15~frames for each \ac{RGS} configuration and
  16~frames for the \ac{BGS}.  To maximise the source density across the
  \ac{FP} and ensure the derivation of a more robust and reliable map, the
  total number of available frames is processed simultaneously for each
  grism-tilt configuration.

\item[CALBLOCK-PV-009:] This field is used to derive the wavelength solution
  (Sect.~\ref{sec:IDS}).  The strategy involves observing a compact \ac{PN}
  characterised by strong emission lines.  For each of the 16~detectors, the
  \ac{PN} is stepped across five different positions, forming a quincunx (i.e.,
  dice-five) pattern.  This results in a total of 80~dithered positions for
  each grism-tilt configuration (see Fig.~\ref{fig:ids_pn_cross}).  To avoid
  uncertainties in the \ac{GWA} repositioning, which would compromise the
  self-consistency of the dispersion solution, all 80~frames are completed
  before moving the~\ac{GWA}.

  \begin{figure}
    \centering
    \includegraphics[width=\linewidth]{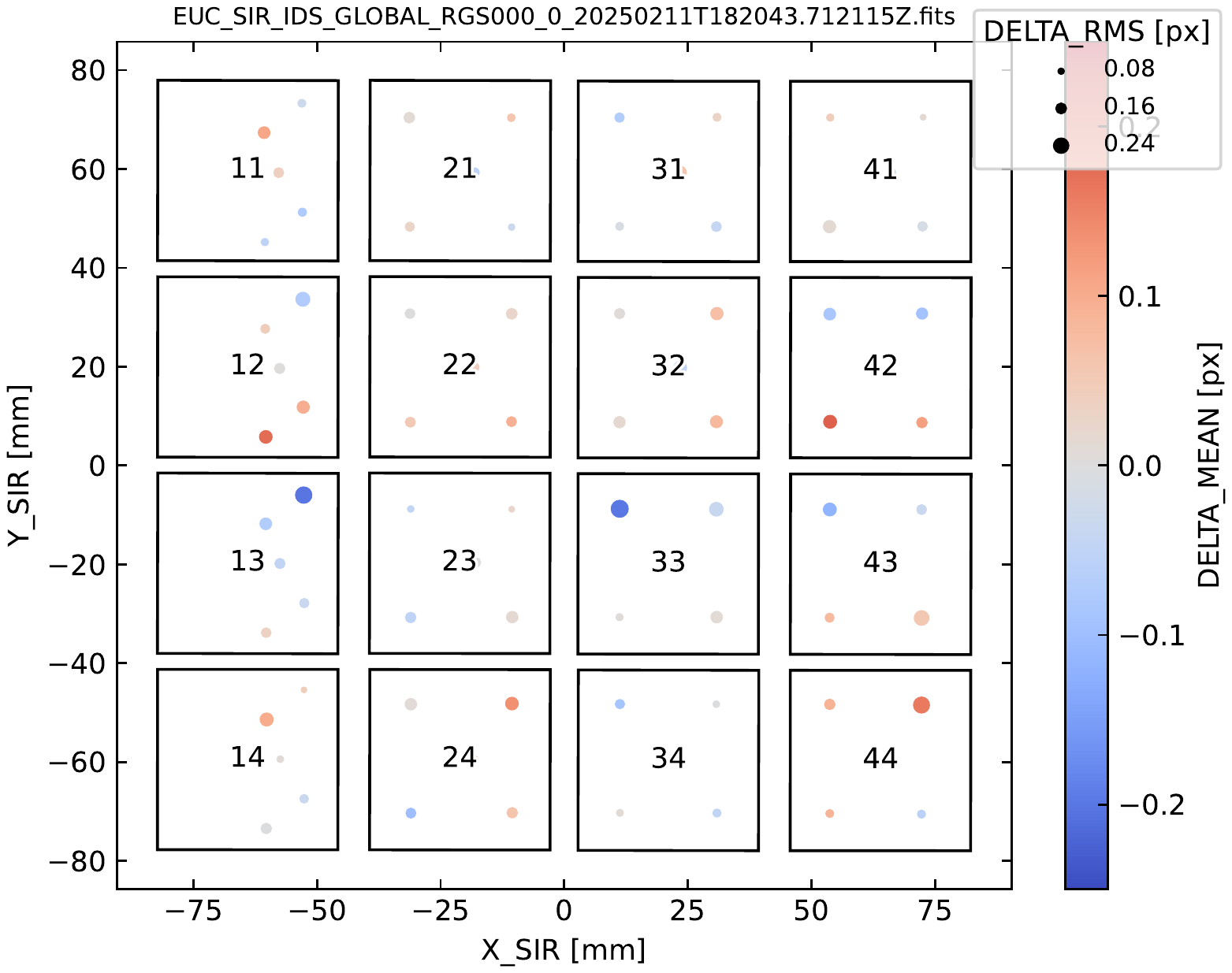}
    \caption{Pattern of observation of the \ac{PN} in CALBLOCK-PV-009, with
      five distinct positions on each detector, following a quincunx pattern,
      for grism-tilt RGS000\texttt{+}0.  The target positions across the
      \ac{FP} during the 80~acquisitions are shown, along with mean (dot
      colour) and \acs*{RMS} (dot size) error of the global inverse dispersion
      solution (IDS), in pixels (see Sect.~\ref{sec:IDS_DQC}).  The quincunx
      patterns on detectors 1x have been `compressed' so that corresponding
      zeroth-order spectrograms (to the left for RGS000\texttt{+}0) still fit
      within the \ac{FP}.}
    \label{fig:ids_pn_cross}
  \end{figure}

  The \ac{PN} SMC-SMP-20~was selected for this purpose due to its compact
  spatial extent, optimal visibility, and lack of significant contamination from
  neighbouring sources \citep{Paterson-EP32}.

\item[CALBLOCK-PV-003:] Following a strategy similar to the PV-009 field, this
  block is dedicated to absolute flux calibration (see Sect.~\ref{sec:AFX}),
  using \ac{WD} GRW+70\,5824 as flux reference, to derive the absolute
  sensitivity curve of the instrumental setup.
\end{description}

We emphasize that \ac{DR1} calibrations rely exclusively on observations of the
\ac{PV} fields, under the implicit assumption of instrument stability, as
discussed in Sect.~\ref{sec:conclusion}; repeated visits of the CALBLOCK-F-001
and CALBLOCK-F-010 are not used in this context.

\section{Geometric calibrations}
\label{sec:Geo}

Geometric models refer to the set of semi-analytical descriptions of the
optical distortions of NISP-S instrument. These models are essential for the
SIR pipelines to properly identify spectrograms on the dispersed science image
(specifically first- and zeroth-orders for \ac{DR1}), and to characterise their
precise location and extent.  Consequently, given the astronomical coordinates
$(\alpha, \delta)$ of a source, as provided by the input MER catalogue
\citep{DR1-TP009}, the geometric calibration allow the accurate location within
the frame of the position at any wavelength of its associated spectrogram.

The geometric description comprises five distinct models: the \ac{FP} array
metrology (FPA, Sect.~\ref{sec:FPA}), optical (OPT, Sect.~\ref{sec:OPT}),
curvature (CRV, Sect.~\ref{sec:CRV}), PSF (Sect.~\ref{sec:PSF}), and inverse
dispersion solution (IDS, Sect.~\ref{sec:IDS}) models, which are used in
conjunction as follows:
  \begin{itemize}
  \item FPA relates physical positions in the \ac{FP} (in~mm) to
    pixel-coordinates of individual detectors;
  \item OPT maps sky coordinates $(\alpha, \delta)$ to reference positions
    $(x_{\text{r}}, y_{\text{r}})$ in the \ac{FP} for zeroth- and first-order
    dispersions;
  \item CRV describes the spectral trace curvature $\delta y$ as a function of
    the position $\delta x$ along the first-order spectrogram for a given
    reference position in the \ac{FP}; similarly, PSF describes the
    point-source width $\sigma$ of this spectrogram;
  \item IDS relates, for first-order spectrograms, the wavelength $\lambda$ to
    position $\delta x$ along the spectral trace.
  \end{itemize}
  Ultimately, these models jointly define, for each of the five grism-tilt
  instrumental configurations, a mapping between the position of a
  source on the sky, the wavelength, and the dispersion order, and the pixel
  coordinates in the \ac{FP}.

The \ac{NISP} \ac{FP} consists of a $4 \times 4$-detector matrix separated by
substantial physical gaps (see Fig.~\ref{fig:ids_pn_cross}).  These regions,
while lacking active sensors, are nonetheless illuminated by the incident beam.
As a result, the grism also disperses light from sources located within these
gaps, producing split spectrogram captured by adjacent detectors.  The exact
geometry of these fragmented spectra depends on the object's position within
the gap.  However, calibration pipelines cannot directly measure these
positions due to the absence of detectors in those regions.

To address this situation, geometric models cannot be defined solely on the
sensitive area of the \ac{FP}: they must also map regions in between and
marginally outside the detector extents.  Consequently, all mentioned models,
except the FPA model, need to be implemented as \emph{global} models (as
opposed to detector-specific \emph{local} models): they provide a continuous
mapping of the full \ac{FP}.  This approach enables the models to predict the
behaviour and position of sources even when located within the gaps or slightly
outside the detector footprint, utilising the FPA model to subsequently convert
these physical coordinates in the \ac{FP} into pixel positions on the relevant
detector.

As mentioned in \citetalias{Q1-TP006}, the SIR pipelines work in a dedicated
coordinate system where the spectrograms are approximately horizontal:
$(X, Y)_{\text{SIR}} \equiv (+Z, -Y)_{\text{MOSAIC}}$ \citep{EuclidSkyNISP}.
Unless otherwise specified, all subsequent geometric formulations follow this
convention.

\subsection{Focal plane assembly (FPA) model}
\label{sec:FPA}

The FPA model (also referred to as detector metrology) describes the physical
arrangement of the detectors in the \ac{FP}.  It serves as the fundamental
link to convert physical \ac{FP} coordinates (in~mm) into detector-specific
coordinates (in pixels), accounting for the imperfect distribution and
alignment of the detectors in the \ac{FP}.  When not properly accounted for,
such misalignments introduce spatial discontinuities between detectors that
continuous analytical models cannot accurately represent.

In practice, the FPA model consists of a collection of 16~linear
transformations $\mathcal{D}_{l}$, mapping, for each detector~$l$, from the
\ac{FP} coordinates (in~mm) to pixel coordinates, and stored as per-detector
\aclp{WCS} \citep[\acsp{WCS},][]{WCS2002}; for DR1 data, the $\mathcal{D}_{l}$
transformations only involve rigid translations of the detectors (potential
rotations and scalings are postponed to DR2).  This model quantifies the
positional deviations of each detector relative to an ideal grid, arising from
three primary factors: translational shifts in \XSIR{} and \YSIR{} within the
\ac{FP}, rotational misalignments relative to the detector grid, and a lack of
perfect coplanarity among the detectors.  The latter could result in slight
pixel-scale discrepancies between adjacent sensors.

The first spectroscopic frames acquired during the \ac{PV} phase revealed that
ground-based metrology, acquired at room temperature, does not fully reflect
the actual in-flight configuration of the \ac{FP}.  Consequently, an effective
FPA model was developed from in-flight data.  The calibration procedure uses
\begin{itemize}
\item a photometric frame where each detector has been astrometrically
  calibrated \citep{Q1-TP003},
\item a catalogue of bright, non-saturated point-like sources at sky
  coordinates $(\alpha_{i}, \delta_{i})$,
\item and the corresponding measured positions $\mathbf{p}^{(l)}_{i}$ of these
  objects for each detector~$l$, in pixels.
\end{itemize}
Starting from ground-based estimates, the procedure establishes an initial
global \acsu{WCS} $\mathcal{G}$, which incorporates distortions via a
fifth-order polynomial \citep{SIP2005} to map the sky objects coordinates
$(\alpha_{i}, \delta_{i})$ into \ac{FP} coordinates.  By composing
$\mathcal{G}$ with $\mathcal{D}_{l}$, the predicted positions of the sources
onto detector~$l$ are derived; a minimisation routine is then applied to adjust
the individual detector translational offsets (specifically updating
\texttt{CRVAL1} and \texttt{CRVAL2} \ac{WCS} keywords only for \ac{DR1}) to
minimise the residuals between the measured pixel positions and those predicted
by the FPA models,
\begin{equation}
  \label{eq:minimization}
  \min_{\text{CRVAL}{\{1,2\}}} \sum_{i} \left\|
    \mathbf{p}^{(l)}_{i} -
    \left( \mathcal{D}_{l} \circ \mathcal{G}\right)(\alpha_{i}, \delta_{i})
  \right\|^{2}\;.
\end{equation}

As this specific procedure was not originally scheduled for \ac{DR1}, an
independent recipe does not yet exist, and its output was directly integrated
into the OPT \texttt{DpdSirOptModel}\footnote{All the SIR data products are
  documented in the \ac{DR1}
  \href{http://st-dm.pages.euclid-sgs.uk/data-product-doc/dmdr1/sirdpd/sirindex.html}{%
    Data Products Description Documentation}.} (see below); consequently, no
stand-alone FPA-specific data product currently exists.  A dedicated metrology
pipeline and data product, including detector rotation and piston, will be
implemented for the forthcoming data releases.

\subsection{Optical (OPT) model}
\label{sec:OPT}

The primary objective of the OPT model is to derive the reference position
$(x_{\text{r}}, y_{\text{r}})$ within a spectroscopic frame of the spectrogram
of a source in the field of view for a reference
wavelength~$\lambda_{\text{r}}$.  For \ac{DR1}, the OPT model accounts for both
the first- and zeroth-dispersion orders; other dispersion orders -- notably
$-1$ and $+2$, with peak flux only a per cent-level of the main dispersion
orders -- are not yet considered in this model, and are therefore ignored in
all subsequent steps (including decontamination).

The OPT model provides a continuous mapping from celestial coordinates
$(\alpha, \delta)$ to physical \ac{FP} SIR coordinates (in~mm).  In the context
of dispersed images, `localisation' of a spectrogram specifically refers to the
identification of a reference wavelength associated with a detectable spectral
feature.  Throughout the following sections, any mention to the object's
reference position within the \ac{FP} implicitly refers to the coordinates of
its reference wavelength.

\begin{figure}
  \centering
  \includegraphics[width=1.0\linewidth]{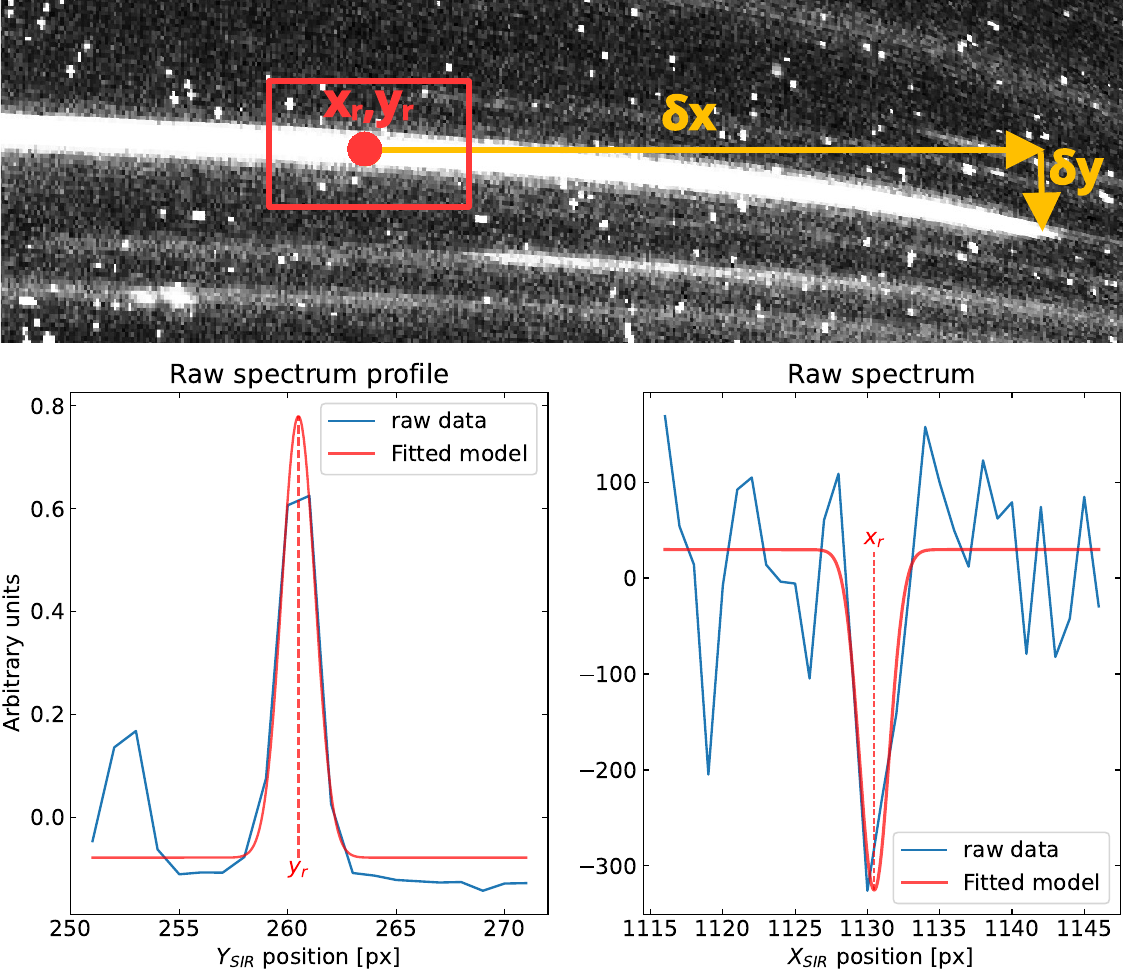}
  \caption{Illustration of the OPT and CRV spectrometric models, jointly used
    to describe the spectrogram geometry.  \emph{Upper panel}: a raw
    spectrogram, with an artificially enhanced curvature to facilitate the
    visualisation of the geometric distortions.  The red box is the area
    selected by the OPT calibration to compute the
    $(x_{\text{r}}, y_{\text{r}})$ spectrogram reference position (red dot);
    cross-dispersion offset $\delta y$ as a function of along-dispersion
    distance $\delta x$ (yellow arrows) allow the characterisation of the
    spectrogram tilt and distortions.  \emph{Lower left panel}: the
    $y_{\text{r}}$ coordinate is determined by fitting the cross-dispersion
    profile averaged over the red box along the dispersion axis.  A similar
    procedure is used to measure $\delta y$ along the trace
    (Sect.~\ref{sec:CRV}).  \emph{Lower right panel}: the $x_{\text{r}}$
    position is derived by fitting reference absorption spectral feature (here
    \ion{Mg}{I}, see text) of the raw spectrum obtained by summing the red box
    along the cross-dispersion axis.  Both panels are from RGS000\texttt{+}0
    observation of a selected bright star ($\JMASS = 15.95$).}
  \label{fig:opt_crv_thumb}
\end{figure}

The first-order reference position is the one of the stellar feature at
$\lambda_{\text{r}}^{\text{RGS}} = \SI{1504}{nm}$ (blended \ion{Mg}{I}) for the
\acp{RGS} (see Fig.~\ref{fig:opt_crv_thumb}), and of the Paschen Pa\,$\beta$
absorption feature at~$\lambda_{\text{r}}^{\text{BGS}} = \SI{1282}{nm}$ for the
\ac{BGS}.  The zeroth-order spectra are concentrated on a few pixels and
exhibit two distinct peaks \citep[e.g.,][]{EP-Gillard}; the OPT model adopts
the bluest peak as a reference for the \acp{RGS}, and the reddest peak for the
\ac{BGS}.

For each grism-tilt combination and spectral order ($+1$ and~$0$), the model
first employs a linear transformation $\mathcal{O}$ to establish an approximate
sky-to-\ac{FP} mapping\footnote{Hereafter, the notation `$\approx$' stands for
  `approximated by'.},
\begin{equation}
  \label{eq:opt_wcs}
  (x_{u}, y_{u}) \approx \mathcal{O}(\alpha, \delta)\;.
\end{equation}
This is subsequently refined by two-dimensional ordinary polynomials
represented by the matrices $\mathsf{P}^{(X)}$ and $\mathsf{P}^{(Y)}$ (of order
$L=5$ in DR1), and accounting for the non-linear optical distortions,
\begin{equation}
  \label{eq:opt_poly}
  x_{\text{r}} \approx \sum_{i,j=0}^{L} p^{(X)}_{i,j}\,x_{u}^{i} y_{u}^{j}\;,
  \qquad
  y_{\text{r}} \approx \sum_{i,j=0}^{L} p^{(Y)}_{i,j}\,x_{u}^{i} y_{u}^{j}\;.
\end{equation}
Each grism-tilt and spectral order has its own set of $\mathsf{P}^{(X, Y)}$
matrices.

The calibration is performed using data from the self-calibration field
CALBLOCK-PV-001; the calibration pipeline works in parallel on each grism-tilt
producing a single \texttt{DpdSirOptModel} for dispersion orders $+1$ and~$0$.
It selects bright yet non-saturated point-like sources from the Two Micron All
Sky Survey catalogue \citep[2MASS,][]{2006AJ....131.1163S} with \textit{J}-band
magnitude between 12 and~16.  The procedure simultaneously utilises multiple
frames (15~per grism-tilt configuration), to ensure as extensive and uniform a
coverage of the \ac{FP} as possible.  The calibration process is organised in
three stages.

\paragraph{Pointing registration.} Using an initial OPT model derived and
validated during the \ac{PV} phase, the pipeline locates zeroth-order spots (on
the four central detectors with a better optical quality) to determine the
effective pointing for each individual frame.  These features are then measured
using the \texttt{DAOStarFinder} algorithm \citep{1987PASP...99..191S}, as
implemented in the \texttt{photutils} package
\citep{larry_bradley_2025_14889440}, with a configuration optimised for the
morphology of zeroth orders.

\paragraph{Zeroth-order calibration.} Once the effective pointing of a frame is
adjusted, the pipeline measures the positions of all zeroth orders on the
\ac{FP} to fit the two global polynomials $\mathsf{P}^{(X,Y)}$, by minimising
the $\chi^{2}$ between the measured and predicted positions.

Since the localisation of the zeroth-orders during the pointing adjustment
stage is dependent on the OPT model, the newly fitted model is utilised to
derive a more accurate pointing correction; this, in turn, allows for the
recalibration of an improved OPT model.  This iterative process is interrupted
when convergence is reached (typically within three iterations).

\paragraph{First-order calibration.} With a stable pointing solution
established, the pipeline calibrates the first-order model, measuring the
positions of stellar absorption features in the first-order spectrograms.  The
absorption feature positions are measured independently along the dispersion
(\XSIR) and cross-dispersion (\YSIR) directions (Fig.~\ref{fig:opt_crv_thumb}).
Similarly to the procedure adopted for the zeroth orders, the pipeline utilises
a pre-calibrated model derived from the \ac{PV} phase to define the regions of
interest where the absorption features are expected to be located.

Within a predefined window, the spectrogram is summed along the
cross-dispersion \YSIR-axis to produce a raw spectrum, whose local continuum is
removed by fitting a linear baseline.  A Gaussian function is then adjusted to
the continuum-subtracted spectrum to localise the reference absorption feature.
The resulting Gaussian centroid $\mu$, converted to SIR coordinates through the
FPA model, is a measurement of $x_{\text{r}}$ for this source
(Fig.~\ref{fig:opt_crv_thumb}, lower-right panel).

Similarly, to determine the position along the cross-dispersion \YSIR-axis, the
spectral region is first untilted according to the grism tilt angle and then
collapsed along the dispersion \XSIR-axis to produce a cross-dispersion
profile.  The centroid of a Gaussian fit on this profile defines the
$y_{\text{r}}$ coordinate of the spectrogram (Fig.\ref{fig:opt_crv_thumb},
lower-left panel).

Fits that do not meet specific quality criteria on amplitude, width, and
minimal number of data points are discarded from further analysis (typically
30\% of the fits).  Finally, the polynomials $\mathsf{P}^{(X, Y)}$ are derived
from $\chi^{2}$-minimisation between observed and expected positions
$(x_{\text{r}}, y_{\text{r}})$ for all selected sources in the \ac{FP}.  The
number of sources used for calibration depends on the specific pointing
sequences and consequently varies across grism configurations, between
\num{4300} and \num{5400} sources for the DR1 calibration.  The model fitting
procedures were performed using $6\times 6$ ($L=5$) $\mathsf{P}$~matrices.

\subsection{Curvature (CRV) model}
\label{sec:CRV}

The CRV pipeline aims at describing their spatial morphology, characterised by
the deviation of the first-order spectrogram traces from an ideal straight line
aligned with the horizontal \XSIR-axis (see Fig.~\ref{fig:opt_crv_thumb}).  The
CRV model describes these deviations and is fundamental for accurately tracing
the signal during the 2D~recti-linear resampling of the spectrograms and the
subsequent spectral extraction process \citepalias[see][]{Q1-TP006}.  Like the
OPT model calibration, this procedure relies on bright, non-saturated point
source spectrograms identified within the self-calibration
field~CALBLOCK-PV-001.

Starting from the first-order reference position of the spectrogram as
delivered by the OPT model, the pipeline then scans along the dispersion
direction \XSIR, extracting and collapsing spectrogram slices at predefined
intervals of 10~px along the spectrogram length -- a choice that captures the
spectrogram curvature while increasing profile \ac{S/N} and providing a
sufficient number of points along the trace.  The overall spectrogram length in
the dispersed images -- defined as the extended and measurable signal --
depends on the grism and on the position in the \ac{FP} (approximately 530~px
for the \acp{RGS} and 350~px for the \ac{BGS}).  The cross-dispersion
integrated profiles are then fitted with a Gaussian error-function (to account
for the marginally sampled profile), to generate
\begin{itemize}
\item a set of $\{\delta x_{i}\}$, the centre of each slice, corresponding to
  the distances along \XSIR-axis with respect to $x_{\text{r}}$ reference
  position,
\item a set of $\{\delta y_{i}\}$ which represent the offsets between the
  resulting Gaussian centroids $\{\mu_{i}\}$ and the $y_{\text{r}}$ reference
  position (Fig.~\ref{fig:opt_crv_thumb}).
\end{itemize}
The associated Gaussian widths $\{\sigma_{i}\}$ are also preserved for the
subsequent \ac{PSF} model (see Sect.~\ref{sec:PSF}).

For each individual source at reference position
$(x_{\text{r}}, y_{\text{r}})$, the set of discrete data points
$\{(\delta x_{i}, \delta y_{i})\}$ is used to model the spectral trace.
Specifically, a $N^{\text{th}}$-order Chebyshev polynomial of the first kind is
fitted to these points to describe the spatial offset $\delta y$ along the
\YSIR-axis as a function of the \XSIR-distance $\delta x$ from the reference
position,
\begin{equation}
  \label{eq:crv-loc}
  \delta y(x_{\text{r}}, y_{\text{r}}, \delta x) \approx
  \sum_{k=0}^{N} \alpha^{(k)}_{x_{\text{r}},y_{\text{r}}}\,T_{k}\left(\delta x\right)\;,
\end{equation}
where $T_{k}$ is the Chebyshev polynomial of order~$k$.  This is the so-called
`local' model, computed individually for selected calibration sources all over
the \ac{FP}.

Due to the optical distortions of the \ac{NISP} instrument, these curvature
coefficients $\alpha^{(k)}_{x_{\text{r}},y_{\text{r}}}$ vary across the
\ac{FP}.  To establish a `global' solution, which maps the entire \ac{FP}, each
local coefficient is itself modelled using a 2D $M^{\text{th}}$-order Chebyshev
polynomial as a function of the object's reference coordinates
$(x_{\text{r}}, y_{\text{r}})$,
\begin{equation}
  \label{eq:crv-glob}
  \alpha^{(k)}_{x_{\text{r}},y_{\text{r}}} \approx
  \sum_{i,j=0}^{M} a^{(k)}_{i,j}\,T_{i}(x_{\text{r}}) T_{j}(y_{\text{r}})\;.
\end{equation}
The $\mathsf{A}^{(k)} = [a^{(k)}_{i,j}]$ matrices for each grism-tilt
configuration are stored in the \texttt{DpdSirCrvModel} data product, with
$N=3$ and $M=2$ for \ac{DR1}.

As mentioned earlier, this global model, based on typically more than
\num{5500} objects over the full~\ac{FP} per grism configuration, allows the
prediction of the local curvature coefficients $\alpha^{(k)}_{x,y}$ for
\emph{any} source within the \ac{FP} -- including sources partially masked by
detector gaps or slightly outside of the \ac{FP} extent, -- ensuring a
continuous and accurate description of the spectroscopic traces; see
Sect.~\ref{sec:CRV_DQC} for illustrations of the CRV chromatic and spatial
performance.

During the \ac{DR1} production, it appeared that the CRV model, as initially
computed from CALBLOCK-PV-001 observations, was sometimes not reaching the
sub-pixel precision required by the optimal extraction procedure; this was
traced back to metrology model inaccuracies and \ac{GWA} sub-degree positioning
jitter, resulting in a small uncertainty on the mean dispersion direction.
This prompted the need for an additional \emph{ad hoc} residual adjustment per
exposure and per detector, based on an analysis of bright star spectrograms
\citepalias[see][]{DR1-TP023}; a more accurate CRV procedure is under
consideration for~\ac{DR2}.

\subsection{Point-spread function (PSF) model}
\label{sec:PSF}

The intrinsic resolution $\sigma$ of NISP-S is estimated from the
cross-dispersion width of point-source spectrograms \citepalias{Q1-TP006} at
the same time as the CRV calibration (Sect.~\ref{sec:CRV}).  The resulting
measurements, covering selected stars and wavelengths, are then robustly
combined into five spectral bins spanning the full spectral range.  At this
stage, the wavelength solution is not yet finalised (see Sect.~\ref{sec:IDS});
consequently, the conversion between detector position along the dispersion
direction ($\delta x$, in pixels) and wavelength $\lambda$ remains approximate,
though without any practical impact.

Similarly to the CRV expansions~(\ref{eq:crv-loc}) and (\ref{eq:crv-glob}), the
variations of the intrinsic resolution $\sigma(x, y, \lambda)$ as function of
the wavelength and of the position in the \ac{FP} are modelled using local
(chromatic) and global (spatial) Chebyshev polynomial expansions,
\begin{equation}
  \label{eq:psf}
  \sigma(x, y, \lambda) \approx \sum_{k=0}^{N} \gamma^{(k)}_{x,y}\,T_{k}(\lambda)
  \hfill\text{with}\hfill
  \gamma^{(k)}_{x,y} \approx \sum_{i,j=0}^{M} c^{(k)}_{i,j}\,T_{i}(x) T_{j}(y)\;.
\end{equation}
The $\mathsf{C}^{(k)} = [c_{i,j}^{(k)}]$ matrices are stored in the
\texttt{DpdSir\-Spectral\-PsfModel} product following the same structural
convention as the CRV model.  As anticipated from the instrument design and its
excellent optical quality, no meaningful differences in spectral resolution are
observed between the various red grism-tilt configurations
\citepalias{Q1-TP006}.

\subsection{Inverse dispersion solution (IDS) model}
\label{sec:IDS}

The final component of the geometric calibrations is the IDS, which provides
the spectroscopic wavelength solution, i.e., the mapping, for a given source,
between a given wavelength $\lambda$ and its corresponding physical
displacement $\delta x$ from the reference position $x_{\text{r}}$ in the~\ac{FP}.

The IDS calibration pipeline follows the fundamental principles of standard
spectroscopic wavelength calibration: by observing sources with known emission
lines, a relationship is established between the observed pixel positions and
the corresponding reference wavelengths.  In the SIR framework, the
transformation is defined from $\lambda$ to \ac{FP} coordinates \emph{at a
  given reference position}, in accordance with the geometric requirements
detailed in Sect.~\ref{sec:FPA}.  Since dispersed imaging cannot use
traditional arc lamps as reference, the procedure relies on astronomical
targets with well-characterised spectral properties, specifically \ac{PN}
SMC-SMP-20, acquired through dedicated observations of the
CALBLOCK-PV-009 field (see Sect.~\ref{sec:CalibObservingStrategy}).

\begin{figure}
  \centering
  \includegraphics[width=1.0\linewidth]{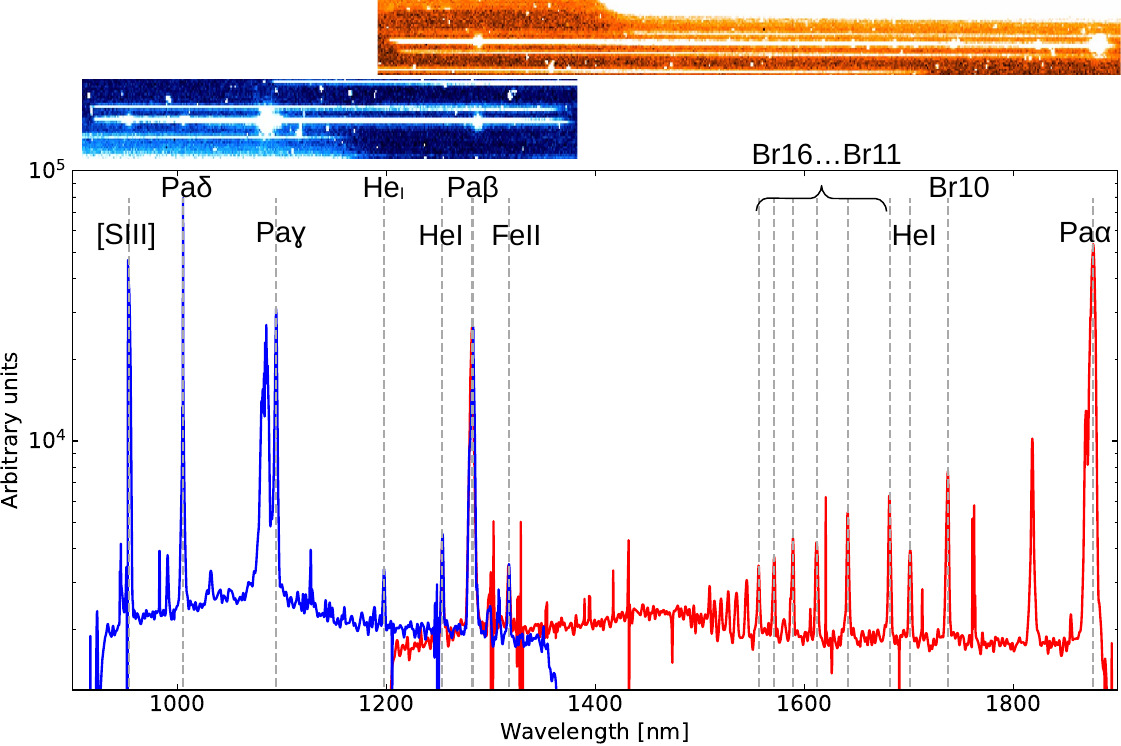}
  \caption{Raw spectrograms (top panel) and corresponding extracted spectra
    (bottom panel) from one observation of the wavelength standard \ac{PN}
    SMC-SMP-20 with the BGS000 and RGS180\texttt{+}0 grisms.  Dashed vertical
    lines correspond to the positions of the tabulated reference emission lines
    defined in the \ac{MDB} and used for wavelength calibration.  Thanks to the
    robustness of the emission line fitting procedure, the presence of a few
    saturated strong emission lines (e.g. Paschen lines) does not significantly
    impact the overall IDS accuracy.}
  \label{fig:ids_pn_example}
\end{figure}

The calibration is performed in parallel for each grism-tilt.  It starts with
the identification within the dispersed image of the reference \ac{PN} target
through a cross-match against the input MER catalogue.  \Euclid's \ac{MDB}
provides sky coordinates and reference vacuum wavelengths \citep{Paterson-EP32}
for selected emission lines of the \ac{PN} both in \BGE and \RGE ranges (see
Fig.~\ref{fig:ids_pn_example}).

Using the previously calibrated OPT and CRV models, along an initial guess of
the IDS derived during the \ac{PV} phase, 80~wavelength-calibrated spectra are
extracted for the \ac{PN} at different positions in the \ac{FP} (see
Fig.~\ref{fig:ids_pn_cross}).  For each spectrum associated with a \ac{PN}
reference position $(x_{\text{r}}, y_{\text{r}})$ in the \ac{FP}, a Gaussian
fit is performed within search windows centred on the reference wavelengths
tabulated in the \ac{MDB}; the resulting Gaussian centroids $\mu$ are assumed
to be the true position of the emission lines.  A local IDS is then derived by
fitting a Chebyshev polynomial of the first kind: the independent variable
corresponds to the tabulated wavelengths $\{\lambda^{R}\}$ (7 for \ac{BGS}, 11
for \acp{RGS}, see Fig.~\ref{fig:ids_pn_example}), while the dependent variable
represents the distance $\{\delta^{R} x\}$ of measured line positions
$\{\mu_{i}\}$ with respect to reference position~$x_{\text{r}}$,
\begin{equation}
  \label{eq:ids-loc}
  \delta^{R} x(x_{\text{r}}, y_{\text{r}}, \lambda^{R}) \approx
  \sum_{k=0}^{N} \beta^{(k)}_{x_{\text{r}},y_{\text{r}}}\,T_{k}\left(\lambda^{R}\right)\;.
\end{equation}
Once the 80~local models are established, a global spatial model is computed
using the same expansion described previously,
\begin{equation}
  \label{eq:ids-glob}
  \beta^{(k)}_{x,y} \approx \sum_{i,j=0}^{M} b^{(k)}_{i,j}\,T_{i}(x)T_{j}(y)\;.
\end{equation}
The resulting global matrices $\mathsf{B}^{k} = [b^{(k)}_{i,j}]$, which
represent the final IDS, mapping the wavelength $\lambda$ to the offset
$\delta x$ along dispersion direction at any reference position within the
entire \ac{FP}, are stored in the \texttt{DpdSirIdsModel} product.  The DR1
calibration assumes $M=4$ and $N=3$ for \acp{RGS}, and $N=2$ for \ac{BGS}.  The
smaller $N$ for \ac{BGS} accounts for the smaller number of reference emission
lines over the smaller wavelength domain.

An illustration of the accuracy -- mean and \ac{RMS} -- of the procedure as
function of position in the \ac{FP} is presented in Fig.~\ref{fig:ids_pn_cross}
for RGS000\texttt{+}0, where the accuracy remains within \si{0.24}{px}; other
grism-tilt configurations present similar behaviours (see
Sect.~\ref{sec:IDS_DQC}).

\section{Spectro-photometric calibrations}
\label{sec:SpectroPhoto}

The second suite of calibration pipelines comprises those dedicated to the flux
calibration process, responsible for converting instrumental signal units into
astrophysical flux units.  By characterising the overall instrumental response
and the variations in throughput, these workflows provide the necessary
sensitivity models and zero-points required to obtain spectra that are accurately
calibrated in flux.

\subsection{Detector-scaling model}
\label{sec:DETScaling}

The detector-scaling procedure aims to compensate for signal variations both
within and across individual detectors, primarily arising from pixel-specific
differences in \ac{QE} \citep{EU-Kubik}.  For the \ac{DR1}, a dedicated
in-flight calibration procedure is not implemented yet; instead, the correction
is still performed by combining ground-based calibration data \citep[stored in
\texttt{DpdSirDetectorScaling}]{EP-Gillard} with an estimate of the zodiacal
light spectrum derived from each individual exposure \citepalias{Q1-TP006}.

\subsection{Relative flux calibration (RFX) model}
\label{sec:RFX}

Large-scale transmission variations across the NISP \ac{FP} arise from a
combination of optical throughput, detector response, and vignetting.  These
variations can locally reach the $\sim 10\%$ level for the tilted-grism
configurations \citep{EuclidSkyOverview} and must be corrected to ensure
spatially uniform spectrophotometry.

\Euclid measures these variations using repeated observations of bright stars
(typically $16 < \HE < 18$) obtained at many dither positions for all
grism-tilt configurations in the self-calibration field.  Because each star is
observed multiple times at different \ac{FP} positions during these
self-calibration runs, the ensemble of observations provides strong statistical
constraints on the large-scale relative flux (RFX) solution that must be
applied.

At a high level, the calibration proceeds in three stages:
\begin{enumerate}
\item gather per-star and per-visit photometric measurements from the extracted
  spectra, in narrow wavelength bins;
\item fit a self-calibration model that simultaneously solves for the
  intrinsic magnitude of each calibration star, a smooth 2D large-scale flat
  pattern per detector, and per-detector \ac{QE}/gain offsets, using
  overlapping regions between detectors;
\item interpolate the fitted solution onto a regular \ac{FP} grid for each
  wavelength bin and grism-tilt configuration, producing a 3D relative flux
  solution with associated uncertainties and quality information.
\end{enumerate}

\begin{figure}
  \centering
  \includegraphics[width=1.0\linewidth]{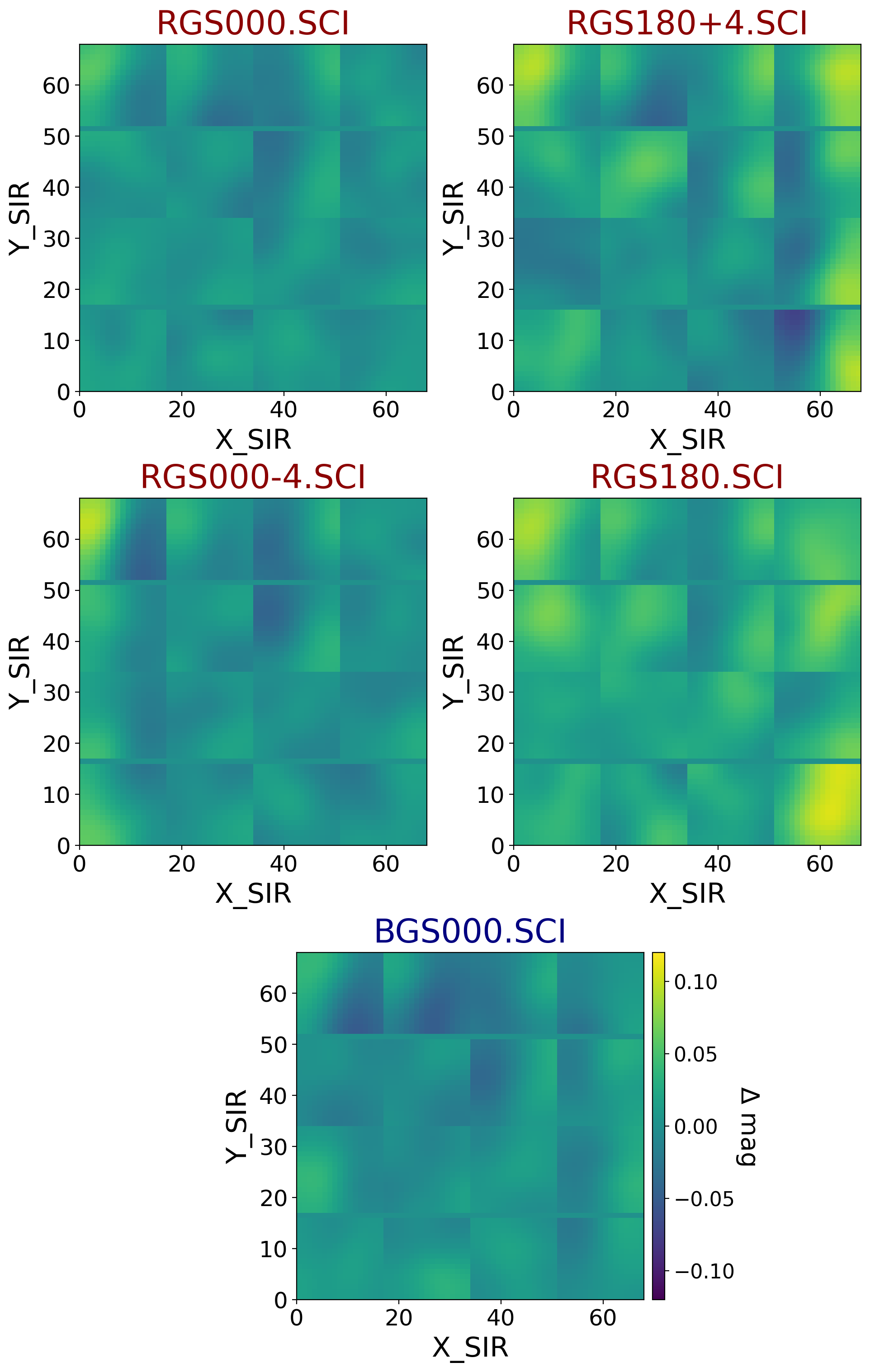}
  \caption{Wavelength-averaged relative flux solutions computed for the
    different grism configurations over the \ac{FP} for \ac{DR1} (all plots
    share the same colour scale).  Expected features such as vignetting at the
    edges in the tilted-grism configurations are apparent.}
  \label{fig:rfx}
\end{figure}

For \ac{DR1}, the relative response is found to be largely achromatic (but see
discussion in Sect.~\ref{sec:conclusion}); to maximise the \ac{S/N} on the
solution, a wavelength-averaged solution is therefore derived that depends only
on \ac{FP} coordinates for each grism-tilt configuration (Fig.~\ref{fig:rfx}).
This solution (stored in \texttt{DpdSirRelativeFluxScaling}) successfully
flattens repeated observations of bright sources and shows clear signatures of
the expected vignetting pattern from optical modelling.  The module derives and
applies large–scale relative flux corrections to NISP slitless spectra so that
repeated observations of the same source yield consistent instrumental fluxes,
independent of detector position, grism-tilt configuration, or epoch.

Potential time variability of the relative flux solution has not been measured
for DR1, and is being investigated for DR2.  However, tests on flux stability
across the survey (see Sect.~\ref{sec:AFX_DQC}) indicate small variation in the
solution.

\subsection{Absolute flux calibration (AFX) model}
\label{sec:AFX}

The absolute flux (AFX) calibration pipeline is designed to create a
sensitivity function that is used to convert instrumental signal units
(electrons per sampling element) into astrophysical flux units
(\si{erg.s^{-1}.cm^{-2}.\AA^{-1}}).  The sensitivity function is created by
first averaging repeated observations of the flux calibration star
GRW+70\,5824, a DA2.4 \ac{WD} \citep{2011ApJ...743..138G} acquired during the
\ac{PV} phase in a quincunx pattern on each detector \citep{EuclidSkyNISP},
independently for each of the five grism-tilt combinations
(RGS000\texttt{+}0/\texttt{-}4, RGS180\texttt{+}0/\texttt{+}4 and BGS000).  For
\ac{DR1}, a separate sensitivity function is created for each grism-tilt
combination (four red and one blue) from the pull-clipped weighted average of
$16 \times 5 = 80$~single-dither spectra (after relative flux scaling) of the
standard star.

\begin{figure}
  \centering
  \includegraphics[width=\linewidth]{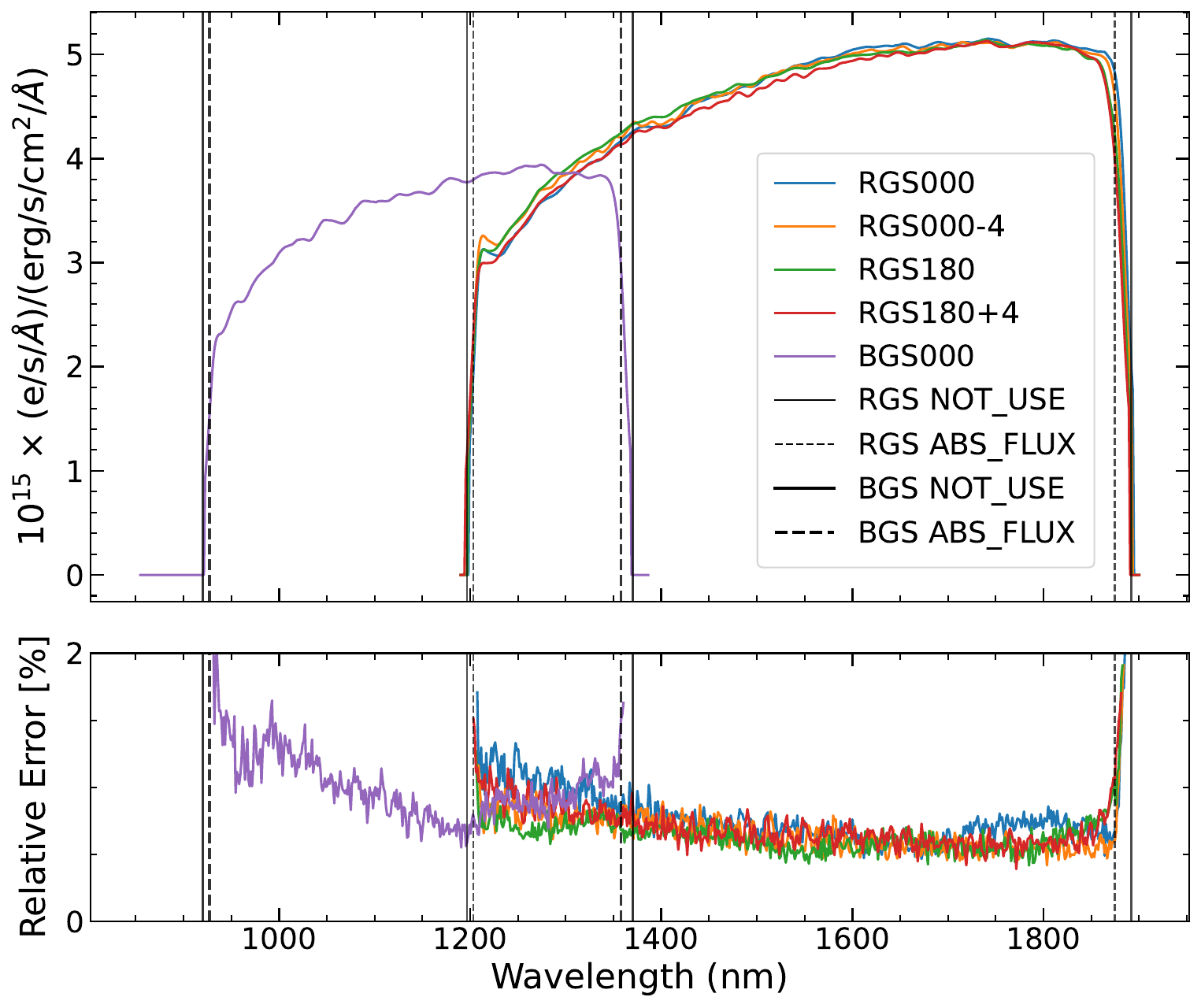}
  % \sidecaption
  % \includegraphics[width=12cm]{all_grisms_sens.png}
  \caption{\emph{Top panel}: sensitivity functions derived from \ac{PV}
    observations of the flux standard \ac{WD} GRW+70\,5824, for the five
    grism-tilt configurations (see legend); vertical solid (resp. dotted) lines
    denote the 25\% (resp.~85\%) throughput masking boundaries, based on the
    local maxima at the two ends of the bandpass.  Spectral pixels outside the
    black lines are flagged as unusable (\protect\path{NOT_USE} bit) in the SIR
    pipeline, and those falling between the black and dotted lines are flagged
    as `suspicious' (\protect\path{ABS_FLUX} bit), since the throughput is
    falling rapidly and deemed uncertain there.  \emph{Bottom panel}: relative
    error in the measured sensitivity functions for the five grism-tilt
    combinations. This plot represents a measure of the scatter in the observed
    optimally-extracted spectra of the bright \ac{WD} used for calibration.}
  \label{fig:sensfunctions}
\end{figure}

The single-dither spectra of the reference star are first extracted (using the
same optimal extraction as for science pipeline, see \citetalias{DR1-TP023})
and averaged, and the instrumental flux units are converted into units of
\si{e^{-}.s^{-1}.\AA^{-1}}.  These spectra are then divided by a suitably
matched reference spectrum, namely model spectrum
\path{grw_70d5824_mod_001.fits} from CALSPEC\footnote{%
  \url{https://archive.stsci.edu/hlsps/reference-atlases/cdbs/current_calspec/}}
\citep{2020AJ....160...21B}, first resampled to \SI{0.1}{nm}, convolved to an
effective spectral resolution of \SI{3.3}{nm} (close to the spectral resolution
of the \ac{RGS} for a point source), and finally rebinned onto the SIR
wavelengths.  The \ac{DR1} sensitivity curves for all five grism-tilt
configurations are stored in the \texttt{DpdSirAbsoluteFluxScaling} and shown
in Fig.~\ref{fig:sensfunctions} (top panel).

This procedure assumes the sensitivity curve is uniform across the \ac{FP};
this is confirmed across most of the spectrum at the $< 1\%$-level through
observations of the \ac{WD}.  The sensitivity function, however, shows
\ac{FP}-dependent variations at the edges of the bands for both the \RGE and
\BGE domains, but this affects only a few spectral pixels.  The net effect
would be increased noise in the extracted (pre-combined) spectrum for a few
pixels near the edge of the band for a minority of \ac{FP} source positions.
This effect is being handled by special masking in future releases of the SIR
pipeline.

\begin{figure}
  \centering
  \includegraphics[width=\linewidth]{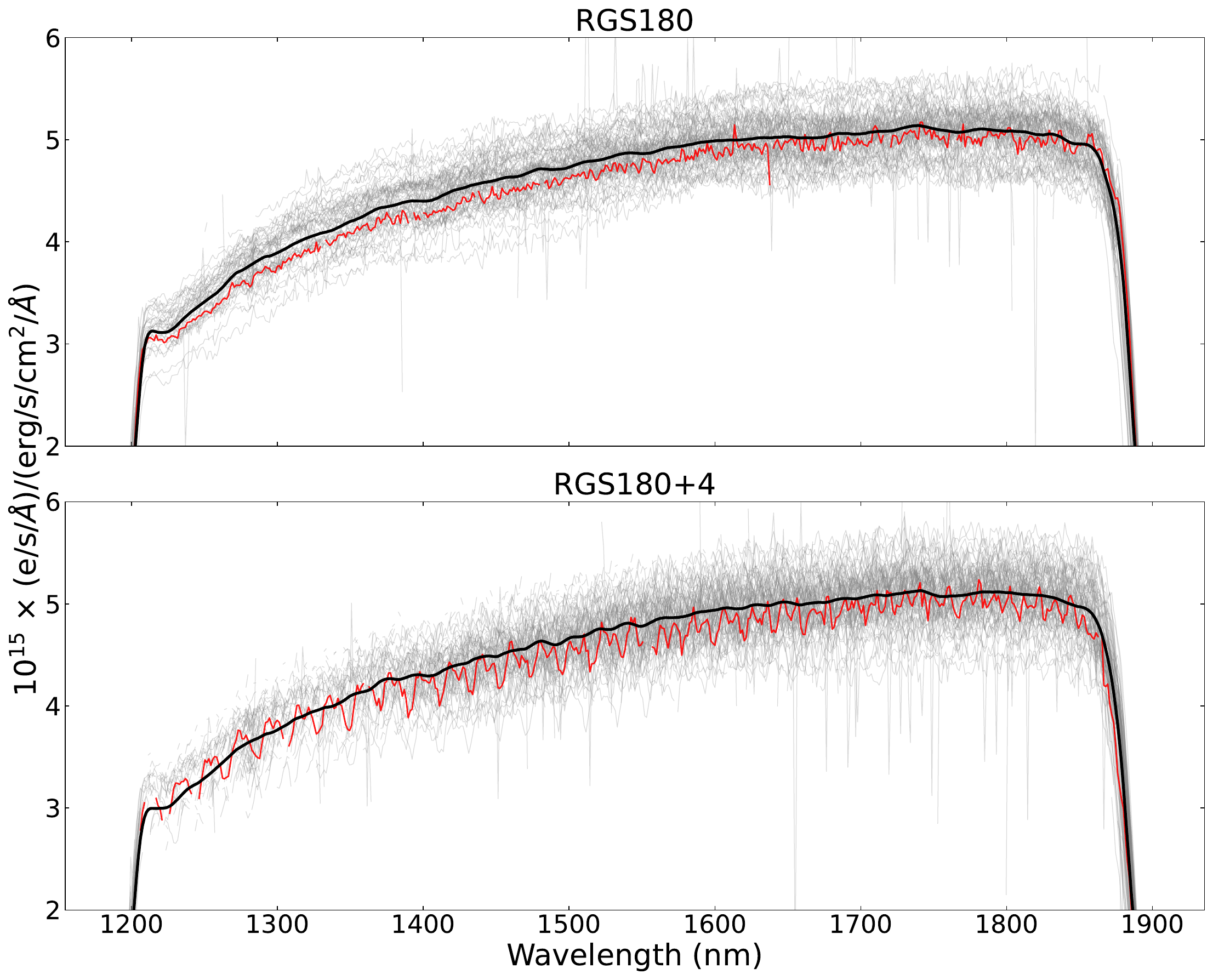}
  % \sidecaption
  % \includegraphics[width=12cm]{all_grisms_sens.png}
  \caption{\emph{Upper panel}: The spread in the input sensitivity functions
    (grey lines) for the RGS180 grism based on 80 white dwarf spectra extracted
    using the optimal extraction technique.  The black line shows the
    pull-clipped average which is used to define the final sensitivity function
    used in \ac{DR1} (see Fig.~\ref{fig:sensfunctions}); the red line
    highlights a typical single example from the full set of sensitivity
    functions before averaging.  \emph{Lower panel}: Same as top panel, but for
    the tilted RGS180\texttt{+}4 grism.  The tilted grisms show low-amplitude
    wiggles cause by resampling Moiré fringing, leading to a very small extra
    ripple in the final sensitivity function at the $< 0.5\%$ level (see text);
    the amplitude and period of the fringing are more obvious on the
    highlighted example of single sensitivity function (red line).}
  \label{fig:sens_distribution}
\end{figure}

It is noted that the spectral extraction, based on a 2D-resampled spectrogram
of a bright (16th magnitude) point source, led to some extra low frequency
pattern being added to the spectra used to calculate the sensitivity function
(see Fig.~\ref{fig:sens_distribution}).  Although this structure was seen on
some individual spectra (especially the tilted configurations), much of it is
suppressed when averaging (for a given grism) the sensitivity functions over
all available observations.  The pattern is the result of Moiré fringing caused
by the under-sampling of the \ac{PSF} of the star and the slight tilt of the
spectrograms \citepalias{DR1-TP023}; we are investigating ways to ameliorate
this effect in future analyses.  However, a small (but statistically
insignificant) ripple is introduced into the final sensitivity functions for
the two tilted \acp{RGS}.

Here we discuss only the formal uncertainties in the sensitivity functions.
Figure~\ref{fig:sensfunctions} (bottom panel) shows the relative error
(\ac{RMS} scatter in the observations divided by the final sensitivity value,
expressed as a percentage) as a function of wavelength for each of the
grism-tilt configuration.  This result is obtained using the optimal extraction
method on the \ac{WD} spectra.  As can be seen from the figure, the relative
error is mostly below 1\%, except for \ac{BGS} for which it reaches 1.5\% for
the bluest wavelengths.  The results are still above mission requirement
(0.5\%), but we think that this does not undermine the validity of the final
sensitivity function for two reasons: (i) the final sensitivity functions are
fitted with a noiseless function to avoid imprinting any noise onto the
calibrated spectra, and (ii) we have shown through further analysis, that the
relative error is inflated because a minority of these very bright \ac{WD}
spectra are slightly displaced from the expected \ac{FP} positions when optimal
extraction is made.  Optimal extraction is very sensitive to small
displacements in the assumed cross-dispersion profile, leading to a systematic
underestimate of the extracted fluxes.  This effect causes a small percentage
($\sim 5\%$) of the spectra to show enhanced scatter compared with the vast
majority of \ac{WD} spectra.  However, since we generate the mean sensitivity
function using a pull-clipping algorithm \citepalias[see][]{Q1-TP006}, the
final result is almost unaffected by this problem, even if it does affect the
formal uncertainty (see Fig.~\ref{fig:sens_distribution}).  SIR is
investigating improvements to the optimal extraction method for very bright
sources, which should bring the final relative error below the formal
requirement.

\section{Quality of calibrated data}
\label{sec:DataQuality}

Upon completion of the calibration processes, a set of \ac{DQC} parameters is
computed to identify potential anomalies, in the generated products and to
verify compliance with the mission requirements.  A dedicated \ac{DQC}
pipeline, using \ac{DQC} values embedded in the calibration products, generates
detailed diagnostic reports, essential for recognizing peculiarities, and
monitors the quality parameters associated with each product.

The overall quality of the calibration products is intrinsically linked to the
final performance and accuracy of the extracted spectra.  Any residual
systematic error in the calibration models directly propagates into the
scientific data, potentially affecting the precision of wavelength and flux
measurements.

\subsection{Focal plane assembly (FPA) model}
\label{FPA_DQC}

\begin{figure}
  \centering
  \includegraphics[width=\linewidth]{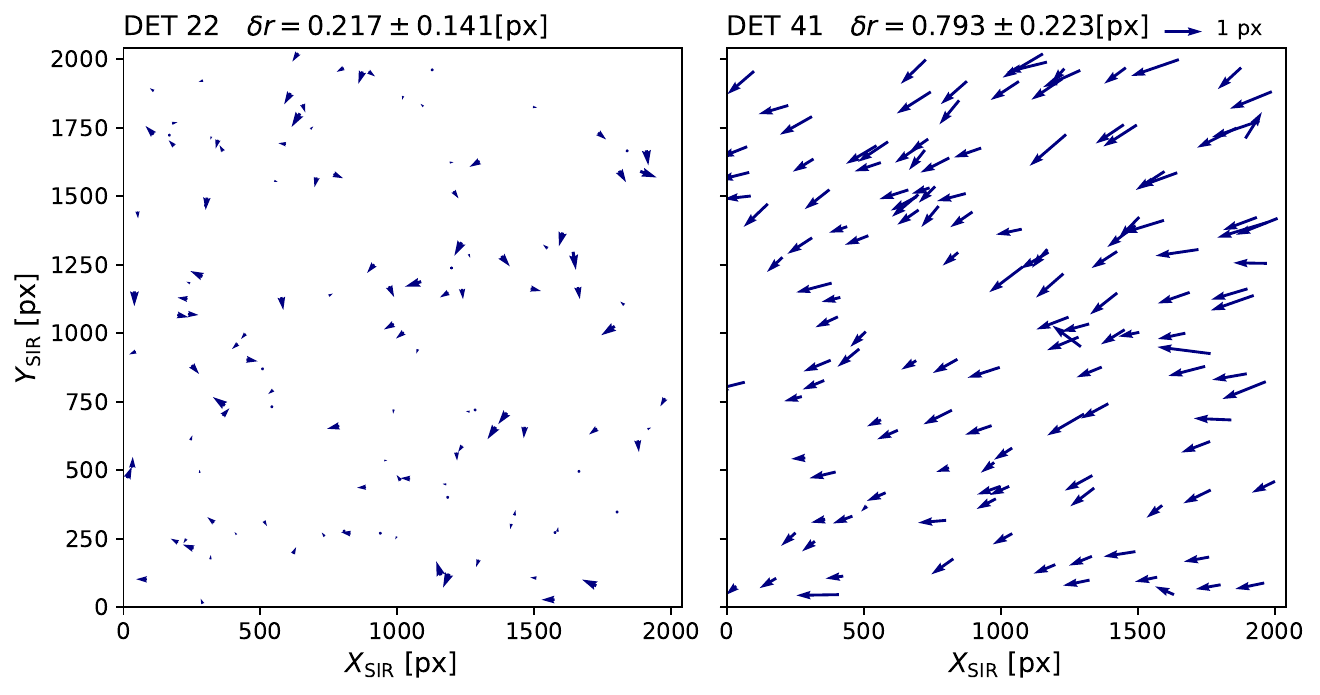}
  \caption{Residuals as a function of object position on two detectors.  The
    \emph{left panel} presents detector 22, in which the residual distribution
    lacks any clear structure.  Conversely, the \emph{right panel} shows
    detector 41 -- located at a corner of the \ac{FP} -- where a structural
    pattern is evident.  Such residuals on the detectors near the edges of the
    \ac{FP} inevitably degrade the accuracy of object localization.  We
    reiterate that this effect is significantly mitigated by the OPT model.
  }
  \label{fig:fpa}
\end{figure}

Although \ac{DR1} still lacks a dedicated calibration pipeline for the FPA
model, evaluating its current performance remains crucial.  To this end, we
used a \texttt{DpdNirCalibratedFrameCatalog} \citep{DR1-TP003} containing
paired celestial $(\alpha, \delta)$ and detector pixel
$(x_{\text{px}}, y_{\text{px}})$ coordinates for a given frame.  We applied the
FPA model to convert pixel coordinates into \ac{FP} positions
$(x_{\text{mm}}, y_{\text{mm}})$ and fit a global \ac{WCS}, $\mathcal{G}$,
mapping the sky and the \ac{FP}.  Concatenating $\mathcal{G}$ with the FPA
model yields predicted pixel positions
\begin{equation}
  \label{eq:fpa}
  (\alpha,\delta)
  \xrightarrow{\mathcal{G}} (\hat{x}_{\text{mm}},\hat{y}_{\text{mm}})
  \xrightarrow{\text{FPA}} (\hat{x}_{\text{px}},\hat{y}_{\text{px}})\;.
\end{equation}
For each detector, we computed the distances $\delta r$ between measured
$(x_{\text{px}}, y_{\text{px}})$ and expected
$(\hat{x}_{\text{px}}, \hat{y}_{\text{px}})$ positions (in pixels).  A deeper
inspection reveals spatial structures in the residuals of certain detectors
that a simple translation in the current model fails to capture (see
Fig.~\ref{fig:fpa}); nevertheless, these remaining uncertainties are
subsequently partially absorbed by the OPT, CRV and IDS models.  Clearly, these
global models cannot fully account for local variations that differ from one
detector to another.  Consequently, certain structures in the $\delta r$
distribution will persist, even if mitigated.

\subsection{Optical (OPT) model}
\label{sec:OPT_DQC}

The OPT calibration pipeline uses the geometric models~(\ref{eq:opt_wcs}) and
(\ref{eq:opt_poly}) to compare the predicted and observed positions of key
spectral reference points, specifically a peak of the zeroth-order spots and a
stellar absorption feature within the first-order spectra (see
Sect.~\ref{sec:OPT}).

The model's accuracy is evaluated using the distribution of residuals
$(\Delta X, \Delta Y)$ to the reference positions, which are pre-processed to
remove outliers via the \ac{IQR} method: data points are considered as outliers
and rejected if they fall outside the interval
$[Q_{25\%} \pm 1.5\;\text{IQR}]$.  This clipping procedure ensures that the
final statistical estimators are representative of the core model performance,
effectively mitigating the impact of spurious spectral measurement or
zeroth-order spot misidentifications.  Table~\ref{tab:opt} summarises the
median and the \ac{sMAD} of these distributions, both in \XSIR{} (along
dispersion) and \YSIR{} (across dispersion) directions, for the zeroth- and
first-order reference positions.

% --- OPT residuals ---
\begin{table*}
  \centering
  \caption{Statistical summary (median $\pm$ \ac{sMAD}, in pixels) of the OPT
    zeroth- and first-order residuals along ($\Delta X$) and across
    ($\Delta Y$) dispersion, averaged over the full \ac{FP} for the five
    different grism-tilt configurations.}
  \label{tab:opt}
  \sisetup{table-format=1.4(1), separate-uncertainty}
  \begin{tabular}{@{}lSSSS@{}}
    \toprule
    & \multicolumn{2}{c}{{Zeroth-order}}
    & \multicolumn{2}{c}{{First-order}} \\
    \cmidrule(lr){2-3} \cmidrule(lr){4-5}
    Configuration
    & {$\Delta X$ [px]}
    & {$\Delta Y$ [px]}
    & {$\Delta X$ [px]}
    & {$\Delta Y$ [px]} \\
    \midrule
    BGS000            & -0.004(0.121) &  0.003(0.130) & -0.029(0.305) &  0.000(0.152) \\
    RGS000\texttt{+}0 &  0.004(0.128) & -0.003(0.131) &  0.013(0.230) & -0.003(0.135) \\
    RGS000\texttt{-}4 &  0.008(0.127) &  0.026(0.143) &  0.013(0.281) & -0.005(0.147) \\
    RGS180\texttt{+}0 &  0.006(0.126) & -0.000(0.136) & -0.012(0.231) & -0.015(0.138) \\
    RGS180\texttt{+}4 &  0.006(0.123) & -0.002(0.130) & -0.011(0.283) &  0.015(0.161) \\
    \bottomrule
  \end{tabular}
\end{table*}

\begin{figure}
  \centering
  \includegraphics[width=\linewidth]{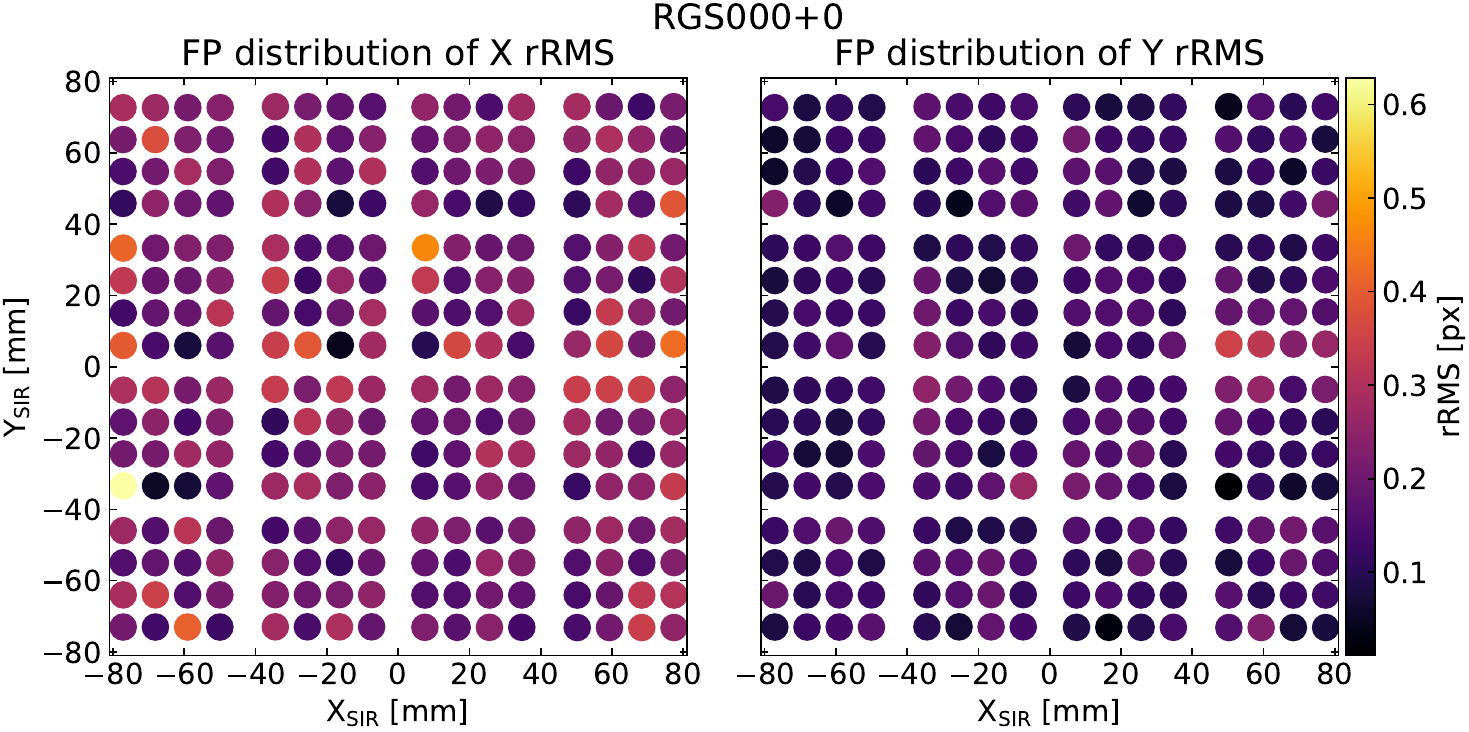}
  \caption{Spatial statistics of the residuals (\emph{left}: $\Delta X$,
    \emph{right}: $\Delta Y$, both in pixels) between measured and OPT-expected
    positions for the first-order reference position of the RGS000\texttt{+}0
    configuration.  For each detector, all valid measurements were collected
    and partitioned into a $4\times 4$ grid of bins.  The robust \ac{RMS}
    (rRMS), defined as $\sqrt{\text{median}^{2} + \text{sMAD}^{2}}$, of the
    residual positions have been calculated.  Other grism-tilt configurations
    show similar results.}
  \label{fig:opt_spatial}
\end{figure}

Figure~\ref{fig:opt_spatial} shows that first-order residuals
$(\Delta X, \Delta Y)$ exhibit no significant trends across the \ac{FP}.  The
larger deviations observed at the edges of few individual detectors are
attributed to current limitations in the FPA model, resulting in fictitious
discontinuities in the global model of the \ac{FP} (see
Sect.~\ref{sec:conclusion}).

The residual statistics for the first-order reference positions demonstrate
that the first stage for the geometrical calibrations, which define the
detailed spectrogram geometry and wavelength solution, is well within the
required tolerances.  It should be further noted that the quoted precision of
the OPT model in the \XSIR{} direction does not represent a hard limit for the
accuracy of the wavelength calibration, since the zero-point term of the IDS
model -- $\beta^{(0)}_{x,y}$ in Eq.~(\ref{eq:ids-glob}) -- can still compensate
for residuals in the reference position along the dispersion axis;
specifically, $\delta^{R}x(\lambda_{\text{r}})$ is not constrained to zero.

\subsection{Curvature (CRV) model}
\label{sec:CRV_DQC}

The CRV model calibration pipeline uses the current model to predict the
expected trace positions for bright, point-like sources.  For each of these
reference objects, the pipeline performs a direct comparison between the
CRV-predicted and measured positions at regular intervals along the dispersion
direction.  Table~\ref{tab:crv} summarises the distribution of these positional
residuals over the full \ac{FP} -- both before ($\Delta Y_{\bullet}$ column)
and after ($\Delta Y_{\star}$ column) the \emph{ad hoc} CRV correction
discussed below --, categorised by grism, highlighting the statistical offset
and the dispersion of the trace modelling accuracy.

% --- CRV residuals ---
\begin{table}
  \centering
  \caption{Same as Table~\ref{tab:opt} but for the CRV first-order
    cross-dispersion residuals ($\Delta Y$), averaged over the full \ac{FP} and
    spectrogram extent. Values are reported both before ($\Delta Y_{\bullet}$)
    and after ($\Delta Y_{\star}$) the \emph{ad hoc} CRV correction.}
  \label{tab:crv}
  \sisetup{table-format=1.4(1), separate-uncertainty}
  \begin{tabular}{@{}lSS@{}}
    \toprule
    Configuration
    & {$\Delta Y_{\bullet}$ [px]} & {$\Delta Y_{\star}$ [px]} \\
    \midrule
    BGS000            & -0.000(0.147) & 0.000(0.104)\\
    RGS000\texttt{+}0 & -0.002(0.136) & 0.000(0.104)\\
    RGS000\texttt{-}4 &  0.009(0.144) & 0.000(0.100)\\
    RGS180\texttt{+}0 &  0.023(0.132) & 0.000(0.101)\\
    RGS180\texttt{+}4 & -0.018(0.164) & 0.000(0.107)\\
    \bottomrule
  \end{tabular}
\end{table}

\begin{figure}
  \centering
  \includegraphics[width=\linewidth]{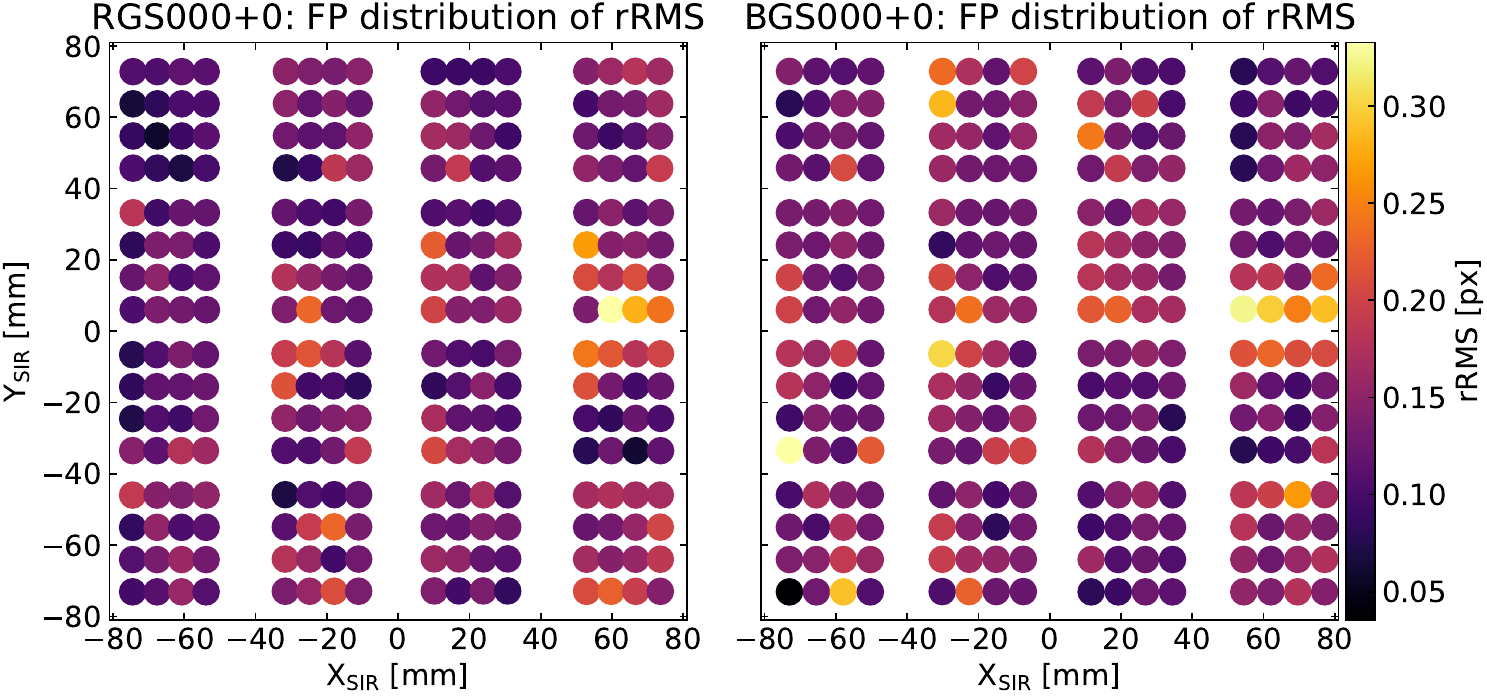}
  \caption{Same as Fig.~\ref{fig:opt_spatial}, but for the CRV rRMS of
    cross-dispersion residuals $\Delta Y$ for the RGS000\texttt{+}0
    (\emph{left}) and BGS000 (\emph{right}) configurations.}
  \label{fig:crv_spatial}
\end{figure}

Figure~\ref{fig:crv_spatial}, analogously to Fig.~\ref{fig:opt_spatial},
illustrates the spatial distribution of the robust \ac{RMS} residuals of
$\Delta Y$ across the \ac{FP}; no significant systematic trends are observed.
The larger residuals noted at the detector edges are likely attributable to the
limitations of the current FPA model, which does not yet fully account for all
detector misalignments (improvements of this model are discussed in the
Sect.~\ref{sec:conclusion}).

\begin{figure}
  \centering
  \includegraphics[width=\linewidth]{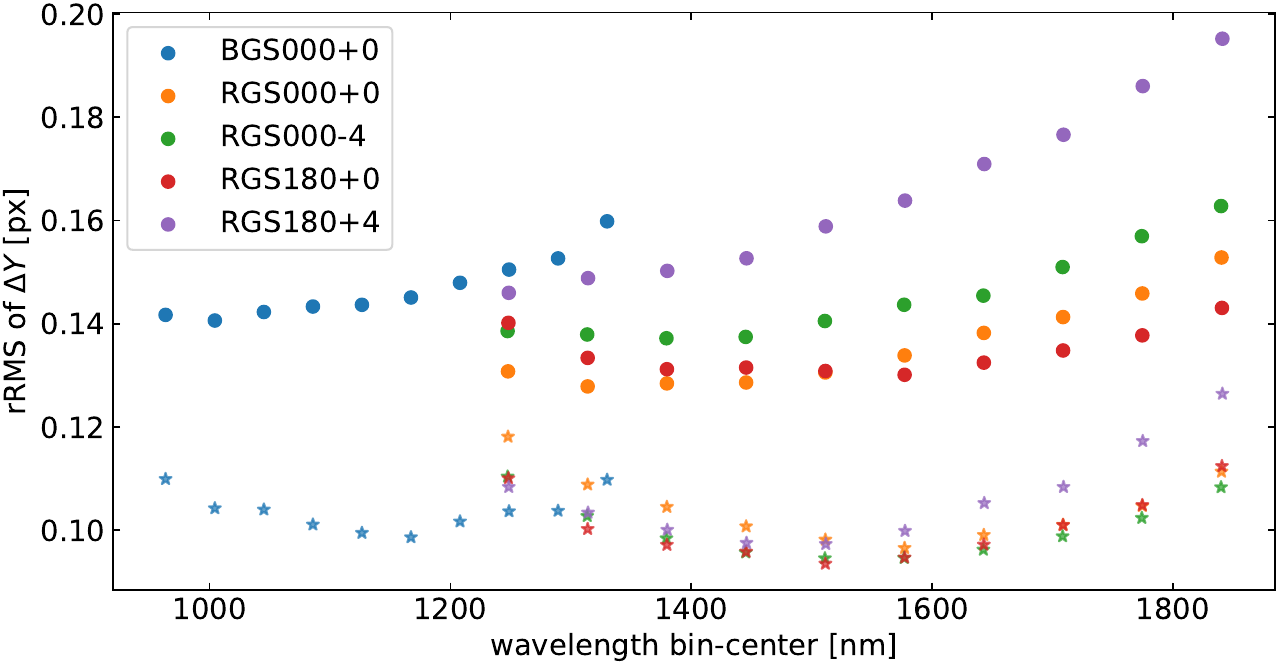}
  \caption{\emph{Circles}: \emph{observed} robust \ac{RMS} of CRV residuals
    between predicted and measured trace positions as a function of wavelength.
    For each bright point-like object used to calibrate the CRV model, the
    cross-dispersion residuals are measured at regular intervals along the
    dispersion direction \XSIR{} (ranging from \SI{-215}{px} to \SI{+275}{px}
    with respect to reference $x_{\text{r}}$ for the \acp{RGS}, and from
    \SI{-275}{px} to \SI{55}{px} for the \ac{BGS}).  The rRMS of these
    quantities is subsequently aggregated into 10~bins and the bin centre is
    converted to approximate wavelength using the \emph{nominal} wavelength
    solution (at this stage a reliable IDS is not computed yet).  \emph{Stars}:
    \emph{estimated} rRMS after the \emph{ad hoc} CRV correction is applied
    during the reduction process (see Sect.~\ref{sec:CRV}).}
  \label{fig:crv_chroma}
\end{figure}

An analysis of the rRMS for each grism as a function of the $\delta x$-distance
from the reference position -- equivalent to evaluating evolutions across the
wavelength range -- reveals that the residuals are consistently distributed
below the mission requirement of~\SI{0.3}{px} at all wavelengths (see
Fig.~\ref{fig:crv_chroma}).  This analysis also reveals a slight colour
dependent deflection limited to a few hundredths of a pixel.

As mentioned in Sect.~\ref{sec:CRV} and detailed in \cite{DR1-TP023}, it should
be noted that an \emph{ad hoc} CRV correction is applied to the DR1 data on a
per-pointing and per-detector basis to account for metrology inaccuracies and
\ac{GWA} jitter.  Consequently, the initial CRV residuals, although already
within the mission requirements, are further improved by this refinement
procedure, as reported in Table~\ref{tab:crv} and in Fig.~\ref{fig:crv_chroma}.

\subsection{Inverse dispersion solution (IDS) model}
\label{sec:IDS_DQC}

\begin{figure}
  \centering
  \includegraphics[width=\linewidth]{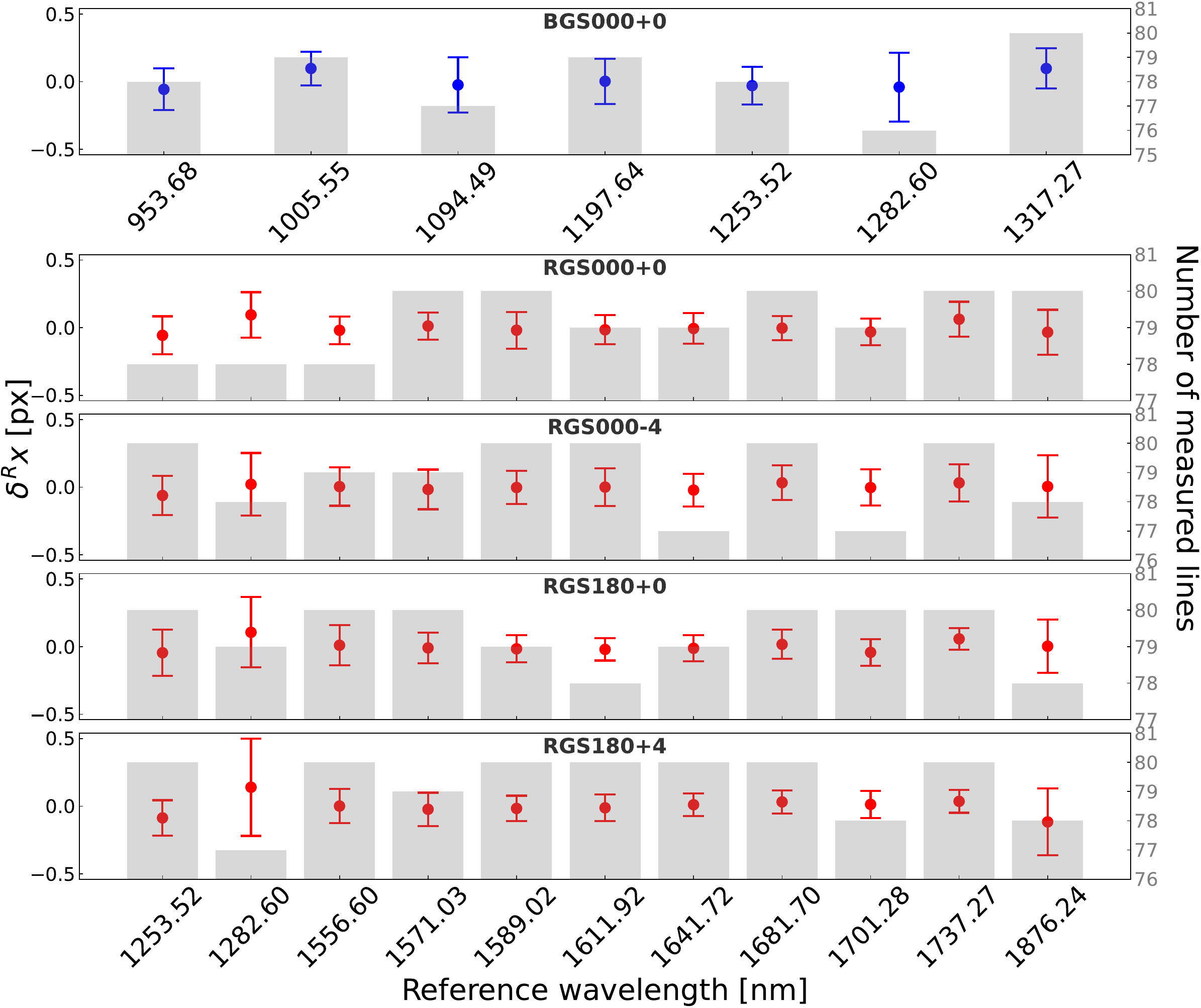}
  \caption{Median (\emph{circles}) and \ac{sMAD} (\emph{error bar}) of the
    $\delta^{R} x$ residuals (in pixel) between expected and observed positions
    for each of the reference lines of the \ac{PN} used in the wavelength
    calibration of the five grism-tilt configurations.  The grey histograms
    (\emph{right axes}) indicate the number of selected positions used for each
    wavelength, out of the 80 initial measurements, as certain line
    measurements may fail due to cosmic rays, detector regions with bad pixels,
    or saturation effects.}
  \label{fig:ids_ALL}
\end{figure}

Similarly to the other geometric models, the performance of the IDS is
evaluated by assessing its accuracy in predicting the positions of the emission
lines for the observed \ac{PN}.  For each \ac{PN} and every emission line
utilised for calibration, the positions on the frame are measured. These
observed positions are then compared with the values predicted by the model,
followed by a detailed analysis of the residuals.  The distribution of these
residuals is presented in Fig.~\ref{fig:ids_ALL} for five grism-tilt
configurations, and a statistical summary averaged over all the reference lines
is presented in Table~\ref{tab:ids}.  We note that all residuals are below the
mission requirement of \SI{0.3}{px}, that is \SI{0.37}{nm} for blue grism and
\SI{0.40}{nm} for red grisms.

\begin{table}
  \centering
  \caption{Statistical summary (median and \ac{sMAD} of rRMS) of the residuals
    between the expected and measured positions for the emission lines utilised
    during the \ac{PN} calibration (see Fig.~\ref{fig:ids_ALL}).  These
    residuals all range within the mission requirement on wavelength
    precision.}
  \label{tab:ids}
  \sisetup{table-format=1.4(1), separate-uncertainty}
  \begin{tabular}{@{}lS@{}}
    \toprule
    Configuration
    & {rRMS [nm]} \\
    \midrule
    BGS000            &  0.205(0.021) \\
    RGS000\texttt{+}0 &  0.152(0.046) \\
    RGS000\texttt{-}4 &  0.189(0.018) \\
    RGS180\texttt{+}0 &  0.146(0.024) \\
    RGS180\texttt{+}4 &  0.137(0.034) \\
    \bottomrule
  \end{tabular}
\end{table}

In the \acp{RGS}, the reference line at \SI{1282.596}{nm} appears to be
problematic, with \ac{sMAD} sometimes larger than 0.3~px; however, this is
attributed to line measurement challenges rather than IDS model inaccuracy, as
this line is blended with a line at \SI{1279.281}{nm}, while the DR1 procedure
does not yet account for blended lines.  Overall, the polynomial local model
shows excellent agreement with the experimental data.

\subsection{Flux calibrations}
\label{sec:AFX_DQC}

The stability of the flux solution was examined by checking the scaling of
measured SIR fluxes for stars with \textit{Gaia} \citep{2023A&A...674A...1G} as a
function of sky position and time.  \Euclid galaxy clustering science puts a
tight constraint on allowable spatial or temporal variations in the flux
solution, with allowable fluctuations below~0.7\%.

\begin{figure*}
  \sidecaption
  \includegraphics[width=12cm]{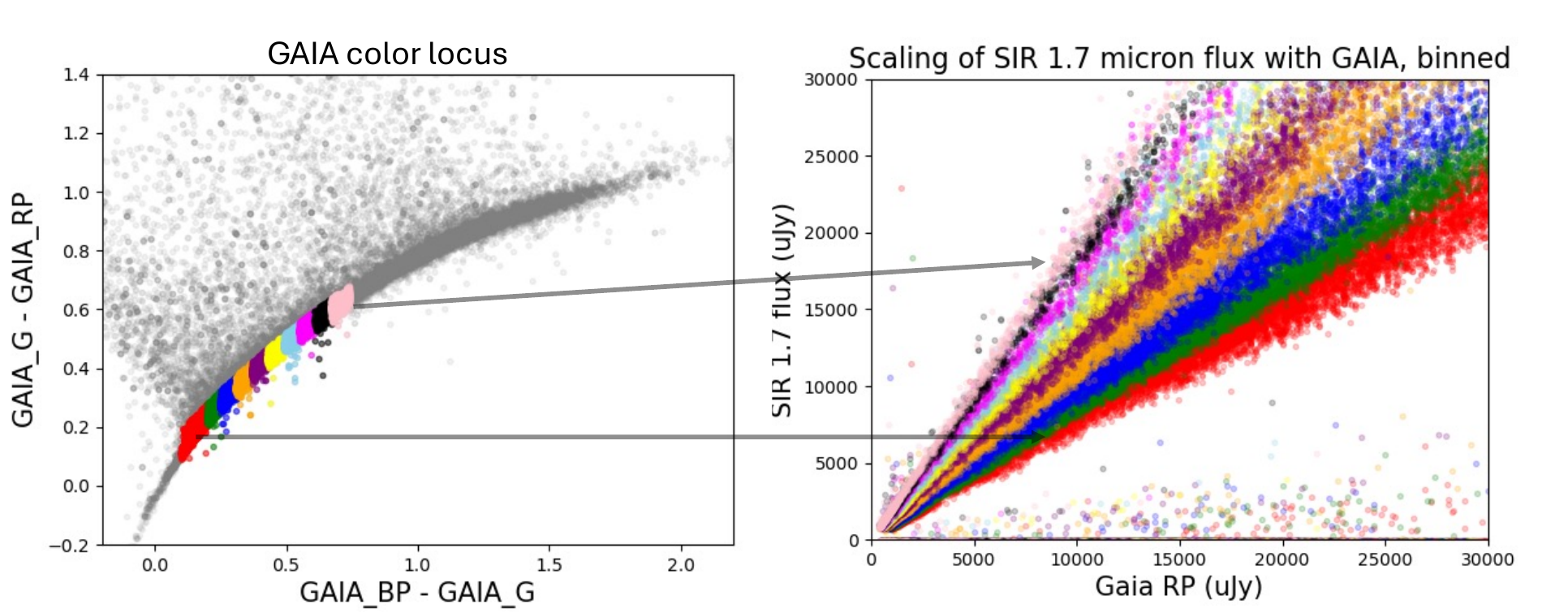}
  \caption{Illustration of how the SIR/\textit{Gaia} flux ratio depends
    strongly on colour locus position.  The different coloured groups, selected
    to fall within relatively narrow bins along the colour locus
    $G-G_{\mathrm{RP}}$ versus $G_{\mathrm{BP}}-G$, correspond to well-defined
    scalings of SIR/\textit{Gaia} flux. This tight scaling is used to infer the
    stability of the SIR fluxes across the DR1 footprint.
  }
  \label{fig:gaia_scaling}
\end{figure*}

\begin{figure*}
    \sidecaption
    \includegraphics[width=12cm]{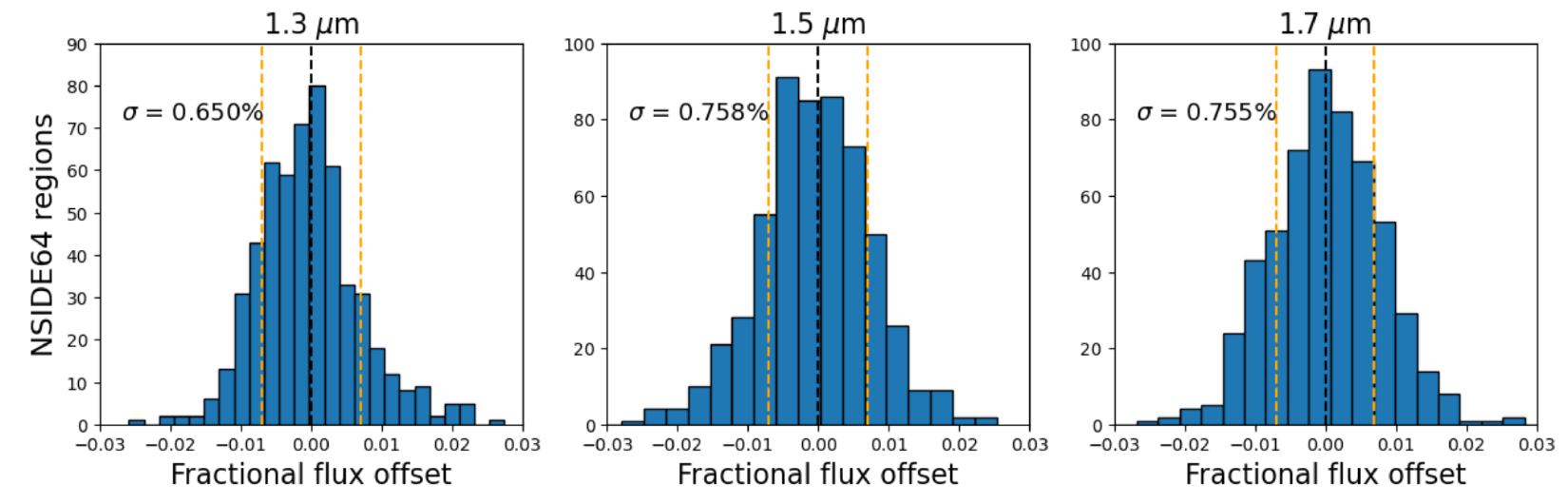}
    \caption{Comparison of the scaling of \textit{Gaia} with SIR flux as a
      function of stellar locus position independently in
      $\sim \SI{0.8}{deg^{2}}$ regions across the DR1 footprint that are
      minimally impacted by dust -- where the average $E(B-V)$ is less
      than~0.02.  Shown here are the distributions of flux scaling offsets
      compared with the average scaling over all fields for three SIR
      wavelength bins.  The spreads in the distributions are $\lesssim 0.7\%$,
      consistent with meeting the flux stability requirement for clustering
      cosmology.}
    \label{fig:flux_stability_results}
\end{figure*}

The flux stability analysis looked at the relation between \textit{Gaia} and
SIR flux as a function of stellar type, as encoded by position on the stellar
locus in $G - G_{\text{RP}}$ versus $G_{\text{BP}} - G$.  The tight relation
between star position on the \textit{Gaia} stellar locus and ratio of SIR flux
to \textit{Gaia} flux is illustrated in Fig.~\ref{fig:gaia_scaling}.  We used
$\sim\num{6e5}$~stars spread over \ac{DR1} to check for deviations in the
relation.  To remove galactic extinction as a confounding factor, we divided
the sky into HEALPix tiles of $\sim \SI{3.36}{deg^{2}}$
($N_{\text{side}} = 32$), computed the average galactic $E(B - V)$ within each
tile, and restricted the analysis to pixels with average $E(B - V) < 0.02$.
The measured stability was quantified by examining the spread in measured
scaling for regions across the \ac{DR1} footprint, which will be sensitive to
positional and temporal variations.  Figure~\ref{fig:flux_stability_results}
indicates that SIR flux variations are of the order of 0.7\%, consistent
with meeting the mission requirement.

\section{Conclusions and planned future work}
\label{sec:conclusion}

In this paper, the SIR geometric and flux calibration pipelines have been
detailed, describing their overall architecture, the nature of the generated
calibration products (part of the \ac{DR1}) and their functional
interdependencies.  The consistency of this calibration chain is the primary
factor determining the final accuracy in both wavelength and flux for the
spectra produced by the SIR science pipelines \citepalias{DR1-TP023}.
Sections~\ref{sec:OPT_DQC} to \ref{sec:IDS_DQC} demonstrated that the geometric
calibrations, which directly influence the wavelength precision, remain within
the mission requirements; correspondingly, Sect.~\ref{sec:AFX_DQC} showed a
comparable level of quality for the spectro-photometric calibrations, which
govern the accuracy of the flux calibration.

While data products delivered for \ac{DR1} meet the baseline requirements
defined in the top-level \Euclid Mission Requirements Document
\citep{DR1-TP026}, 
our analysis identifies specific areas where further optimisation would
significantly enhance the spectral quality for \ac{DR2}.  Our current
development efforts for the upcoming release are focused on the following key
areas.

To improve spectral localisation and ensure a more precise wavelength
zero-point calibration, the interplay between the FPA and OPT/CRV models is
currently being refined.  Key developments are underway:
\begin{description}
\item[FPA model optimisation:] a dedicated calibration pipeline for the FPA
  model, based on NIR photometric data, is being implemented.  This pipeline
  could go beyond rigid translations and also derives rotation and scale
  factors per detector.  This approach aims to enhance the overall detector
  metrology in the \ac{FP} and geometric \ac{FP}-based model accuracy.
\item[\ac{GWA} jitter analysis:] the impact of \ac{GWA} positioning
  uncertainty, estimated to be $\lesssim \ang{0.1;;}$-RMS, on the OPT and CRV
  models is being quantified.

  As mentioned in Sect.~\ref{sec:CRV}, it was observed in the \ac{DR1}
  processing that the CRV model still required a frame-by-frame sub-pixel
  adjustment at the detector level \citepalias{DR1-TP023}; this correction
  appeared essential to ensure that 2D spectrograms are accurately located,
  enabling the rigorous application of optimal extraction, very sensitive to
  the miscentering of the assumed cross-dispersion profile, and ultimately
  preventing flux inaccuracies.  Research is therefore underway on the proper
  way to model and include this \ac{GWA} jitter correction into the OPT and CRV
  geometric models.
\end{description}

Wavelength calibration (IDS) accuracy is being improved through two
complementary strategies:
\begin{description}
\item[Model stabilisation:] due to the sparse distribution of calibration
  sources -- approximately 80~\ac{PN} positions in the \ac{FP} per grism-tilt
  configuration (see Fig.~\ref{fig:ids_pn_cross}) --, global polynomial
  expansions (such as Eq.~\ref{eq:ids-glob}) are susceptible to Runge's
  phenomenon, leading to unphysical oscillations at the edges of the
  interpolation domain.  To mitigate this, stabilisation techniques and
  regularisation methods are being introduced to ensure a smooth and reliable
  solution across the entire~\ac{FP}.
\item[Emission-line centroiding:] to improve the measurement of the \ac{PN}
  emission lines, cosmic-ray rejection techniques and optimal extraction of the
  spectra are being deployed.  These refinements are expected to yield more
  precise reference line positions, leading to a better wavelength solution.
\end{description}

Although no specific updates are planned for the relative flux calibration
(RFX), this pipeline component will directly benefit from scheduled
improvements to the science pipeline \citepalias{DR1-TP023}, namely a better
background subtraction and a more precise masking of zeroth-order spectrograms.

For the absolute flux calibration (AFX), several improvements are planned for
\ac{DR2}, including:
\begin{description}
\item[Variation of the grism passband:] based on an analysis of the way in
  which the actual wavelengths of the edges of the sensitivity curve are
  dependent on the position in the \ac{FP}, we have developed a grism-dependent
  strategy to mask (as `unusable' and `suspicious') domains near the edge of
  the band where this effect is most prominent.
\item[`Suspicious' AFX-scaling masking:] the spectral domain over which the AFX
  scaling is deemed `suspicious' will change from \SIrange{25}{85}{\%} of the
  band-edge throughput for \ac{DR1}, to \SIrange{25}{50}{\%} in \ac{DR2}.  When
  combined with the extra masking discussed in the previous bullet, the net
  effect is to free-up a few spectroscopic samples across all grisms at the
  edges of the band. It was considered that the DR1 limits were more
  conservative than they needed to be.
\item[AFX calibration cross-checks:] check the absolute calibration, based on
  the primary calibrator GRW+70\,5824, with other spectrophotometric standard
  stars, including several faint \acp{WD} which lie in the self-cal/EDF-N and
  EDF-S fields (Deustua et al. in prep.).  This is similar to the strategy used
  by OU-NIR.
\end{description}

It is important to highlight that the \ac{DR1} calibrations are exclusively
computed from \ac{PV} observations, and therefore do not account for any
potential time evolution in these models since then.  For the geometric part,
no major source of variability in the intrinsic optical properties -- directly
related to the physical characteristics of lenses and grisms, e.g. distortion
pattern, dispersive power -- has been identified.  This is supported by the
fact that the overall wavelength accuracy -- a direct consequence of the
quality of the geometric models -- is well controlled and does not degrade over
time \citepalias[see][]{DR1-TP023}.  Regarding flux calibration, the situation
is less straightforward, as some temporal evolution of the detectors has been
identified, mostly because of ice contamination which primarily affects
\ac{BGS}, but has some impact on \acp{RGS} as well \citep{DR1-TP003,
  DR1-TP026}.  However, the results shown in
Fig.~\ref{fig:flux_stability_results} indicate that the effect on the \acp{RGS}
does not induce a coherent and significant flux calibration offset over the DR1
data footprint.  Nevertheless, temporally evolving zero points due to ice
contamination are being monitored and will be accounted for in future data
releases.  We estimate that the impact on \ac{RGS} spectra is $\lesssim 1\%$
over \ac{DR1} (consistent with the results in
Fig~\ref{fig:flux_stability_results}), with variations potentially as large as
5 to 15\% for \ac{BGS}, particularly at the shortest wavelengths.

A temporal analysis of the calibration products will be carried out as part of
\ac{DR2}, based on the monthly self-calibration observations (for all but IDS
and AFX calibrations).  Regarding wavelength solution (IDS), we will conduct a
comparison between initial \ac{PN} calibration sequence (CALBLOCK-PV-009) and
more recent one (CALBLOCK-F-010); as mentioned above, absolute flux calibration
(AFX) will be cross-checked with regularly observed secondary calibrators.

The refinements currently being implemented for the \ac{DR2} calibration
products, coupled with upcoming enhancements to the SIR Science pipeline, are
expected to yield a significantly more stable and accurate spectroscopic data
set.  These improvements will minimise systematic uncertainties, ensuring that
the \ac{DR2} products reach the full scientific potential of the \Euclid
mission.

% Add the acknowledgement using the achnowledgements environment.
% Do not use \acknowledgement{....} as this affects the formatting
% of the references.

\begin{acknowledgements}
  Funded by the European Union – Next Generation EU, Mission 4 Component 1
  Large Scale Lab (LaScaLa), CUP C53D23001390006. 

  \AckDRone                      % DR1 release
  \AckEC                        % Standard Euclid

  This publication makes use of data products from the Two Micron All Sky
  Survey, which is a joint project of the University of Massachusetts and the
  Infrared Processing and Analysis Center/California Institute of Technology,
  funded by the National Aeronautics and Space Administration and the National
  Science Foundation.

  In the development of our pipeline, we acknowledge use of the Python
  libraries Numpy 
  \citep{2020Natur.585..357H}, Scipy 
  \citep{2020NatMe..17..261V},
  Matplotlib 
  \citep{2007CSE.....9...90H}, Astropy 
  \citep{astropy_2013, astropy_2018, astropy_2022} and
  Pandas 
  \citep{pandas-devpandas_2024}.
\end{acknowledgements}

% Here comes the reference list, generated via bibtex from
% your bibfile my.bib and Euclid.bib.  Please make sure that
% the same paper is not referenced twice, one from your my.bib
% file, and once from Euclid.bib.

\bibliography{Euclid, sir}

@ARTICLE{DR1-TP003,
        author ={{Euclid Collaboration: Polenta}, G. and {Frailis}, M. and {Alavi}, A. and others},
        title = "{Euclid Data Release 1 (DR1). Near-infrared processing and data products}",
        journal = {A\&A, submitted},
        year = 2026
        }

@ARTICLE{DR1-TP009,
        author ={{Euclid Collaboration: K\"ummel}, M. and {Caro}, F. and {Dimauro}, P. and others},
        title = "{Euclid Data Release (DR1). The MERge processing}",
        journal = {A\&A, submitted},
        year = 2026
        }

@ARTICLE{DR1-TP023,
        author ={{Euclid Collaboration: Scodeggio}, M. and {Copin}, Y. and others},
        title = "{Euclid Data Release 1 (DR1). From spectrograms to spectra: The SIR spectroscopic processing function}",
        journal = {in preparation},
        year = 2026
        }

@ARTICLE{DR1-TP026,
        author ={{Euclid Collaboration: Schirmer}, M. and {Paterson}, K. and {Alavi}, A. and others},
        title = "{Euclid Data Release (DR1). Calibration strategy and observations}",
        journal = {A\&A, submitted},
        year = 2026
        }

@ARTICLE{DR1-TP027,
        author ={{Euclid Collaboration: Tereno}, I. and {Dinis}, J. and {Dupac}, X. and others},
        title = "{The Euclid surveys at DR1}",
        journal = {arXiv, submitted},
        year = 2026
        }

@ARTICLE{DR1-TP033,
        author ={{Euclid Collaboration: Ruffenach}, M. and {Frailis}, M. and {Zacchei}, A. and others},
        title = "{Euclid Data Release 1 (DR1). An overview of the Science Ground Segment architecture and operations}",
        journal = {A\&A, submitted},
        year = 2026
        }

@ARTICLE{Q1-TP003,
       author = {{Euclid Collaboration: Polenta}, G. and {Frailis}, M. and {Alavi}, A. and others},
       title = "{Euclid Quick Data Release (Q1). III. NIR processing and data products}",
      journal = {\aap},
         year = 2026,
        month = jun,
       volume = {711},
          eid = {A3},
        pages = {A3},
          doi = {10.1051/0004-6361/202554657},
       adsurl = {https://ui.adsabs.harvard.edu/abs/2026A&A...711A...3E}
}

@ARTICLE{Q1-TP006,
       author = {{Euclid Collaboration: Copin}, Y. and {Fumana}, M. and {Mancini}, C. and others},
       title = "{Euclid Quick Data Release (Q1). VI. From spectrograms to spectra: The SIR spectroscopic processing function}",
      journal = {\aap},
         year = 2026,
        month = jun,
       volume = {711},
          eid = {A6},
        pages = {A6},
          doi = {10.1051/0004-6361/202554627},
       adsurl = {https://ui.adsabs.harvard.edu/abs/2026A&A...711A...6E}
}

@ARTICLE{EuclidSkyOverview,
author = {{Euclid Collaboration: Mellier}, Y. and {Abdurro'uf} and {Acevedo~Barroso}, J.A. and others},
	title = {Euclid - I. Overview of the Euclid mission},
	DOI= "10.1051/0004-6361/202450810",
	url= "https://doi.org/10.1051/0004-6361/202450810",
	journal = {A\&A},
	year = 2025,
	volume = 697,
	pages = "A1",
}

@ARTICLE{EuclidSkyNISP,
author = {{Euclid Collaboration: Jahnke}, K. and {Gillard}, W. and {Schirmer}, M. and others},
	title = {Euclid - III. The NISP Instrument},
	DOI= "10.1051/0004-6361/202450786",
	url= "https://doi.org/10.1051/0004-6361/202450786",
	journal = {A\&A},
	year = 2025,
	volume = 697,
	pages = "A3",
}

@misc{DR1cite,
author = "{Euclid Data Release 1}",
howpublished = "\url{https://doi.org/10.57780/esa-d079c51}",
year = 2026
}

@ARTICLE{EU-Kubik,
       author = {{Euclid Collaboration: Kubik}, B. and {Barbier}, R. and {Clemens}, J. and others},
       title = "{Euclid preparation: LXXX. Overview of Euclid infrared detector performance from ground tests}",
      journal = {\aap},
         year = 2026,
        month = mar,
       volume = {707},
          eid = {A230},
        pages = {A230},
          doi = {10.1051/0004-6361/202555777},
archivePrefix = {arXiv},
       eprint = {2507.11326},
 primaryClass = {astro-ph.IM},
       adsurl = {https://ui.adsabs.harvard.edu/abs/2026A&A...707A.230E}
}

@ARTICLE{EP-Gillard,
       author = {{Euclid Collaboration: Gillard}, W. and {Maciaszek}, T. and {Prieto}, E. and others}, 
        title = "{Euclid preparation: LXXVII. The NISP spectroscopy channel, on ground performance and calibration}",
      journal = {\aap},
         year = 2026,
        month = mar,
       volume = {707},
          eid = {A227},
        pages = {A227},
          doi = {10.1051/0004-6361/202555859},
archivePrefix = {arXiv},
       eprint = {2506.08378},
 primaryClass = {astro-ph.IM},
       adsurl = {https://ui.adsabs.harvard.edu/abs/2026A&A...707A.227E}
}

@ARTICLE{Paterson-EP32,
       author = {{Euclid Collaboration: Paterson}, K. and {Schirmer}, M. and {Copin}, Y. and others},
        title = "{Euclid preparation. XXVII. A UV-NIR spectral atlas of compact planetary nebulae for wavelength calibration}",
      journal = {\aap},
         year = 2023,
        month = jun,
       volume = {674},
          eid = {A172},
        pages = {A172},
          doi = {10.1051/0004-6361/202346252},
archivePrefix = {arXiv},
       eprint = {2303.15525},
 primaryClass = {astro-ph.GA},
       adsurl = {https://ui.adsabs.harvard.edu/abs/2023A&A...674A.172E}
}

@ARTICLE{Scaramella-EP1,
       author = {{Euclid Collaboration: Scaramella}, R. and {Amiaux}, J. and {Mellier}, Y. and others},
        title = "{Euclid preparation. I. The Euclid Wide Survey}",
      journal = {\aap},
         year = 2022,
        month = jun,
       volume = {662},
          eid = {A112},
        pages = {A112},
          doi = {10.1051/0004-6361/202141938},
archivePrefix = {arXiv},
       eprint = {2108.01201},
 primaryClass = {astro-ph.CO},
       adsurl = {https://ui.adsabs.harvard.edu/abs/2022A&A...662A.112E}
}

@ARTICLE{2006AJ....131.1163S,
       author = {{Skrutskie}, M.~F. and {Cutri}, R.~M. and {Stiening}, R. and {Weinberg}, M.~D. and {Schneider}, S. and {Carpenter}, J.~M. and {Beichman}, C. and {Capps}, R. and {Chester}, T. and {Elias}, J. and {Huchra}, J. and {Liebert}, J. and {Lonsdale}, C. and {Monet}, D.~G. and {Price}, S. and {Seitzer}, P. and {Jarrett}, T. and {Kirkpatrick}, J.~D. and {Gizis}, J.~E. and {Howard}, E. and {Evans}, T. and {Fowler}, J. and {Fullmer}, L. and {Hurt}, R. and {Light}, R. and {Kopan}, E.~L. and {Marsh}, K.~A. and {McCallon}, H.~L. and {Tam}, R. and {Van Dyk}, S. and {Wheelock}, S.},
        title = "{The Two Micron All Sky Survey (2MASS)}",
      journal = {\aj},
         year = 2006,
        month = feb,
       volume = {131},
       number = {2},
        pages = {1163-1183},
          doi = {10.1086/498708},
       adsurl = {https://ui.adsabs.harvard.edu/abs/2006AJ....131.1163S}
}

@ARTICLE{2011ApJ...743..138G,
       author = {{Gianninas}, A. and {Bergeron}, P. and {Ruiz}, M.~T.},
        title = "{A Spectroscopic Survey and Analysis of Bright, Hydrogen-rich White Dwarfs}",
      journal = {\apj},
         year = 2011,
        month = dec,
       volume = {743},
       number = {2},
          eid = {138},
        pages = {138},
          doi = {10.1088/0004-637X/743/2/138},
archivePrefix = {arXiv},
       eprint = {1109.3171},
 primaryClass = {astro-ph.SR},
       adsurl = {https://ui.adsabs.harvard.edu/abs/2011ApJ...743..138G}
}

@ARTICLE{2020NatMe..17..261V,
       author = {{Virtanen}, Pauli and {Gommers}, Ralf and {Oliphant}, Travis E. and {Haberland}, Matt and {Reddy}, Tyler and {Cournapeau}, David and {Burovski}, Evgeni and {Peterson}, Pearu and {Weckesser}, Warren and {Bright}, Jonathan and {van der Walt}, St{\'e}fan J. and {Brett}, Matthew and {Wilson}, Joshua and {Millman}, K. Jarrod and {Mayorov}, Nikolay and {Nelson}, Andrew R.~J. and {Jones}, Eric and {Kern}, Robert and {Larson}, Eric and {Carey}, C.~J. and {Polat}, {\.I}lhan and {Feng}, Yu and {Moore}, Eric W. and {VanderPlas}, Jake and {Laxalde}, Denis and {Perktold}, Josef and {Cimrman}, Robert and {Henriksen}, Ian and {Quintero}, E.~A. and {Harris}, Charles R. and {Archibald}, Anne M. and {Ribeiro}, Ant{\^o}nio H. and {Pedregosa}, Fabian and {van Mulbregt}, Paul and {SciPy 1. 0 Contributors}},
        title = "{SciPy 1.0: fundamental algorithms for scientific computing in Python}",
      journal = {Nature Methods},
         year = 2020,
        month = feb,
       volume = {17},
        pages = {261-272},
          doi = {10.1038/s41592-019-0686-2},
archivePrefix = {arXiv},
       eprint = {1907.10121},
 primaryClass = {cs.MS},
       adsurl = {https://ui.adsabs.harvard.edu/abs/2020NatMe..17..261V}
}

@ARTICLE{2020AJ....160...21B,
       author = {{Bohlin}, Ralph C. and {Hubeny}, Ivan and {Rauch}, Thomas},
        title = "{New Grids of Pure-hydrogen White Dwarf NLTE Model Atmospheres and the HST/STIS Flux Calibration}",
      journal = {\aj},
         year = 2020,
        month = jul,
       volume = {160},
       number = {1},
          eid = {21},
        pages = {21},
          doi = {10.3847/1538-3881/ab94b4},
archivePrefix = {arXiv},
       eprint = {2005.10945},
 primaryClass = {astro-ph.SR},
       adsurl = {https://ui.adsabs.harvard.edu/abs/2020AJ....160...21B}
}

@ARTICLE{2020Natur.585..357H,
       author = {{Harris}, Charles R. and {Millman}, K. Jarrod and {van der Walt}, St{\'e}fan J. and {Gommers}, Ralf and {Virtanen}, Pauli and {Cournapeau}, David and {Wieser}, Eric and {Taylor}, Julian and {Berg}, Sebastian and {Smith}, Nathaniel J. and {Kern}, Robert and {Picus}, Matti and {Hoyer}, Stephan and {van Kerkwijk}, Marten H. and {Brett}, Matthew and {Haldane}, Allan and {del R{\'\i}o}, Jaime Fern{\'a}ndez and {Wiebe}, Mark and {Peterson}, Pearu and {G{\'e}rard-Marchant}, Pierre and {Sheppard}, Kevin and {Reddy}, Tyler and {Weckesser}, Warren and {Abbasi}, Hameer and {Gohlke}, Christoph and {Oliphant}, Travis E.},
        title = "{Array programming with NumPy}",
      journal = {\nat},
         year = 2020,
        month = sep,
       volume = {585},
       number = {7825},
        pages = {357-362},
          doi = {10.1038/s41586-020-2649-2},
archivePrefix = {arXiv},
       eprint = {2006.10256},
 primaryClass = {cs.MS},
       adsurl = {https://ui.adsabs.harvard.edu/abs/2020Natur.585..357H}
}

@ARTICLE{2023A&A...674A...1G,
       author = {{Gaia Collaboration: Vallenari}, A. and {Brown}, A.~G.~A. and {Prusti}, T. and {de Bruijne}, J.~H.~J. and {Arenou}, F. and {Babusiaux}, C. and {Biermann}, M. and {Creevey}, O.~L. and {Ducourant}, C. and {Evans}, D.~W. and {Eyer}, L. and {Guerra}, R. and {Hutton}, A. and {Jordi}, C. and {Klioner}, S.~A. and {Lammers}, U.~L. and {Lindegren}, L. and {Luri}, X. and {Mignard}, F. and {Panem}, C. and {Pourbaix}, D. and {Randich}, S. and {Sartoretti}, P. and {Soubiran}, C. and {Tanga}, P. and {Walton}, N.~A. and {Bailer-Jones}, C.~A.~L. and {Bastian}, U. and {Drimmel}, R. and {Jansen}, F. and {Katz}, D. and {Lattanzi}, M.~G. and {van Leeuwen}, F. and {Bakker}, J. and {Cacciari}, C. and {Casta{\~n}eda}, J. and {De Angeli}, F. and {Fabricius}, C. and {Fouesneau}, M. and {Fr{\'e}mat}, Y. and {Galluccio}, L. and {Guerrier}, A. and {Heiter}, U. and {Masana}, E. and {Messineo}, R. and {Mowlavi}, N. and {Nicolas}, C. and {Nienartowicz}, K. and {Pailler}, F. and {Panuzzo}, P. and {Riclet}, F. and {Roux}, W. and {Seabroke}, G.~M. and {Sordo}, R. and {Th{\'e}venin}, F. and {Gracia-Abril}, G. and {Portell}, J. and {Teyssier}, D. and {Altmann}, M. and {Andrae}, R. and {Audard}, M. and {Bellas-Velidis}, I. and {Benson}, K. and {Berthier}, J. and {Blomme}, R. and {Burgess}, P.~W. and {Busonero}, D. and {Busso}, G. and {C{\'a}novas}, H. and {Carry}, B. and {Cellino}, A. and {Cheek}, N. and {Clementini}, G. and {Damerdji}, Y. and {Davidson}, M. and {de Teodoro}, P. and {Nu{\~n}ez Campos}, M. and {Delchambre}, L. and {Dell'Oro}, A. and {Esquej}, P. and {Fern{\'a}ndez-Hern{\'a}ndez}, J. and {Fraile}, E. and {Garabato}, D. and {Garc{\'\i}a-Lario}, P. and {Gosset}, E. and {Haigron}, R. and {Halbwachs}, J.-L. and {Hambly}, N.~C. and {Harrison}, D.~L. and {Hern{\'a}ndez}, J. and {Hestroffer}, D. and {Hodgkin}, S.~T. and {Holl}, B. and {Jan{\ss}en}, K. and {Jevardat de Fombelle}, G. and {Jordan}, S. and {Krone-Martins}, A. and {Lanzafame}, A.~C. and {L{\"o}ffler}, W. and {Marchal}, O. and {Marrese}, P.~M. and {Moitinho}, A. and {Muinonen}, K. and {Osborne}, P. and {Pancino}, E. and {Pauwels}, T. and {Recio-Blanco}, A. and {Reyl{\'e}}, C. and {Riello}, M. and {Rimoldini}, L. and {Roegiers}, T. and {Rybizki}, J. and {Sarro}, L.~M. and {Siopis}, C. and {Smith}, M. and {Sozzetti}, A. and {Utrilla}, E. and {van Leeuwen}, M. and {Abbas}, U. and {{\'A}brah{\'a}m}, P. and {Abreu Aramburu}, A. and {Aerts}, C. and {Aguado}, J.~J. and {Ajaj}, M. and {Aldea-Montero}, F. and {Altavilla}, G. and {{\'A}lvarez}, M.~A. and {Alves}, J. and {Anders}, F. and {Anderson}, R.~I. and {Anglada Varela}, E. and {Antoja}, T. and {Baines}, D. and {Baker}, S.~G. and {Balaguer-N{\'u}{\~n}ez}, L. and {Balbinot}, E. and {Balog}, Z. and {Barache}, C. and {Barbato}, D. and {Barros}, M. and {Barstow}, M.~A. and {Bartolom{\'e}}, S. and {Bassilana}, J.-L. and {Bauchet}, N. and {Becciani}, U. and {Bellazzini}, M. and {Berihuete}, A. and {Bernet}, M. and {Bertone}, S. and {Bianchi}, L. and {Binnenfeld}, A. and {Blanco-Cuaresma}, S. and {Blazere}, A. and {Boch}, T. and {Bombrun}, A. and {Bossini}, D. and {Bouquillon}, S. and {Bragaglia}, A. and {Bramante}, L. and {Breedt}, E. and {Bressan}, A. and {Brouillet}, N. and {Brugaletta}, E. and {Bucciarelli}, B. and {Burlacu}, A. and {Butkevich}, A.~G. and {Buzzi}, R. and {Caffau}, E. and {Cancelliere}, R. and {Cantat-Gaudin}, T. and {Carballo}, R. and {Carlucci}, T. and {Carnerero}, M.~I. and {Carrasco}, J.~M. and {Casamiquela}, L. and {Castellani}, M. and {Castro-Ginard}, A. and {Chaoul}, L. and {Charlot}, P. and {Chemin}, L. and {Chiaramida}, V. and {Chiavassa}, A. and {Chornay}, N. and {Comoretto}, G. and {Contursi}, G. and {Cooper}, W.~J. and {Cornez}, T. and {Cowell}, S. and {Crifo}, F. and {Cropper}, M. and {Crosta}, M. and {Crowley}, C. and {Dafonte}, C. and {Dapergolas}, A. and {David}, M. and {David}, P. and {de Laverny}, P. and {De Luise}, F. and {De March}, R.},
        title = "{Gaia Data Release 3. Summary of the content and survey properties}",
      journal = {\aap},
         year = 2023,
        month = jun,
       volume = {674},
          eid = {A1},
        pages = {A1},
          doi = {10.1051/0004-6361/202243940},
archivePrefix = {arXiv},
       eprint = {2208.00211},
 primaryClass = {astro-ph.GA},
       adsurl = {https://ui.adsabs.harvard.edu/abs/2023A&A...674A...1G}
}

@ARTICLE{2007CSE.....9...90H,
       author = {{Hunter}, John D.},
        title = "{Matplotlib: A 2D Graphics Environment}",
      journal = {Computing in Science and Engineering},
         year = 2007,
        month = may,
       volume = {9},
       number = {3},
        pages = {90-95},
          doi = {10.1109/MCSE.2007.55},
       adsurl = {https://ui.adsabs.harvard.edu/abs/2007CSE.....9...90H}
}

@article{astropy_2013,
  title = {Astropy: {{A}} Community {{Python}} Package for Astronomy},
  shorttitle = {Astropy},
  author = {{Astropy Collaboration: Robitaille}, Thomas P. and Tollerud, Erik J. and Greenfield, Perry and Droettboom, Michael and Bray, Erik and Aldcroft, Tom and Davis, Matt and Ginsburg, Adam and {Price-Whelan}, Adrian M. and Kerzendorf, Wolfgang E. and Conley, Alexander and Crighton, Neil and Barbary, Kyle and Muna, Demitri and Ferguson, Henry and Grollier, Fr{\'e}d{\'e}ric and Parikh, Madhura M. and Nair, Prasanth H. and Unther, Hans M. and Deil, Christoph and Woillez, Julien and Conseil, Simon and Kramer, Roban and Turner, James E. H. and Singer, Leo and Fox, Ryan and Weaver, Benjamin A. and Zabalza, Victor and Edwards, Zachary I. and Azalee Bostroem, K. and Burke, D. J. and Casey, Andrew R. and Crawford, Steven M. and Dencheva, Nadia and Ely, Justin and Jenness, Tim and Labrie, Kathleen and Lim, Pey Lian and Pierfederici, Francesco and Pontzen, Andrew and Ptak, Andy and Refsdal, Brian and Servillat, Mathieu and Streicher, Ole},
  year = {2013},
  month = oct,
  journal = {\aap},
  volume = {558},
  pages = {A33},
  issn = {0004-6361},
  doi = {10.1051/0004-6361/201322068},
  url = {http://adsabs.harvard.edu/abs/2013A%26A...558A..33A},
  urldate = {2016-10-20}
}

@article{astropy_2018,
  title = {The {{Astropy Project}}: {{Building}} an {{Open-science Project}} and {{Status}} of the v2.0 {{Core Package}}},
  shorttitle = {The {{Astropy Project}}},
  author = {{Astropy Collaboration: Price-Whelan}, A. M. and Sip{\H o}cz, B. M. and G{\"u}nther, H. M. and Lim, P. L. and Crawford, S. M. and Conseil, S. and Shupe, D. L. and Craig, M. W. and Dencheva, N. and Ginsburg, A. and VanderPlas, J. T. and Bradley, L. D. and {P{\'e}rez-Su{\'a}rez}, D. and {de Val-Borro}, M. and Aldcroft, T. L. and Cruz, K. L. and Robitaille, T. P. and Tollerud, E. J. and Ardelean, C. and Babej, T. and Bach, Y. P. and Bachetti, M. and Bakanov, A. V. and Bamford, S. P. and Barentsen, G. and Barmby, P. and Baumbach, A. and Berry, K. L. and Biscani, F. and Boquien, M. and Bostroem, K. A. and Bouma, L. G. and Brammer, G. B. and Bray, E. M. and Breytenbach, H. and Buddelmeijer, H. and Burke, D. J. and Calderone, G. and Cano Rodr{\'i}guez, J. L. and Cara, M. and Cardoso, J. V. M. and Cheedella, S. and Copin, Y. and Corrales, L. and Crichton, D. and D'Avella, D. and Deil, C. and Depagne, {\'E}. and Dietrich, J. P. and Donath, A. and Droettboom, M. and Earl, N. and Erben, T. and Fabbro, S. and Ferreira, L. A. and Finethy, T. and Fox, R. T. and Garrison, L. H. and Gibbons, S. L. J. and Goldstein, D. A. and Gommers, R. and Greco, J. P. and Greenfield, P. and Groener, A. M. and Grollier, F. and Hagen, A. and Hirst, P. and Homeier, D. and Horton, A. J. and Hosseinzadeh, G. and Hu, L. and Hunkeler, J. S. and Ivezi{\'c}, {\v Z}. and Jain, A. and Jenness, T. and Kanarek, G. and Kendrew, S. and Kern, N. S. and Kerzendorf, W. E. and Khvalko, A. and King, J. and Kirkby, D. and Kulkarni, A. M. and Kumar, A. and Lee, A. and Lenz, D. and Littlefair, S. P. and Ma, Z. and Macleod, D. M. and Mastropietro, M. and McCully, C. and Montagnac, S. and Morris, B. M. and Mueller, M. and Mumford, S. J. and Muna, D. and Murphy, N. A. and Nelson, S. and Nguyen, G. H. and Ninan, J. P. and N{\"o}the, M. and Ogaz, S. and Oh, S. and Parejko, J. K. and Parley, N. and Pascual, S. and Patil, R. and Patil, A. A. and Plunkett, A. L. and Prochaska, J. X. and Rastogi, T. and Reddy Janga, V. and Sabater, J. and Sakurikar, P. and Seifert, M. and Sherbert, L. E. and {Sherwood-Taylor}, H. and Shih, A. Y. and Sick, J. and Silbiger, M. T. and Singanamalla, S. and Singer, L. P. and Sladen, P. H. and Sooley, K. A. and Sornarajah, S. and Streicher, O. and Teuben, P. and Thomas, S. W. and Tremblay, G. R. and Turner, J. E. H. and Terr{\'o}n, V. and {van Kerkwijk}, M. H. and {de la Vega}, A. and Watkins, L. L. and Weaver, B. A. and Whitmore, J. B. and Woillez, J. and Zabalza, V. and {Astropy Contributors}},
  year = {2018},
  month = sep,
  journal = {\aj},
  volume = {156},
  pages = {123},
  issn = {0004-6256},
  doi = {10.3847/1538-3881/aabc4f},
  url = {http://cdsads.u-strasbg.fr/abs/2018AJ....156..123A},
  urldate = {2018-10-17}
}

@misc{astropy_2022,
  title = {The {{Astropy Project}}: {{Sustaining}} and {{Growing}} a {{Community-oriented Open-source Project}} and the {{Latest Major Release}} (v5.0) of the {{Core Package}}},
  shorttitle = {The {{Astropy Project}}},
  author = {{Astropy Collaboration: Price-Whelan}, Adrian M. and Lian Lim, Pey and Earl, Nicholas and Starkman, Nathaniel and Bradley, Larry and Shupe, David L. and Patil, Aarya A. and Corrales, Lia and Brasseur, C. E. and N{\"o}the, Maximilian and Donath, Axel and Tollerud, Erik and Morris, Brett M. and Ginsburg, Adam and Vaher, Eero and Weaver, Benjamin A. and Tocknell, James and Jamieson, William and {van Kerkwijk}, Marten H. and Robitaille, Thomas P. and Merry, Bruce and Bachetti, Matteo and G{\"u}nther, H. Moritz and Aldcroft, Thomas L. and {Alvarado-Montes}, Jaime A. and Archibald, Anne M. and B{\'o}di, Attila and Bapat, Shreyas and Barentsen, Geert and Baz{\'a}n, Juanjo and Biswas, Manish and Boquien, M{\'e}d{\'e}ric and Burke, D. J. and Cara, Daria and Cara, Mihai and E Conroy, Kyle and Conseil, Simon and Craig, Matthew W. and Cross, Robert M. and Cruz, Kelle L. and D'Eugenio, Francesco and Dencheva, Nadia and Devillepoix, Hadrien A. R. and Dietrich, J{\"o}rg P. and Davis Eigenbrot, Arthur and Erben, Thomas and Ferreira, Leonardo and {Foreman-Mackey}, Daniel and Fox, Ryan and Freij, Nabil and Garg, Suyog and Geda, Robel and Glattly, Lauren and Gondhalekar, Yash and Gordon, Karl D. and Grant, David and Greenfield, Perry and Groener, Austen M. and Guest, Steve and Gurovich, Sebastian and Handberg, Rasmus and Hart, Akeem and {Hatfield-Dodds}, Zac and Homeier, Derek and Hosseinzadeh, Griffin and Jenness, Tim and Jones, Craig K. and Joseph, Prajwel and Bryce Kalmbach, J. and Karamehmetoglu, Emir and Ka{\l}uszy{\'n}ski, Miko{\l}aj and Kelley, Michael S. P. and Kern, Nicholas and Kerzendorf, Wolfgang E. and Koch, Eric W. and Kulumani, Shankar and Lee, Antony and Ly, Chun and Ma, Zhiyuan and MacBride, Conor and Maljaars, Jakob M. and Muna, Demitri and Murphy, N. A. and Norman, Henrik and O'Steen, Richard and Oman, Kyle A. and Pacifici, Camilla and Pascual, Sergio and {Pascual-Granado}, J. and Patil, Rohit R. and Perren, Gabriel I and Pickering, Timothy E. and Rastogi, Tanuj and Roulston, Benjamin R. and Ryan, Daniel F and Rykoff, Eli S. and Sabater, Jose and Sakurikar, Parikshit and Salgado, Jes{\'u}s and Sanghi, Aniket and Saunders, Nicholas and Savchenko, Volodymyr and Schwardt, Ludwig and {Seifert-Eckert}, Michael and Shih, Albert Y. and Shrey Jain, Anany and Shukla, Gyanendra and Sick, Jonathan and Simpson, Chris and Singanamalla, Sudheesh and Singer, Leo P. and Singhal, Jaladh and Sinha, Manodeep and Sip{\H o}cz, Brigitta M. and Spitler, Lee R. and Stansby, David and Streicher, Ole and {\v S}umak, Jani and Swinbank, John D. and Taranu, Dan S. and Tewary, Nikita and Tremblay, Grant R. and {de Val-Borro}, Miguel and Van Kooten, Samuel J. and Vasovi{\'c}, Zlatan and Verma, Shresth and Cardoso, Jos{\'e} Vin{\'i}cius de Miranda and Williams, Peter K. G. and Wilson, Tom J. and Winkel, Benjamin and {Wood-Vasey}, W. M. and Xue, Rui and Yoachim, Peter and ZHANG, Chen and Zonca, Andrea},
  year = {2022},
  month = jun,
  journal = {arXiv e-prints},
  url = {https://ui.adsabs.harvard.edu/abs/2022arXiv220614220T},
  urldate = {2022-08-26}
}

@misc{pandas-devpandas_2024,
  title = {Pandas-Dev/Pandas: {{Pandas}}},
  shorttitle = {Pandas-Dev/Pandas},
  author = {{The Pandas development team}},
  year = {2024},
  month = sep,
  doi = {10.5281/zenodo.3509134},
  url = {https://zenodo.org/records/13819579},
  urldate = {2025-01-06},
  copyright = {BSD 3-Clause "New" or "Revised" License, Open Access},
  publisher = {Zenodo}
}

@misc{larry_bradley_2025_14889440,
  author       = {Larry Bradley and Brigitta Sip{\H o}cz and Thomas Robitaille
                  and Erik Tollerud and Z\`e Vin{\'{\i}}cius and Christoph Deil
                  and Kyle Barbary and Tom J Wilson and Ivo Busko and Axel
                  Donath and Hans Moritz G{\"u}nther and Mihai Cara and
                  P. L. Lim and Sebastian Me{\ss}linger and Zach Burnett and
                  Simon Conseil and Michael Droettboom and Azalee Bostroem and
                  E. M. Bray and Lars Andersen Bratholm and William Jamieson
                  and Adam Ginsburg and Geert Barentsen and Matt Craig and
                  Sergio Pascual and Shivangee Rathi and Marshall Perrin and
                  Brett M. Morris},
  title        = {astropy/photutils: 2.2.0},
  month        = feb,
  year         = 2025,
  publisher    = {Zenodo},
  version      = {2.2.0},
  doi          = {10.5281/zenodo.14889440},
  url          = {https://doi.org/10.5281/zenodo.14889440},
  swhid        = {swh:1:dir:11159107f27a28985192ed1118b1f2055709d093
                   ;origin=https://doi.org/10.5281/zenodo.596036;visi
                   t=swh:1:snp:ae8c4a55d349d43e53cfe9ce92e678fcfe840f
                   3b;anchor=swh:1:rel:0117f67e8888adcdfc85308287dd9c
                   854b466389;path=astropy-photutils-ffb96c5
                  },
}

@ARTICLE{1987PASP...99..191S,
       author = {{Stetson}, Peter B.},
        title = "{DAOPHOT: A Computer Program for Crowded-Field Stellar Photometry}",
      journal = {\pasp},
         year = 1987,
        month = mar,
       volume = {99},
        pages = {191},
          doi = {10.1086/131977},
       adsurl = {https://ui.adsabs.harvard.edu/abs/1987PASP...99..191S}
}

@INPROCEEDINGS{SIP2005,
       author = {{Shupe}, D.~L. and {Moshir}, Mehdrdad and {Li}, J. and {Makovoz}, D. and {Narron}, R. and {Hook}, R.~N.},
        title = "{The SIP Convention for Representing Distortion in FITS Image Headers}",
    booktitle = {Astronomical Data Analysis Software and Systems XIV},
         year = 2005,
       editor = {{Shopbell}, P. and {Britton}, M. and {Ebert}, R.},
       series = {Astronomical Society of the Pacific Conference Series},
       volume = {347},
        month = dec,
        pages = {491},
       adsurl = {https://ui.adsabs.harvard.edu/abs/2005ASPC..347..491S}
}

@ARTICLE{WCS2002,
       author = {{Calabretta}, M.~R. and {Greisen}, E.~W.},
        title = "{Representations of celestial coordinates in FITS}",
      journal = {\aap},
         year = 2002,
        month = dec,
       volume = {395},
        pages = {1077-1122},
          doi = {10.1051/0004-6361:20021327},
archivePrefix = {arXiv},
       eprint = {astro-ph/0207413},
 primaryClass = {astro-ph},
       adsurl = {https://ui.adsabs.harvard.edu/abs/2002A&A...395.1077C}
}

% Now you can add appendices.
% In this example, the appendices are in one column mode.
% If that is not requires, comment out \onecolumn
% Note that appendices in A\&A come {\it after\/} the references.

\begin{acronym}
  % Only capitalize as needed
  \acro{BAO}{baryon acoustic oscillations} %
  \acro{BGS}{blue grism} %
  \acro{DQC}{data quality control} %
  \acro{DR1}{Data Release~1}
  \acro{DR2}{Data Release~2}
  \acro{EDS}{Euclid Deep Survey} %
  \acro{EWS}{Euclid Wide Survey} %
  \acro{FP}{focal plane} %
  \acro{IQR}{inter-quartile range} %
  \acro{GWA}{grism-wheel assembly} %
  \acro{MDB}{mission database} %
  \acro{NISP}{Near-Infrared Spectrometer and Photometer}
  \acro{PN}{planetary nebula} %
  \acro{PV}{performance-verification} %
  \acro{PSF}{point-spread function} %
  \acro{QE}{quantum efficiency} %
  \acro{sMAD}{normally-scaled median absolute deviation} %
  \acro{RGS}{red grism} %
  \acro{RMS}{root mean square} %
  \acro{ROS}{reference observing sequence} %
  \acro{RSD}{redshift-space distortions} %
  \acro{S/N}{signal-to-noise ratio} %
  \acro{WCS}{world coordinate system} %
  \acro{WD}{white dwarf}
\end{acronym}

\begin{appendix}
\end{appendix}

\label{LastPage}
\end{document}